\documentclass[fleqn,usenatbib]{mnras}

\usepackage{newtxtext,newtxmath}

\usepackage[T1]{fontenc}

\DeclareRobustCommand{\VAN}[3]{#2}
\let\VANthebibliography\thebibliography
\def\thebibliography{\DeclareRobustCommand{\VAN}[3]{##3}\VANthebibliography}

\usepackage{graphicx}	
\usepackage{amsmath}	
\usepackage{listings}
\usepackage{color}
\definecolor{dkgreen}{rgb}{0,0.6,0}
\definecolor{gray}{rgb}{0.5,0.5,0.5}
\definecolor{mauve}{rgb}{0.58,0,0.82}
\definecolor{golden}{rgb}{0.86,0.65,0.01}
\title[Birth of the ELMs]{Birth of the ELMs: a ZTF survey for evolved cataclysmic variables turning into extremely low-mass white dwarfs}

\author[El-Badry et al.]{
Kareem El-Badry,$^{1,2,3,6}$\thanks{E-mail: kareem.el-badry@cfa.harvard.edu}
Hans-Walter Rix,$^{3}$
Eliot Quataert,$^{4}$
Thomas Kupfer,$^{5}$
Ken J. Shen$^{6}$
\\
$^{1}$Center for Astrophysics | Harvard \& Smithsonian, 60 Garden Street, Cambridge, MA 02138, USA \\
$^{2}$Harvard Society of Fellows, Harvard University, 78 Mt. Auburn Street, Cambridge, MA 02138, USA \\
$^{3}$Max-Planck Institute for Astronomy, K\"onigstuhl 17, D-69117 Heidelberg, Germany\\
$^{4}$Department of Astrophysical Sciences, Princeton University, Princeton, NJ 08544, USA\\
$^{5}$Department of Physics \& Astronomy, Texas Tech University, P.O. Box 41051, Lubbock, TX 79409, USA\\
$^{6}$Department of Astronomy and Theoretical Astrophysics Center, University of California Berkeley, Berkeley, CA 94720, USA\\
}

\date{Submitted to MNRAS}

\pubyear{2021}

\begin{document}
\label{firstpage}
\pagerange{\pageref{firstpage}--\pageref{lastpage}}
\maketitle

\begin{abstract}
We present a systematic survey for mass-transferring and recently-detached cataclysmic variables (CVs) with evolved secondaries, which are progenitors of extremely low mass white dwarfs (ELM WDs), AM CVn systems, and detached ultracompact binaries. We select targets below the main sequence in the {\it Gaia} color-magnitude diagram with ZTF light curves showing large-amplitude ellipsoidal variability and orbital period $P_{\rm orb} < 6$\,hr. This yields 51 candidates brighter than $G=18$, of which we have obtained many-epoch spectra for 21. We confirm all 21 to be completely-- or nearly--Roche lobe filling close binaries. 13 show evidence of ongoing mass transfer, which has likely just ceased in the other 8. Most of the secondaries are hotter than any previously known CV donors, with temperatures $4700<T_{{\rm eff}}/{\rm K}<8000$. Remarkably, all secondaries with $T_{\rm eff} \gtrsim 7000\,\rm K$ appear to be detached, while all cooler secondaries are still mass-transferring. This transition likely marks the temperature where magnetic braking becomes inefficient due to loss of the donor's convective envelope. Most of the proto-WD secondaries have masses near $0.15\,M_{\odot}$; their companions have masses near $0.8\,M_{\odot}$.  We infer a space density of $\sim 60\,\rm kpc^{-3}$, roughly 80 times lower than that of normal CVs and three times lower than that of ELM WDs. The implied Galactic birth rate, $\mathcal{R}\sim 60\,\rm Myr^{-1}$, is half that of AM CVn binaries. Most systems are well-described by MESA models for CVs in which mass transfer begins only as the donor leaves the main sequence. All are predicted to reach minimum periods $5\lesssim P_{{\rm orb}}/{\rm min}\lesssim30$ within a Hubble time, where they will become AM CVn binaries or merge. This sample triples the known evolved CV population and offers broad opportunities for improving understanding of the compact binary population.   
\end{abstract}




\section{Introduction}
\label{sec:intro}
Cataclysmic variables (CVs) are short-period binaries in which a low-mass star transfers mass to a white dwarf (WD) through stable Roche lobe overflow \citep[RLOF; see][for a review]{Warner_2003}. At orbital periods $P_{\rm orb}\lesssim 6$ hours, the mass-losing stars in most CVs fall on a tight ``donor sequence'' of mass, radius, spectral type, and luminosity as a function of period \citep{Patterson_1984, Beuermann_1998, Smith_1998, Knigge_2006, Knigge_2011, Abrahams2020}. This sequence, and the distribution of CV periods along it, has proved foundational for the development and testing of models for mass transfer in interacting binaries. 

By definition, the donor stars in CVs have radii equal to their Roche lobe radii. This imposes a (nearly) deterministic relation between the orbital period, $P_{\rm orb}$, and the donor's mean density, $\overline{\rho}_{\rm donor}$:
\begin{align}
    \label{eq:rhostar}
    \overline{\rho}_{\rm donor} \approx107\,{\rm g\,cm^{-3}}\left(P_{\rm orb}/{\rm hr}\right)^{-2},
\end{align}
with very weak dependence on the mass of either component.\footnote{Equation~\ref{eq:rhostar} is calculated from the \citet{Eggleton_1983} fitting formula for the Roche lobe equivalent radius and is accurate within 6\% for mass ratios $0.01 <q < 1$.} Here $\overline{\rho}_{\rm donor}=3M_{\rm donor} /\left (4\pi R_{\rm donor}^3\right)$, with $M_{\rm donor}$ and $R_{\rm donor}$ the mass and equivalent radius of the mass-losing star. Most CV donors are only mildly out of thermal equilibrium due to mass loss, so their radii and effective temperatures differ only modestly from those of single main sequence stars of the same mass \citep{Knigge_2006}. The long-term evolution of CVs is governed foremost by angular momentum loss through magnetic braking and gravitational radiation, which shrink the orbits of CVs and in turn reduce the mass of the donors. The CV donor sequence is thus an evolutionary track, with initially massive donors in longer-period  CVs evolving along the sequence to short periods and low masses. 

A small fraction of CV donors are observed to deviate significantly from the donor sequence, with unusually high effective temperatures and luminosities for their orbital period \citep[e.g.][]{Thorstensen_2002, Thorstensen_2002b}. These objects are rare compared to normal CVs, but about a dozen have been identified over the last two decades, mostly with K-type donor stars ($T_{\rm eff}\approx 4,500\,\rm K$, at periods where ordinary CV donors have $T_{\rm eff} \lesssim 3,500\,\rm K$). A particularly extreme member of this evolved CV population is the recently discovered binary LAMOST J0140, with an F-type donor at a period of 3.81 hours \citep{ElBadry2021}. 

Such overluminous donors are thought to form from CVs in which the donor's first Roche lobe overflow only occurs near the end of its main-sequence evolution, as it is beginning to form a helium core.  By the time they reach short periods, these ``evolved CV'' donors are predicted to consist of a helium core with a thick (few hundredths of a solar mass) hydrogen envelope, whose luminosity comes primarily from hydrogen shell burning. Their expected mass transfer rates are much lower than in normal CVs of similar period, and the donor effective temperatures are higher. Eventually, the most evolved donors are expected to detach from their Roche lobes, contracting and heating at near constant luminosity, until they become extremely low-mass (ELM) white dwarfs \citep[e.g.][]{Sun_2018}. The timescale for this contraction and evolution toward the WD cooling track is set by the remaining mass of burnable hydrogen in the envelope of the proto-ELM WD and is typically about a Gyr \citep[e.g.][]{Istrate2014}. 

The formation and evolution of such detached ELM WDs -- which have attracted attention as companions to millisecond pulsars \citep[e.g.][]{Driebe1998}, progenitors of ultracompact binaries \citep[e.g.][]{Tutukov1985, Tutukov1987, Podsiadlowski_2003} and thermonuclear supernovae \citep[e.g.][]{Livne_1990, Bildsten_2007}, and as some of the loudest gravitational wave sources in the LISA band \citep[e.g.][]{Hermes2012, Kupfer2018} -- has been an area of active study in the last decade. Much of what is known observationally about the ELM WD population is due to the ELM survey \citep{Brown2010}, which carried out targeted spectroscopic follow-up of ELM candidates based on SDSS colors. The survey's final sample \citep{Brown2020} contained 98 spectroscopically vetted double white dwarf binaries, including 79 in which at least one component has an inferred mass below $0.3\,M_{\odot}$. Essentially all ELM WDs appear to be in short-period binaries. Those discovered by the ELM survey have periods ranging from 0.0089 to 1.5 days, with a median period of 0.25 days. All of their companions are consistent with being other WDs, though some could also be neutron stars. 

The objects detected by the ELM survey all have effective temperatures $T_{\rm eff}> 8000\,\rm K$, and most have surface gravities $\log\left[g/\left({\rm cm\,s^{-2}}\right)\right] > 5.5$. This is largely a consequence of the color selection the survey employed. Cooler and more bloated ELM WDs -- objects that were until recently CVs with ongoing mass transfer -- are expected to exist. However, their temperatures and surface gravities are predicted to be similar to those of main sequence A and F stars, so they are not easily identified with color cuts alone \citep[e.g.][]{Pelisoli2018}.

This paper introduces a systematic search for these cooler and more bloated ELM WDs, and the evolved CVs from which they formed.  Although their colors are usually indistinguishable from those of main sequence stars, their smaller masses and radii cause them to fall below the main sequence. Periodic light curve variability due to tidal distortion allows us to distinguish them from low-metallicity main-sequence stars, which inhabit a similar region of the color-magnitude diagram. Throughout the paper, we refer to such nascent ELM WDs and unusually warm CV donors as ``proto-WDs'' and ``evolved CV donors'' interchangeably: we are interested in both objects with ongoing mass transfer and those that have recently terminated mass transfer.
 
The remainder of this paper is organized as follows. Section~\ref{sec:sample_selection} describes our selection of proto-ELM WD candidates, which is based on CMD position and ellipsoidal light curve variability. Section~\ref{sec:spec_sample} describes our follow-up spectroscopic observations and modeling to constrain system masses, radii, and luminosities. In most cases, joint modeling of light curves, spectra, radial velocities, parallaxes, and broadband spectral energy distributions allows us to obtain precise constraints on the masses and radii of the donor stars, and somewhat weaker constraints on the companion masses. Section~\ref{sec:results} compares the inferred physical parameters of the binaries in our sample to those of objects from other surveys and to binary evolution models. We discuss our survey's selection function and the space density of evolved CVs and proto-ELM WDs in Section~\ref{sec:discussion}, and we summarize our findings in Section~\ref{sec:conclusions}.

The Appendices provide additional details. Appendix~\ref{sec:rl_prior} describes the priors adopted in our inference of donor masses and radii. Appendix~\ref{sec:halpha} shows the H$\alpha$ line profiles of objects in our sample. Appendix~\ref{sec:evolution_long_term} explores the long-term brightness evolution and sources of photometric scatter in the observed light curves. Appendix~\ref{sec:nospec_sample} presents objects in the light curve-selected sample for which we have not yet obtained spectroscopic follow-up. Appendix~\ref{sec:overlap} explores the overlap of our sample with other surveys.

\section{Sample selection}
\label{sec:sample_selection}

\begin{figure*}
    \centering
    \includegraphics[width=\textwidth]{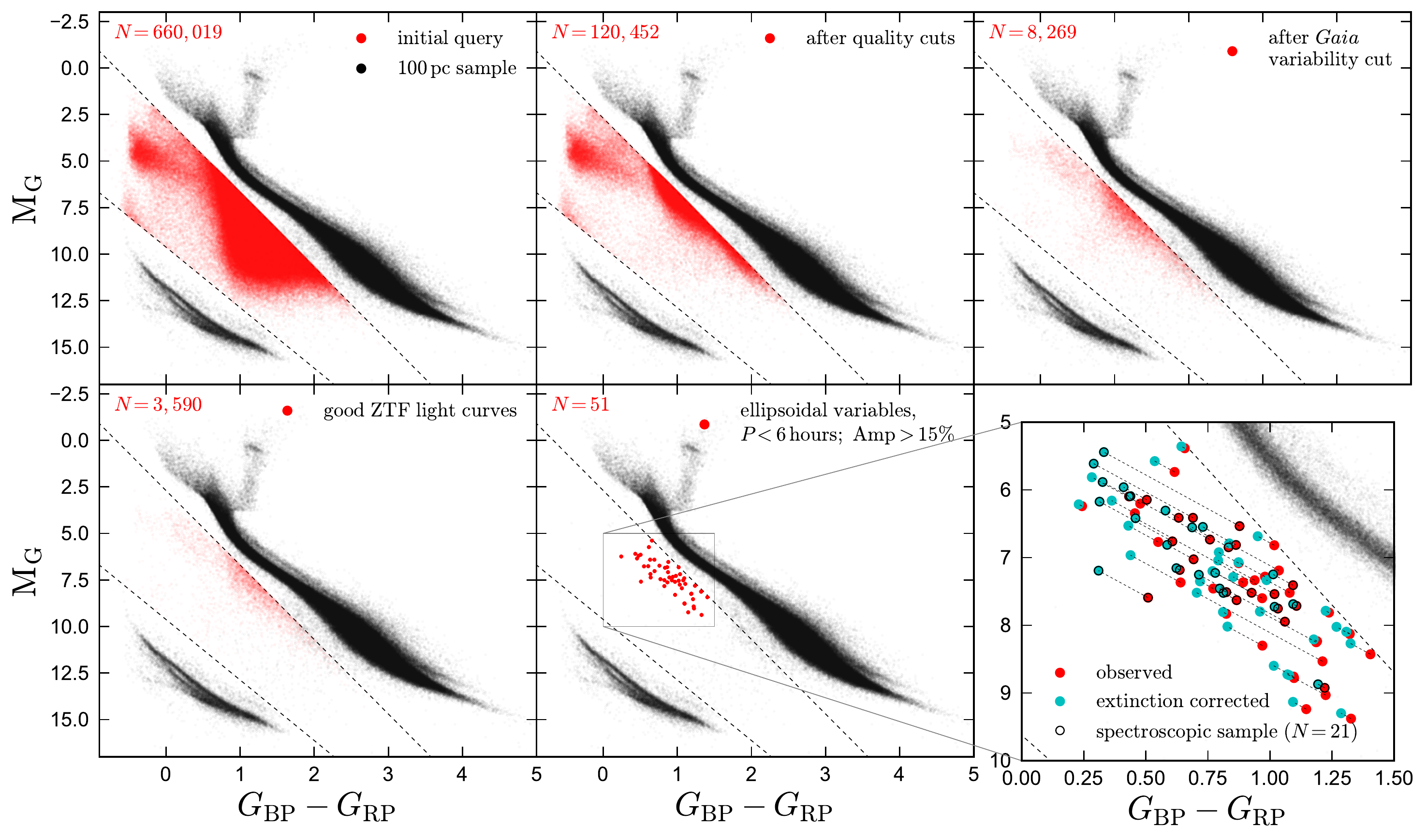}
    \caption{Sample selection in the color-magnitude diagram. In all panels, black points show the {\it Gaia} 100 pc sample for context. Red points show our candidates (not limited to 100 pc) at various stages of cleaning. We first (upper left) select all sources with precise parallaxes that fall between the main sequence and WD cooing sequence. Many of these sources have spurious parallaxes or colors; we apply astrometric and photometric quality cuts to remove these (upper middle). We then require significant scan-to-scan brightness variability as estimated from the {\it Gaia} flux RMS (upper right), and that sources have a good ZTF light curve (bottom left). Finally, we inspect the ZTF light curves individually and select objects dominated by ellipsoidal variability with $P_{\rm orb} < 6$ hours. Bottom right panel shows the final sample with (cyan) and without (red) an extinction/reddening correction. Objects for which we have obtained spectroscopic follow-up are marked with a border.}  
    \label{fig:sample_selection}
\end{figure*}

The objects we hope to find have similar temperatures and surface gravities to main sequence AFGK stars, but with lower masses and smaller radii. They are thus expected to fall below the main sequence in the color--absolute magnitude diagram (CMD). Normal CVs with unevolved donors also fall in this region of the CMD \citep[e.g.][]{Abril2020, Abrahams2020} -- not because their donors are below the main sequence, but because a disk around the accreting WD dominates the optical photometry. To filter out normal CVs, we target objects whose optical light curves are dominated by ellipsoidal variability of the donor star \citep[e.g.][]{WilsonSofia1976}, induced by tidal distortion from the companion. At short periods ($P_{\rm orb} < 6$ hours), this is only expected to occur if the donor is warmer and more luminous than in a normal CV \citep[e.g.][]{Rebassa_2014, Wakamatsu2021}. Our spectroscopic follow-up (Section~\ref{sec:spec_sample}) demonstrates that this approach efficiently eliminates normal CVs and other evolved-CV imposters, such as detached main sequence + WD binaries and metal-poor main sequence stars that naturally fall below the main sequence. 

\subsection{Gaia CMD selection}
\label{sec:gaia_cmd}
We began by selecting from {\it Gaia} eDR3 \citep{GaiaCollaboration2021} relatively bright sources ($G < 18$) with precise parallaxes ($\varpi/\sigma_{\varpi} > 5$) that fall between the main sequence and WD cooling sequence in the color-magnitude diagram. This was accomplished with the following ADQL query,

\begin{lstlisting}
select * from gaiaedr3.gaia_source
where parallax_over_error > 5
and phot_g_mean_mag < 18
and phot_g_mean_mag + 5 * log10(parallax / 100) < (3.25*bp_rp + 9.625)
and phot_g_mean_mag + 5 * log10(parallax / 100) > (4*bp_rp + 2.7)
and phot_rp_mean_flux_over_error > 5
and phot_bp_mean_flux_over_error > 5
and astrometric_sigma5d_max < 1
\end{lstlisting}
which returned 660,019 sources. These are plotted in the upper left panel of Figure~\ref{fig:sample_selection}, where for context we also plot a clean sample of {\it Gaia} sources within 100 pc \citep{GaiaCollaboration2021b}. The photometry and astrometry of the 100 pc sample has higher SNR and is cleaner (i.e., subject to more quality cuts) than a large majority of sources returned by our query. It is also limited to a smaller volume, and thus is less biased toward intrinsically bright sources. 

A large fraction of sources between the main sequence and WD cooling sequence -- a region of the CMD which, astrophysically, is expected to be sparsely populated -- are spurious. A tiny fraction of normal main sequence stars being scattered into this region may thus dominate the initial selection. We applied several quality cuts to minimize contamination from sources with unreliable parallaxes and/or colors. Within the initial query, the requirement of \texttt{astrometric\_sigma5d\_max < 1} removes sources for which a good astrometric fit could not be achieved with a single-star astrometric model   \citep[see][]{Fabricius2021, GaiaCollaboration2021b}, and the cuts of  \texttt{phot\_bp/rp\_mean\_flux\_over\_error > 5} ensure that the colors are not too noisy. 

To further reduce the number of sources with spurious astrometric solutions, we retrieved the astrometric fidelity parameter calculated by \citet[][``\texttt{fidelity\_v1}'']{Rybizki2021} for all sources returned by the initial query. This parameter is calculated based on 14 quantities reported in {\it Gaia} eDR3 using a neural network classifier that is trained on sources with both high-quality and manifestly incorrect parallaxes. We remove all sources for which \texttt{fidelity\_v1 < 0.75}, corresponding approximately to a $< 25\%$ probability of a spurious astrometric solution. This left us with 161,614 sources.\footnote{Our results are not very sensitive to the adopted threshold: we also tried 0.5, which gives an additional 21,060 candidates at this stage, but does not add a single good candidate to the final sample.}

Besides having spurious astrometry, sources can also appear below the main sequence if they have unreliable colors. {\it Gaia} colors are calculated by integrating over low-resolution spectra that are dispersed over a $3.5\times 2.1$ arcsec data acquisition window; they are therefore quite susceptible to blending by nearby sources. Various photometric flags are provided as part of {\it Gaia} eDR3 to identify sources with unreliable colors \citep{Evans2018, Riello2021}, but these also remove some sources in sparse regions of the sky with apparently good photometry. After some experimentation, we opted to remove all sources that have a companion within 6 arcsec with brighter $G$-band magnitude than the source itself. This removed an additional 41,162 sources, leaving us with 120,452 sources (upper middle panel of Figure~\ref{fig:sample_selection}). We verified that both this cut and the astrometric fidelity cut remove only a small fraction ($\sim 5\%$) of sources with reliable astrometry and photometry as judged by their CMD position (i.e., sources on the main sequence). We can thus safely use them to eliminate ``junk'' without seriously affecting the selection function of our search \citep[e.g.][]{Rix2021}.

We next wish to identify ellipsoidal variables. Before querying light curves for all sources, we applied a crude cut to filter out most non-variable sources using {\it Gaia} variability flags. Photometrically variable sources can be identified based on the flux ``error'' reported in the {\it Gaia} archive, which is calculated empirically from the scatter in flux measurements across multiple scans.\footnote{Specifically, the parameter \texttt{phot\_g\_mean\_flux\_error} represents the standard deviation of the single-epoch $G$-band fluxes divided by the square root of the number of visits, which is reported as \texttt{phot\_g\_n\_obs}.} Following \citet{Guidry2020}, we calculate a $G$-band variability metric, 
\begin{equation}
    \label{eq:varindex}
    V_{G}=\frac{\sigma_{F}}{\left\langle F\right\rangle }\sqrt{n_{{\rm obs}}},
\end{equation}
where $\sigma_{F}$, $\left\langle F\right\rangle $, and $n_{{\rm obs}}$ refer respectively to the parameters \texttt{phot\_g\_mean\_flux\_error}, \texttt{phot\_g\_mean\_flux}, and \texttt{phot\_g\_n\_obs} in the {\it Gaia} archive. Not all scan-to-scan flux variability is astrophysical: for faint sources, scatter due to photon noise is significant. We therefore compute a modified variability metric, 

\begin{equation}
    \tilde{V}_{G}=V_{G}-f\left(G\right),
\end{equation}
where $f(G)$ is a fitting function taken from \citet[][their Equation E1]{Guidry2020} that approximately represents the 99th percentile of $V_G$ at a given $G$ magnitude. We only retain sources with $\tilde{V}_G > 0.02$.\footnote{For bright sources with negligible photon noise, this represents an RMS flux variability greater than 2\%. For a well-sampled light curve and a sinusoidally varying source, the RMS variablility is $\sqrt{2}/4\approx 0.35$ times the peak-to-peak variability amplitude, so $\tilde{V}_G>0.02$ selects peak-to-peak variability amplitudes greater than 5.6\%. Given the {\it Gaia} eDR3 photometric precision, the cut of $\tilde{V}_G>0.02$ at $G=17$ ($G=18$) corresponds approximately to a peak-to-peak amplitude greater than 11\% (14.5\%) for a sinusoidally-varying source. } This cuts the sample to 8,269 sources, which are shown in the upper right panel of Figure~\ref{fig:sample_selection}.

\subsection{Light curve selection}
\label{sec:ztf_selection}
We next searched for ellipsoidal variables among these candidates. Of the 8,269 remaining sources, 4,819 have declination $\delta > -28$\,deg, such that they could potentially fall within the footprint of the Zwicky Transient Facility  \citep[ZTF;][]{Bellm2019, Graham2019, Masci2019}. We queried the public ZTF 4th data release, which was the most recent release at the time of our sample selection, for $g-$ and $r-$ band light curves of all these sources. This yielded light curves for 4,066 objects. Investigating the 753 sources that did {\it not} have light curves, we found that a majority of them are faint stars that are close (6-15 arcsec) to a very bright star ($G \lesssim 9$). We have no reason to expect genuine CVs or ELM WDs to be preferentially found in such a configuration, so we suspect that these are simply sources with blended {\it Gaia} photometry that are not really below the main sequence. 

In order to be able to reliably identify periodicity, we only considered sources for which the ZTF DR4 light curves contained at least 40 clean visits in at least one bandpass. This left us with 3,590 sources (bottom left panel of Figure~\ref{fig:sample_selection}). 

We inspected the ZTF light curves of these sources individually. About half show clear evidence of variability, with uncertainty-normalized RMS variability greater than 1.5 (where the expected value for a non-variable source is $\approx 1$). The fact that a significant number of apparently non-variable sources enter the sample reflects the fact that the {\it Gaia} flux errors are an imperfect proxy for variability. We find that increasing the $\tilde{V}_G$ threshold can significantly reduce the number of non-variable sources that enter the sample, but at the expense of completeness for true variable sources. 

To find periodic variables, we computed Lomb–Scargle periodograms \citep[e.g.][]{VanderPlas2018} on a period grid from 1 to 24 hours. We visually inspected the periodograms as well as the effects of folding the light curves to the 6 most significant periods. Many CVs and related objects are known to exhibit quasi-regular outbursts (``dwarf novae'') due to disk instability \citep[e.g.][]{Hameury_2020}. To reduce the chance of such outbursts confounding our period search, we also removed the brightest 5\% of points in each light curve and inspected periodograms computed from the remaining photometric points. 

\begin{figure*}
    \centering
    \includegraphics[width=\textwidth]{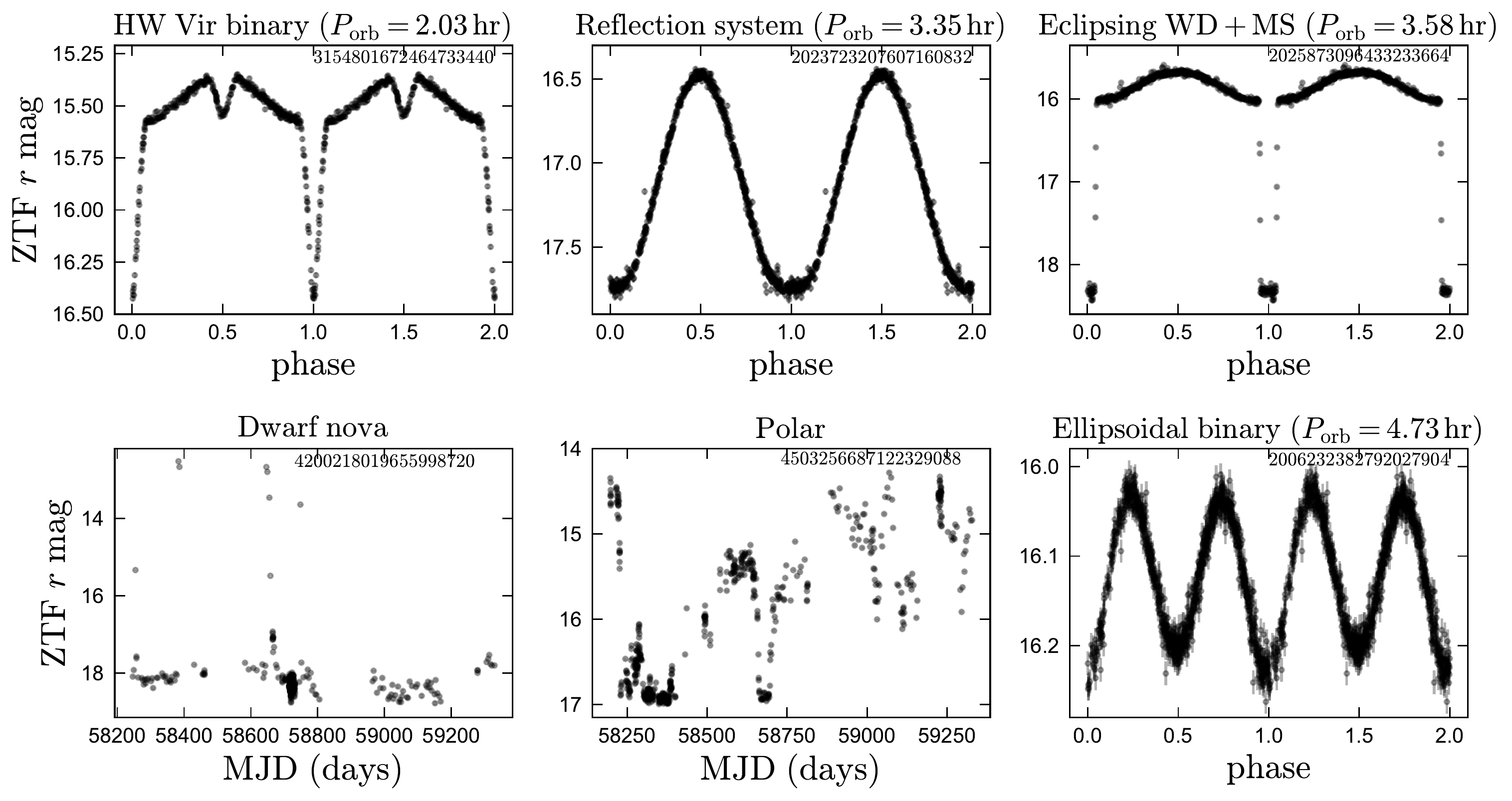}
    \caption{Examples of various classes of variables found between the main sequence and WD cooling track in the CMD (lower left panel of Figure~\ref{fig:sample_selection}). Upper left: HW Vir type binary, in which an M dwarf orbits a core helium burning hot subdwarf nearly edge-on. Upper center: reflection effect binary, in which an M dwarf orbits a hot white dwarf and is heated on one side. Upper right: eclipsing white dwarf + main sequence binary, also displaying a reflection effect. Bottom left: dwarf nova (CV) displaying outbursts and other variability due to changes in disk structure.  Bottom center: polar (i.e., a magnetic CV) displaying irregular variability on a range of timescales. Bottom right: ellipsoidal variable, our primary targets of interest. These can be distinguished from reflection systems by their unequal light curve minima. All objects are labeled by their {\it Gaia} eDR3 IDs. }
    \label{fig:various_lcs}
\end{figure*}

\begin{figure*}
    \centering
    \includegraphics[width=\textwidth]{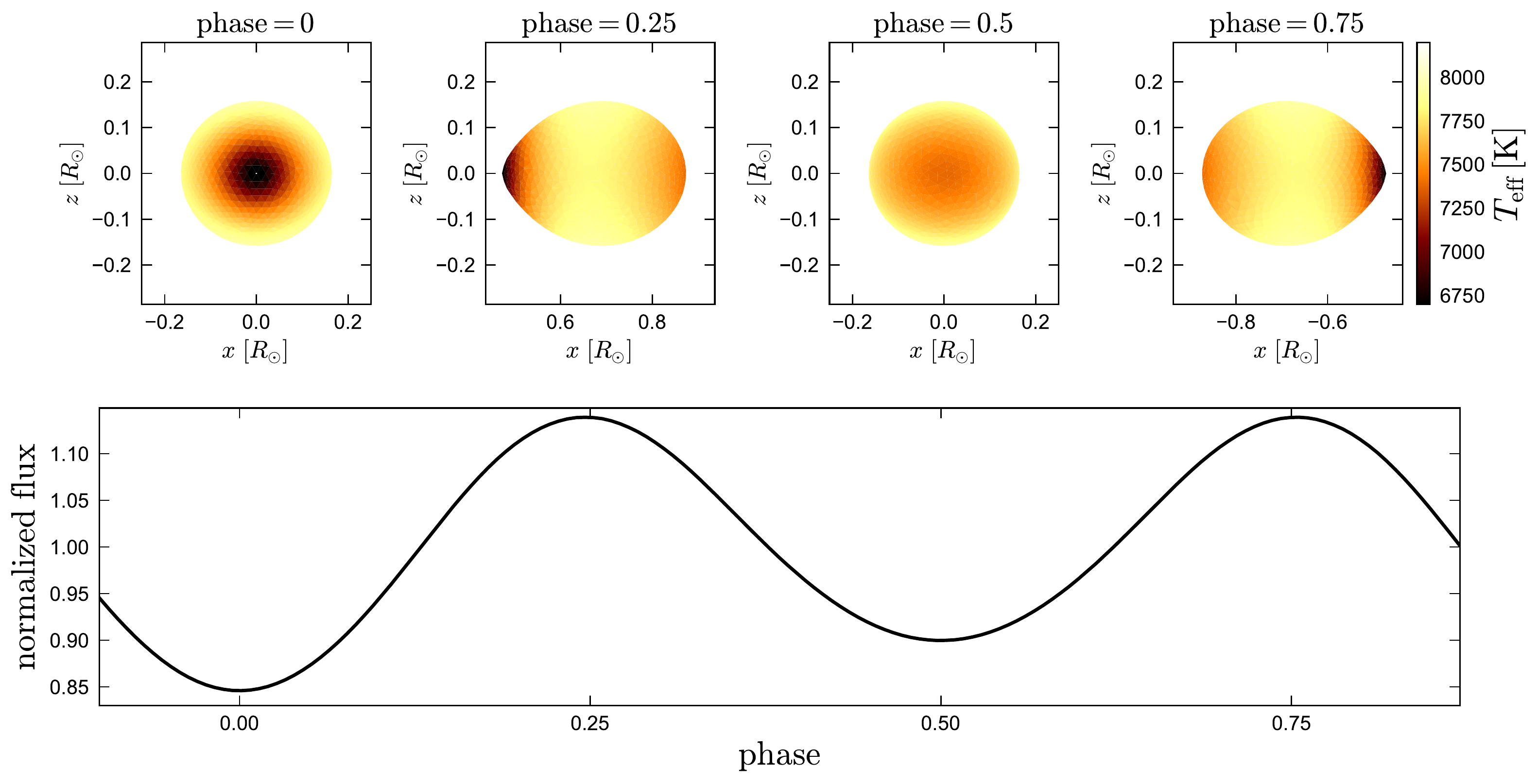}
    \caption{Schematic illustration of ellipsoidal variability using a \texttt{Phoebe} model for a Roche-filling star with similar parameters to objects in our sample. The geometric cross-section of the tidally-distorted donor is larger when it is viewed edge-on (phase 0.25 and 0.75) than when it is viewed end-on (phase 0 or 0.5). This gives rise to quasi-sinusoidal variability with maximum power at half the orbital period. Regions of the photosphere with lower surface gravity are cooler than higher-gravity regions; this ``gravity darkening'' gives rise to unequal light curve minima. The maximum peak-to-peak amplitude expected for pure ellipsoidal variation is about 30\%, with some dependence on the gravity- and limb-darkening law. }
    \label{fig:schematic}
\end{figure*}

The search revealed a menagerie of  light curve types, of which a few examples are shown in Figure~\ref{fig:various_lcs}.  Many sources show irregular variability on a range of timescales, characteristic of CVs and nova-like variables \citep[see][for a study of CVs with ZTF]{Szkody2020}. Other sources show deep, boxy eclipses; in this region of the CMD, these are mainly detached WD + main sequence binaries in which the WD is hot \citep[e.g.][]{Parsons2010}. Ellipsoidal variables -- what we are searching for -- show approximately sinusoidal variability but with brightness minima of alternating depths due to gravity darkening (see Figure~\ref{fig:schematic}). Non-eclipsing binaries containing a WD or hot subdwarf and a cool, main-sequence companion often show a quasi-sinusoidal reflection effect \citep[e.g.][]{Ratzloff2020}. These light curves appear similar to ellipsoidal variables but lack the alternating minima caused by gravity darkening. We find many such systems with $\rm M_{G} \lesssim 5$ and very blue colors; these likely contain hot subdwarfs (stripped core He-burning stars) rather than WDs. Among these are both pure reflection systems and ``HW Vir''-type systems showing eclipses \citep[e.g.][]{Schaffenroth2019, Enenstein2021}.  

To isolate sources likely to be evolved CVs and recently detached proto-ELM WDs, we searched specifically for sources  meeting the following criteria: 

\begin{enumerate}
    \item Evidence of ellipsoidal variability, with brightness minima of alternating depths (Figure~\ref{fig:schematic}). This ensures that the donor star, not an accretion disk, dominates the optical light curve, and thus selects systems with overluminous donors and/or underluminous disks. We do not insist that light curves shown {\it only} ellipsoidal variability. Some of our targets also show evidence of eclipses (likely from an accretion disk around a WD companion), long-term brightness evolution, and other scatter not captured by a pure ellipsoidal model.  
    
    \item Orbital periods $P_{\rm orb} < 6$ hours. This limit is  chosen to allow us to straightforwardly distinguish CVs with the most evolved donors from ``normal'' CVs, since it is only at $P_{\rm orb}\lesssim 6$ hours that most CV donors fall on a tight and well-defined donor sequence (see \citealt{Knigge_2006}; at longer periods, {\it all} donors are evolved). This limit also ensures that both stars are relatively small, and (in most cases) evolved, because binaries containing two main-sequence stars with orbital periods shorter than 5 hours are very rare \citep[e.g.][]{Rucinski2007, Jayasinghe2020}. This prevents, e.g., main-sequence contact binaries with spurious colors or parallaxes from entering the sample.

    \item Peak-to-peak variability amplitude greater than 15\% in the $g-$band. This selects binaries in which the tidally distorted star is nearly or completely Roche-lobe filling (Section~\ref{sec:light_curve_models}); in practice, it corresponds to a filling factor $R_{\rm donor}/R_{\rm Roche\,lobe} \gtrsim 0.8$.
    
    \item Similar variability amplitude in the $g$ and $r$ bands (within $20\%$). This excludes most systems in which a significant fraction of the optical light is contributed by a disk or by a hot WD companion; such systems have more dilution, and thus lower variability amplitude, in the $g$ band. 
    
    \item Not a hot subdwarf: we excluded a few ellipsoidal systems in the hot-subdwarf clump of the CMD, with $M_{G}<5$ and $G_{\rm BP}-G_{\rm RP} < 0.2$. Such systems are likely binaries containing a WD and a core helium burning star, with different evolutionary histories from the objects we seek.
    
\end{enumerate}

\begin{figure*}
    \centering
    \includegraphics[width=\textwidth]{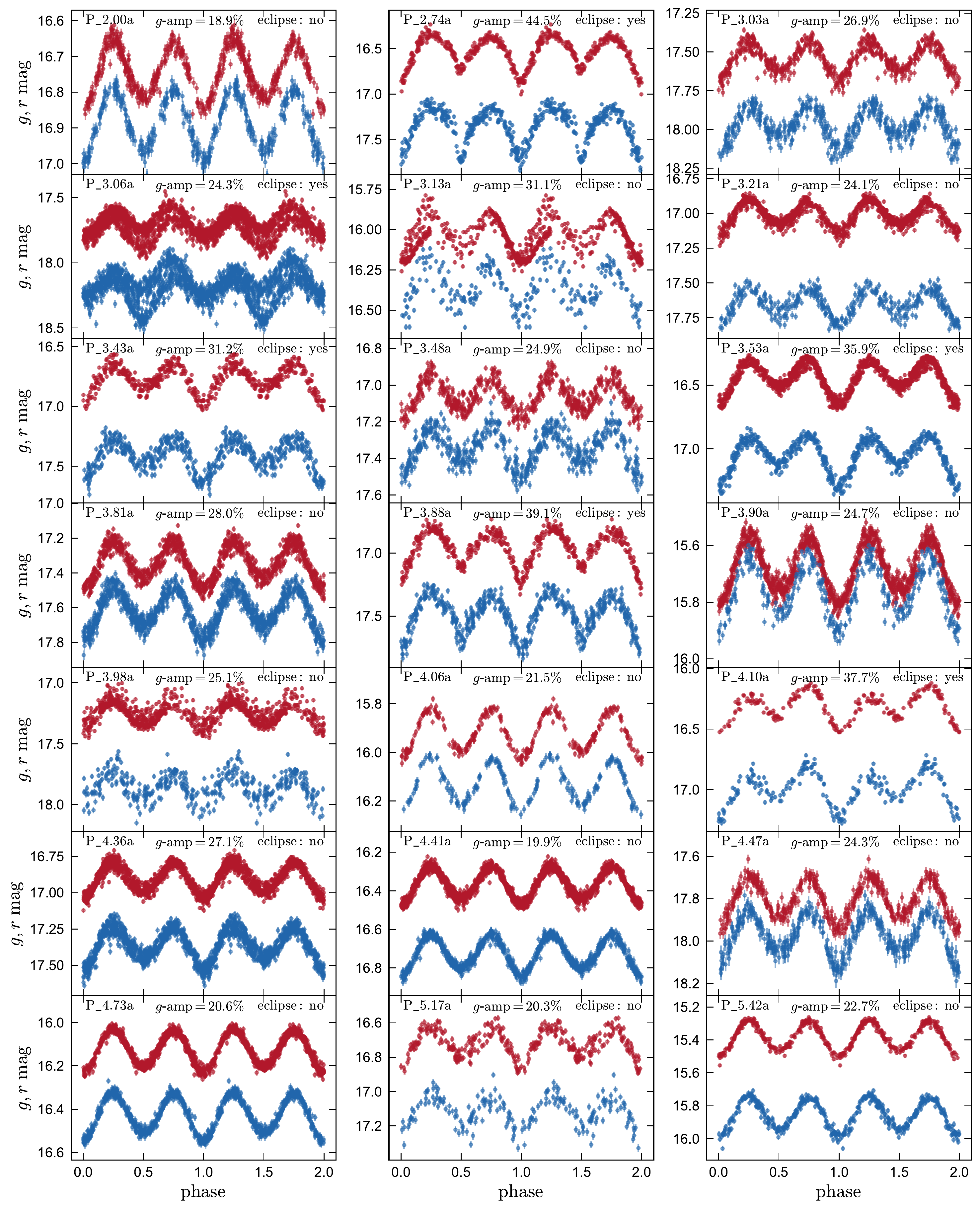}
    \caption{Phased ZTF $g$ (blue) and $r$ (red)  light curves for all targets in the spectroscopic sample. Targets are sorted by increasing orbital period. A diversity of light curve morphologies is evident. Some targets have light curves suggestive of ellipsoidal a variation with little intrinsic scatter (e.g. \texttt{P\_2.00a}). Others show evidence of eclipses (e.g. \texttt{P\_2.74a}). A majority show significant intrinsic scatter (e.g. \texttt{P\_3.03a}). }
    \label{fig:light_cuves}
\end{figure*}

Given our selection criteria, the most common false-positives with similar light curve shapes to ellipsoidal variables are non-eclipsing reflection effect binaries consisting of a hot, compact star and a cooler companion (typically an M dwarf). These can be distinguished from ellipsoidal variables because (a) their variability amplitudes are significantly larger in the $r$-band than in the $g$-band, and (b) they lack the unequal minima of ellipsoidal variables. Some single-mode pulsators also have light curve shapes similar to ellipsoidal variables, but these also lack unequal minima. 

\subsection{Final sample}
\label{sec:full_smap}

\begin{table*}
	\centering
	\caption{Basic properties of sources in the spectroscopic sample. ``ID'' (internal to this work) is based on orbital period in hours; we also list {\it Gaia} eDR3 IDs. $l$ and $b$ are Galactic longitude and latitude. $G$ is the {\it Gaia} mean magnitude. $P_{\rm orb}$ is the spectroscopically-confirmed orbital period. $\varpi_{\rm corrected}$ is the {\it Gaia} eDR3 parallax after correcting for the parallax zeropoint and underestimated parallax uncertainties. $E(g-r)$ is the reddening from the 3D dustmap from \citet{Green2019}. $g-$amplitude and $g-$scatter represent the flux-space peak-to-peak variability amplitude, and the intrinsic scatter on top of this variability, as determined by fitting the ZTF $g-$band light curve with a harmonic series (Section~\ref{sec:lightcurves}). The final columns indicates whether the light curve shows evidence of eclipses, and whether emission lines are apparent in our Kast spectra (Section~\ref{sec:spectra}). } 
	\label{tab:observables}
	\begin{tabular}{clllccccccc} 
		\hline
		ID & {\it Gaia} eDR3 ID & $l$ & $b$ &  $G$ & $P_{\rm orb}$ &  $\varpi_{\rm corrected}$ &  $E(g-r)$ & $g$-amplitude & $g$-scatter  & eclipse? \\
		& & [deg] & [deg] & [mag] & [hours] &  [mas] & [mag] & [\%] & [\%]  &  \\
\texttt{P\_2.00a} & 4393660804037754752 & 27.186816 & 24.692143 & 16.71 & 2.00 & $1.53 \pm 0.06$ & $0.15 \pm 0.01$ & $18.9 \pm 0.5$ & 0.5 & no \\
\texttt{P\_2.74a} & 861540207303947776 & 144.679729 & 51.707772 & 16.59 & 2.74 & $2.96 \pm 0.04$ & $0.02 \pm 0.01$ & $44.5 \pm 1.3$ & 5.8 & yes \\
\texttt{P\_3.03a} & 1030236970683510784 & 164.923089 & 37.466722 & 17.55 & 3.03 & $1.01 \pm 0.11$ & $0.02 \pm 0.01$ & $26.9 \pm 1.2$ & 3.2 & no \\
\texttt{P\_3.06a} & 2133938077063469696 & 80.998856 & 19.945344 & 17.82 & 3.06 & $0.94 \pm 0.08$ & $0.04 \pm 0.00$ & $24.3 \pm 1.0$ & 6.7 & yes \\
\texttt{P\_3.13a} & 4228735155086295552 & 48.406966 & -24.364330 & 16.06 & 3.13 & $1.59 \pm 0.06$ & $0.08 \pm 0.00$ & $31.1 \pm 1.7$ & 5.3 & no \\
\texttt{P\_3.21a} & 45968897530496384 & 176.054947 & -25.099680 & 17.15 & 3.21 & $1.23 \pm 0.09$ & $0.42 \pm 0.01$ & $24.1 \pm 1.0$ & 2.8 & no \\
\texttt{P\_3.43a} & 184406722957533440 & 170.832489 & 0.668255 & 16.79 & 3.43 & $1.73 \pm 0.08$ & $0.26 \pm 0.02$ & $31.2 \pm 1.1$ & 4.2 & yes \\
\texttt{P\_3.48a} & 4002359459116052864 & 226.959802 & 81.563599 & 17.13 & 3.48 & $1.05 \pm 0.09$ & $0.01 \pm 0.01$ & $24.9 \pm 1.0$ & 3.7 & no \\
\texttt{P\_3.53a} & 1965375973804679296 & 85.310555 & -7.402658 & 16.56 & 3.53 & $1.76 \pm 0.05$ & $0.01 \pm 0.03$ & $35.9 \pm 0.7$ & 2.5 & yes \\
\texttt{P\_3.81a} & 373857386785825408 & 123.946735 & -23.414504 & 17.35 & 3.81 & $0.68 \pm 0.11$ & $0.04 \pm 0.01$ & $28.0 \pm 0.6$ & 2.4 & no \\
\texttt{P\_3.88a} & 1077511538271752192 & 138.807482 & 40.385317 & 17.01 & 3.88 & $1.29 \pm 0.06$ & $0.11 \pm 0.01$ & $39.1 \pm 1.3$ & 4.2 & yes \\
\texttt{P\_3.90a} & 3053571840222008192 & 224.903771 & 4.361578 & 15.68 & 3.90 & $1.25 \pm 0.04$ & $0.08 \pm 0.02$ & $24.7 \pm 0.8$ & 2.1 & no \\
\texttt{P\_3.98a} & 896438328413086336 & 184.067846 & 20.533027 & 17.24 & 3.98 & $1.11 \pm 0.08$ & $0.06 \pm 0.00$ & $25.1 \pm 2.3$ & 6.3 & no \\
\texttt{P\_4.06a} & 4358250649810243584 & 13.966485 & 28.872606 & 15.88 & 4.06 & $1.54 \pm 0.04$ & $0.22 \pm 0.01$ & $21.5 \pm 0.8$ & 0.0 & no \\
\texttt{P\_4.10a} & 3064766376117338880 & 230.299867 & 17.978975 & 16.42 & 4.10 & $1.85 \pm 0.06$ & $0.01 \pm 0.01$ & $37.7 \pm 1.6$ & 3.9 & yes \\
\texttt{P\_4.36a} & 2126361067562200320 & 76.768521 & 12.207378 & 16.99 & 4.36 & $0.95 \pm 0.05$ & $0.10 \pm 0.01$ & $27.1 \pm 0.5$ & 2.5 & no \\
\texttt{P\_4.41a} & 2171644870571247872 & 94.656028 & -0.208116 & 16.37 & 4.41 & $1.21 \pm 0.04$ & $0.24 \pm 0.02$ & $19.9 \pm 0.3$ & 0.0 & no \\
\texttt{P\_4.47a} & 1315840437462118400 & 45.465896 & 45.091351 & 17.80 & 4.47 & $0.49 \pm 0.08$ & $0.07 \pm 0.01$ & $24.3 \pm 0.7$ & 0.0 & no \\
\texttt{P\_4.73a} & 2006232382792027904 & 102.761386 & -0.526714 & 16.12 & 4.73 & $1.16 \pm 0.04$ & $0.30 \pm 0.01$ & $20.6 \pm 0.3$ & 0.0 & no \\
\texttt{P\_5.17a} & 4382957882974327040 & 17.266003 & 28.142421 & 16.86 & 5.17 & $1.02 \pm 0.07$ & $0.11 \pm 0.00$ & $20.3 \pm 1.7$ & 3.8 & no \\
\texttt{P\_5.42a} & 286337708620373632 & 149.823713 & 14.253607 & 15.41 & 5.42 & $1.71 \pm 0.04$ & $0.41 \pm 0.02$ & $22.7 \pm 0.5$ & 1.2 & no \\
\hline
		\hline
	\end{tabular}
\end{table*}

Our search, concluding in visual light curve inspection, yielded 51 candidate evolved CVs and bloated proto-ELM WDs; these are shown on the CMD in the bottom center panel of Figure~\ref{fig:sample_selection}. The bottom right panel shows the same sample but includes corrections for extinction and reddening, which are modest but often not negligible ($E(B-V) \lesssim 0.4$). In contrast to the sample at earlier stages of the selection, most of the final sample does not run up against the upper boundary in the CMD that separates our selection region from the main sequence. The objects that do crowd against the boundary are a mix of non-variable contaminants and longer-period binaries. Nevertheless, our selection  does likely miss some evolved CVs closer to the main sequence (see Appendix~\ref{sec:overlap}).

Thus far, we have carried out spectroscopic follow-up of 21 of these 51 candidates. Their light curves are shown in Figure~\ref{fig:light_cuves}; light curves of the other 30 candidates are shown in Appendix~\ref{sec:nospec_sample}. {\it All} 21 objects for which we have obtained follow-up spectra appear to be genuine evolved CVs or recently detached proto-ELM WDs: there is no contamination from low-metallicity main sequence stars, detached main sequence + WD binaries, or normal CVs on the donor sequence. The purity of our sample is thus high. Its completeness is more complicated to quantity, as it is affected by the magnitude limit, CMD selection, light curve analysis, and targeting for spectroscopic follow-up.  We discuss the selection function of our sample further in Section~\ref{sec:selection_function}. 

We have also cross-matched our sample with published surveys for CVs and ELM WDs. We describe the results in Appendix~\ref{sec:overlap}. In brief, one of the 51 objects in our light curve selected sample has previously been recognized to be an ELM WD, and one to be an evolved CV. In addition, one of the objects in our spectroscopic sample, \texttt{P\_3.81a}, is the object Lamost J0140, which we analyzed in detail in \citet{ElBadry2021}. We retain the object in our sample here, since it was recovered by the CMD + light curve search described above.

\subsection{Light curves}
\label{sec:lightcurves}
Figure~\ref{fig:light_cuves}  shows that the light curves of objects in the final sample still exhibit some diversity. Some show pure ellipsoidal variability, with little evidence of intrinsic scatter. Others show the same shape expected for ellipsoidal variables, but with additional scatter; that is, the photometric uncertainties are smaller than the observed dispersion at a given phase. Several objects show eclipses. In most cases, these are manifest as a brightness minimum at phase 0 that is deeper than expected due to ellipsoidal variability; this occurs when a disk obscures the donor. In one case (\texttt{P\_2.74a}), an eclipse of the disk and/or WD companion is also apparent, at phase 0.5. 
 
To constrain the peak-to-peak variability amplitude and scatter for each target, we fit the ZTF $g-$band light curves with a 6-term Fourier model, which is sufficiently flexible to capture all the mean light curve shapes represented in the sample. Along with the Fourier amplitudes, we fit a scatter term, which is added in quadrature to the observational uncertainties in the likelihood function. This represents discrepancies between the model and data due to underestimated uncertainties, short-timescale scatter (i.e., pulsations or disk flickering), or long-term light curve evolution. The derived amplitudes and scatter values are reported in Table~\ref{tab:observables}, and in Table~\ref{tab:nospec} for objects not in the spectroscopic sample. A majority of objects have nonzero scatter. We think this is real, not a consequence of underestimated uncertainties, since most other ZTF sources with similar magnitude, color, and sky position have scatter consistent with 0 when we model them the same way.

In a few cases, the light curves show clear non-periodic evolution. One example is shown in Figure~\ref{fig:evolution_light_curve}, which shows an object that spontaneously becomes fainter by $\approx 0.07$ mag and changes its phased light curve shape,  presumably due to a change in the structure of the accretion disk. Two objects in the spectroscopic sample have been observed in outburst; their light curves are shown in Figure~\ref{fig:outburst_light_curve}. The outbursts resemble those commonly found in dwarf novae, but their recurrence timescale must be much longer than the typical $\approx$ month outburst timescale found in ordinary CVs at similar periods. 

\subsubsection{Source of light curve scatter}
\label{sec:signatures_emission}

\begin{figure*}
    \centering
    \includegraphics[width=\textwidth]{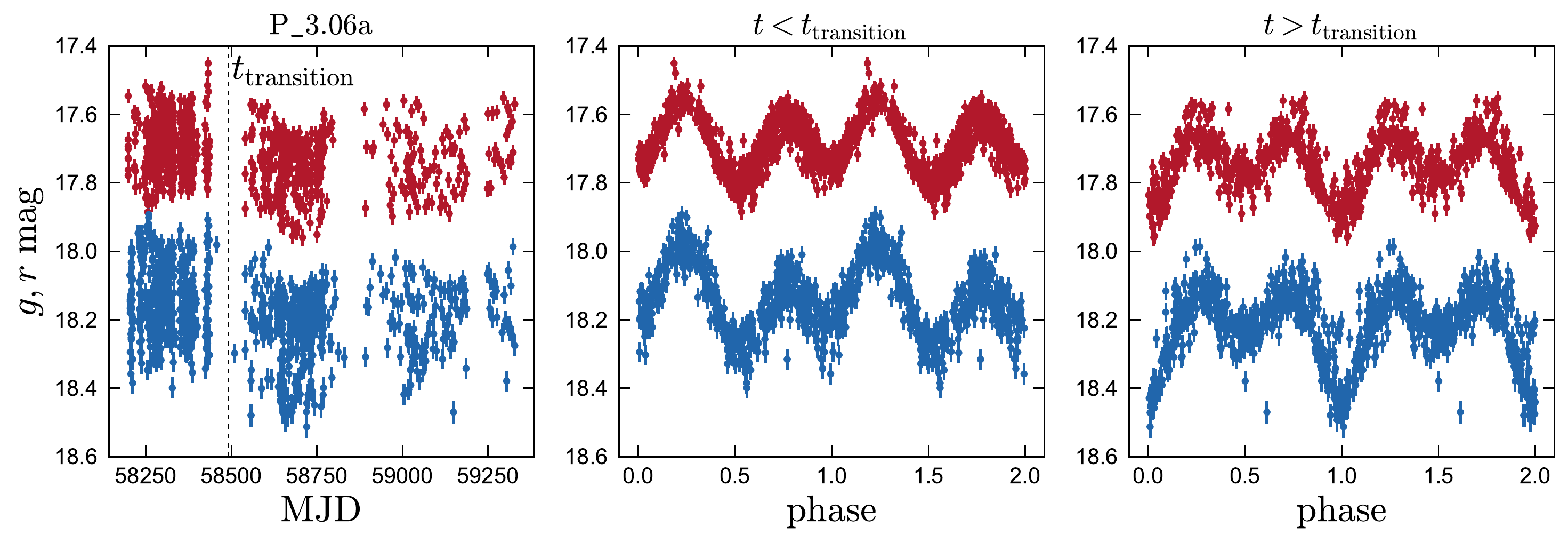}
    \caption{Light curve of \texttt{P\_3.06a}, the object with the clearest long-term evolution in the spectroscopic sample. Left panel shows the raw ZTF $g$- and $r$-band light curves.  A transition is evident at $\rm MJD\approx 58500$, when the mean brightness decreases by $\approx 0.1$ mag. The middle and right panels show phased light curves from before and after this transition, with phase defined the same way in both panels. An eclipse is evident in the right panel, presumably when the donor is obscured by the disk. The disk was likely brighter before the transition, when no eclipse was evident. } 
    \label{fig:evolution_light_curve}
\end{figure*}

\begin{figure*}
    \centering
    \includegraphics[width=\textwidth]{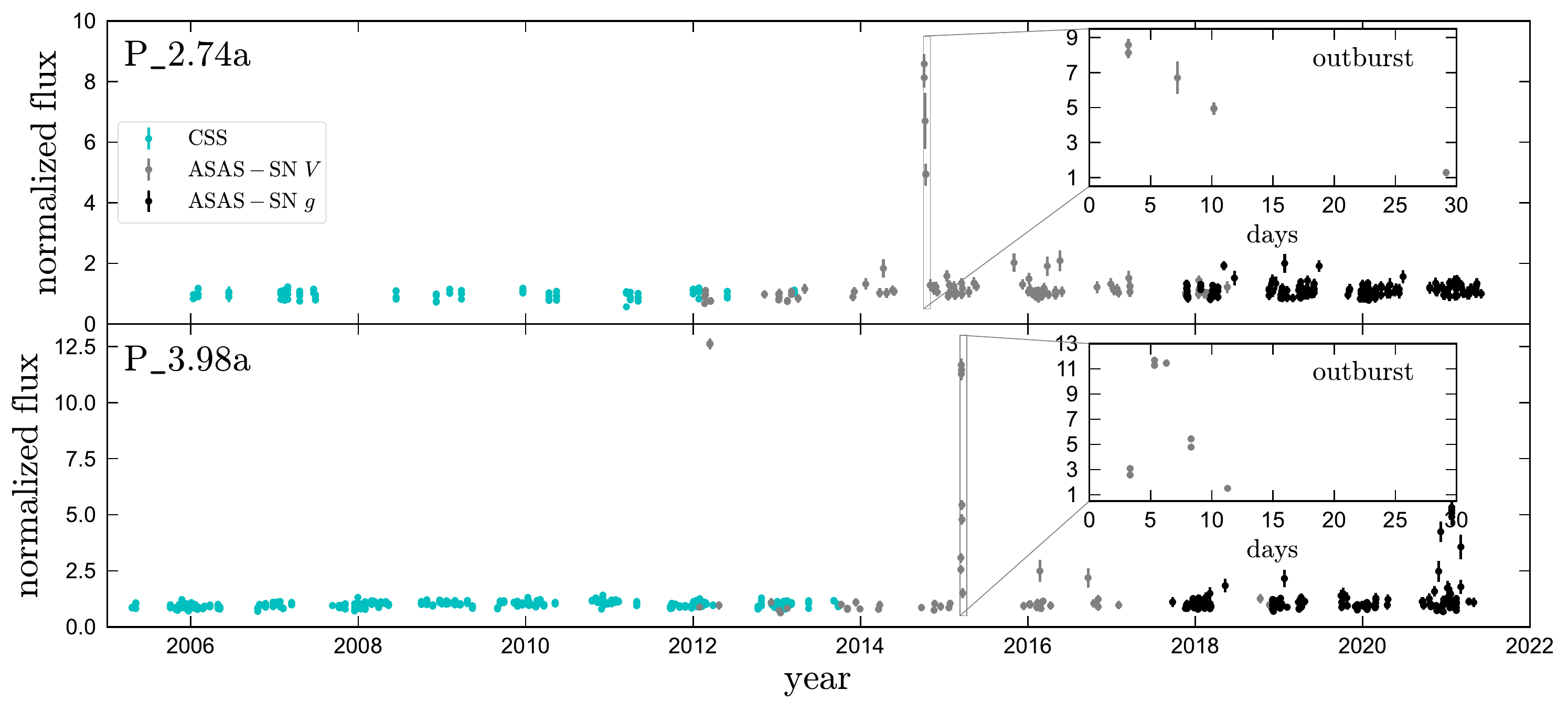}
    \caption{Light curves of two targets with outburst-like events in their observed light curves. Both targets, which are among the coolest (``least evolved'') in our sample, show only one unambiguous dwarf nova-like outburst in 16 years of data. The light curves, from ASAS-SN and CSS, are described in Section~\ref{sec:ephemerides}. } 
    \label{fig:outburst_light_curve}
\end{figure*}

The excess scatter found in the light curves of most objects in our sample could in principle originate from long-term brightness variations (e.g. due to changes in disk structure; Figure~\ref{fig:evolution_light_curve}), from short-timescale flickering of the disk, or from pulsations in the donor star. We investigate the origin of scatter in Appendix~\ref{sec:evolution_long_term}. We conclude that long-term brightness variations due to changes in disk structure dominate for most objects with scatter: when a Fourier model is subtracted from the observed light curves, few-percent brightness variations on month-to-year timescales are evident in about half of the sample. Where available, we also investigate observations from the ZTF high-cadence Galactic plane survey \citep[][]{Kupfer2021}. As expected, most objects have significantly less scatter in their phased light curves when only a few hours of continuous observations are considered than when we model the full $\sim$3 year ZTF baseline. The short-timescale scatter is still nonzero in some cases, suggesting that short-timescale variations due to flickering or pulsations also contribute.

\section{Spectroscopic sample}
\label{sec:spec_sample}

To prioritize objects with hot and evolved donors, we obtained follow-up spectra preferentially for candidates that are relatively luminous, with $M_{\rm G} \lesssim 8$ (see lower right panel of Figure~\ref{fig:sample_selection}). For context, ordinary CV donors with $P_{\rm orb} = 5$ hours and $P_{\rm orb} = 3$ hours have $M_G \approx 9.5$ and $M_G \approx 12$, respectively. We intentionally selected objects with a range of light curve morphologies. Most of the remaining analysis in this paper is focused on the objects in this spectroscopic sample; properties of the remaining 30 objects can be found in Appendix~\ref{sec:nospec_sample}. 

Table~\ref{tab:observables} lists basic properties of the sample. Source IDs, Galactic coordinates, apparent magnitudes, and parallaxes are taken from {\it Gaia} eDR3 \citep{GaiaCollaboration2021}. Orbital periods are measured from the ZTF light curves and verified with spectroscopic RVs. We ``correct'' the reported parallaxes using the position-, color-, and magnitude-dependent parallax zeropoint derived by \citet{Lindegren2021}, and we inflate the reported parallax uncertainties using the empirical inflation formula derived using wide binaries by \citet[][their Equation 16]{El-Badry2021_gaia}. We take reddening values $E(g-r)$ from the \citet{Green2019} 3D dust map; the listed uncertainties account for both reported uncertainties in $E(g-r)$ and for distance uncertainties.  

\subsection{Kast spectra}
\label{sec:spectra}
We obtained multi-epoch spectra for the 21 targets listed in Table~\ref{tab:observables} using the Kast double spectrograph \citep{Miller1994} on the 3\,m Shane telescope at Lick observatory. Most of the observations were taken during bright time between February and June of 2021; observations for J0140 (\texttt{P\_3.81a}) were taken in July and August of 2020 \citep[see][]{ElBadry2021}.  We used the 600/7500 grating on the red side and the 600/4310 grism on the blue side, with the D55 dichroic and a 2 arcsec slit. This results in wavelength coverage of 3300–8400 \AA\,\,with typical resolution (FWHM) of 4 \AA\,\,on the blue side and 5 \AA\,\,on the red side. 

Individual exposure times ranged from 5 to 10 minutes, with shorter exposures for brighter targets and objects with shorter orbital periods. The typical single-visit SNR at 6500\,\AA\,\,ranged from $\approx 10$ for the faintest targets to $\approx 40$ for the brightest. At  4000\,\AA, the same range was 5 to 20. The SNR of the final coadded spectra is a factor of 3-4 higher. We covered at least half an orbit on almost all targets in order to be able to constrain the radial velocity solution. When possible, all spectra for a given target were obtained in a continuous 2-4 hour period on one night. 

To minimize effects of flexure on the wavelength solution, we took a new set of arcs on-sky while tracking each target every $\sim$30 minutes. We reduced the spectra using \texttt{pypeit} \citep{Prochaska_2020}, which performs bias and flat field correction, cosmic ray removal, wavelength calibration, flexure correction using sky lines, sky subtraction, extraction of 1d spectra, and heliocentric RV corrections. 

We measured RVs in two steps via cross-correlation. We first cross-correlated an ATLAS/SYNTHE model spectrum \citep{Kurucz_1970, Kurucz_1979, Kurucz_1993} with an effective temperature of $T_{\rm eff} = 6500\,\rm K$, surface gravity $\log( g/{\rm cm\,s^{-1}}) = 4.5$, and metallicity [Fe/H]=0  with the observed single-epoch spectra, used the thus-measured RVs to shift all spectra to rest frame, and coadded them. We then fit the resulting coadded spectrum (section~\ref{sec:spec_fitting}) to obtain atmospheric parameters for a more accurate model. Finally, we use the best-fit model spectrum to repeat the RV measurement and coadding, yielding more accurate RVs. We also re-fit the coadded spectrum to obtain final atmospheric parameters.

The spectra of most objects are dominated by absorption lines from the donor, but more than half show evidence of weak emission, particularly in H$\alpha$, presumably from an accretion disk around the WD. Several objects have clear H$\alpha$ emission lines that rise above the stellar continuum. In others, the H$\alpha$ absorption line is simply weaker than in the best-fit model spectrum, such that emission is visible only in the residuals. In measuring RVs (and in fitting atmospheric parameters), we mask regions of the spectrum with obvious emission lines. 

It is worth noting that spectra can change with orbital phase. Indeed, in a few objects with emission lines and eclipses evident in their light curves, we do observe that the emission becomes weaker at phase 0.5, when the disk is obscured by the donor. We do not attempt to correct for such variation in this work, as it primarily affects spectral regions with emission lines, which we mask during fitting.

\subsection{Spectral fitting}
\label{sec:spec_fitting}

\begin{table*}
	\centering
	\caption{Parameter constraints from fitting spectra (Section~\ref{sec:spec_fitting}; first 4 columns) and spectral energy distributions (Section~\ref{sec:ang_diameter}; last 3 columns). The $\log g_{\rm spec+period}$ constraint includes a prior based on the fact that the donors/proto-WDs nearly fill their Roche lobes (Equation~\ref{eq:logg_roche}). The SED fitting takes the spectroscopic effective temperature as a prior; we adopt the spec+SED constraint as the fiducial $T_{\rm eff}$ of the proto-WD in the rest of the paper.}
	\label{tab:sed_table}
	\begin{tabular}{lccccccc} 
		\hline
		ID & emission? & $T_{\rm eff,\,spec}$ & $\log g_{\rm spec+period}$ & $\rm [Fe/H]_{\rm spec}$ & $T_{\rm eff,\,spec+SED}$ & $\Theta_{\rm spec+SED}$ & $E(B-V)_{\rm SED+spec}$ \\
          & & [K] & [dex] & [dex] & [K] & [$\mu$\,as] & [mag] \\
		\hline
        \texttt{P\_2.00a} & no & $7646 \pm 100$ &  $5.09 \pm 0.10$ &  $-0.58 \pm 0.20$ &  $7734 \pm 73$ &  $2.44 \pm 0.04$ &  $0.14 \pm 0.01$ \\
        \texttt{P\_2.74a} & yes & $5023 \pm 300$ &  $4.77 \pm 0.10$ &  $0.12 \pm 0.40$ &  $4726 \pm 52$ &  $6.37 \pm 0.11$ &  $0.02 \pm 0.01$ \\
        \texttt{P\_3.03a} & yes & $6096 \pm 200$ &  $4.85 \pm 0.10$ &  $-0.40 \pm 0.30$ &  $5910 \pm 77$ &  $2.44 \pm 0.05$ &  $0.02 \pm 0.01$ \\
        \texttt{P\_3.06a} & yes & $5776 \pm 200$ &  $4.70 \pm 0.10$ &  $-0.02 \pm 0.30$ &  $5862 \pm 70$ &  $2.34 \pm 0.05$ &  $0.04 \pm 0.01$ \\
        \texttt{P\_3.13a} & no & $6331 \pm 100$ &  $4.65 \pm 0.10$ &  $0.25 \pm 0.20$ &  $6662 \pm 54$ &  $4.11 \pm 0.06$ &  $0.08 \pm 0.01$ \\
        \texttt{P\_3.21a} & no & $7051 \pm 100$ &  $4.85 \pm 0.10$ &  $0.25 \pm 0.20$ &  $7022 \pm 72$ &  $3.47 \pm 0.06$ &  $0.41 \pm 0.01$ \\
        \texttt{P\_3.43a} & yes & $5836 \pm 200$ &  $4.84 \pm 0.10$ &  $0.01 \pm 0.30$ &  $6193 \pm 83$ &  $4.12 \pm 0.07$ &  $0.25 \pm 0.02$ \\
        \texttt{P\_3.48a} & yes & $6412 \pm 100$ &  $4.65 \pm 0.10$ &  $-0.26 \pm 0.20$ &  $6444 \pm 62$ &  $2.46 \pm 0.04$ &  $0.01 \pm 0.01$ \\
        \texttt{P\_3.53a} & yes & $5830 \pm 200$ &  $4.70 \pm 0.10$ &  $-0.20 \pm 0.30$ &  $5324 \pm 66$ &  $5.14 \pm 0.07$ &  $0.03 \pm 0.02$ \\
        \texttt{P\_3.81a} & yes & $6665 \pm 100$ &  $4.86 \pm 0.10$ &  $-0.30 \pm 0.20$ &  $6689 \pm 72$ &  $2.12 \pm 0.04$ &  $0.04 \pm 0.01$ \\
        \texttt{P\_3.88a} & yes & $5699 \pm 200$ &  $4.72 \pm 0.10$ &  $-0.00 \pm 0.30$ &  $5875 \pm 102$ &  $3.37 \pm 0.08$ &  $0.09 \pm 0.01$ \\
        \texttt{P\_3.90a} & no & $7396 \pm 100$ &  $4.78 \pm 0.10$ &  $0.18 \pm 0.20$ &  $7442 \pm 87$ &  $3.87 \pm 0.06$ &  $0.06 \pm 0.01$ \\
        \texttt{P\_3.98a} & yes & $5334 \pm 300$ &  $4.58 \pm 0.10$ &  $-0.28 \pm 0.40$ &  $5122 \pm 53$ &  $4.05 \pm 0.08$ &  $0.06 \pm 0.01$ \\
        \texttt{P\_4.06a} & no & $7410 \pm 100$ &  $4.56 \pm 0.10$ &  $-0.48 \pm 0.20$ &  $7587 \pm 72$ &  $3.95 \pm 0.06$ &  $0.21 \pm 0.01$ \\
        \texttt{P\_4.10a} & yes & $5318 \pm 200$ &  $4.60 \pm 0.10$ &  $-0.20 \pm 0.30$ &  $4846 \pm 38$ &  $6.42 \pm 0.09$ &  $0.01 \pm 0.01$ \\
        \texttt{P\_4.36a} & yes & $6139 \pm 200$ &  $4.74 \pm 0.10$ &  $0.15 \pm 0.30$ &  $5998 \pm 80$ &  $3.43 \pm 0.07$ &  $0.10 \pm 0.01$ \\
        \texttt{P\_4.41a} & no & $7315 \pm 200$ &  $4.53 \pm 0.10$ &  $-0.40 \pm 0.30$ &  $7013 \pm 105$ &  $3.94 \pm 0.06$ &  $0.25 \pm 0.02$ \\
        \texttt{P\_4.47a} & no & $6855 \pm 200$ &  $4.71 \pm 0.10$ &  $-0.79 \pm 0.30$ &  $7171 \pm 108$ &  $1.57 \pm 0.04$ &  $0.06 \pm 0.01$ \\
        \texttt{P\_4.73a} & no & $7753 \pm 100$ &  $4.59 \pm 0.10$ &  $-0.32 \pm 0.20$ &  $7619 \pm 85$ &  $4.04 \pm 0.06$ &  $0.31 \pm 0.01$ \\
        \texttt{P\_5.17a} & yes & $6171 \pm 200$ &  $4.42 \pm 0.10$ &  $-0.35 \pm 0.30$ &  $6187 \pm 61$ &  $3.46 \pm 0.06$ &  $0.11 \pm 0.01$ \\
        \texttt{P\_5.42a} & no & $7075 \pm 100$ &  $4.48 \pm 0.10$ &  $-0.52 \pm 0.20$ &  $7326 \pm 78$ &  $6.29 \pm 0.09$ &  $0.34 \pm 0.01$ \\
		\hline
	\end{tabular}
\end{table*}

\begin{figure*}
    \centering
    \includegraphics[width=\textwidth]{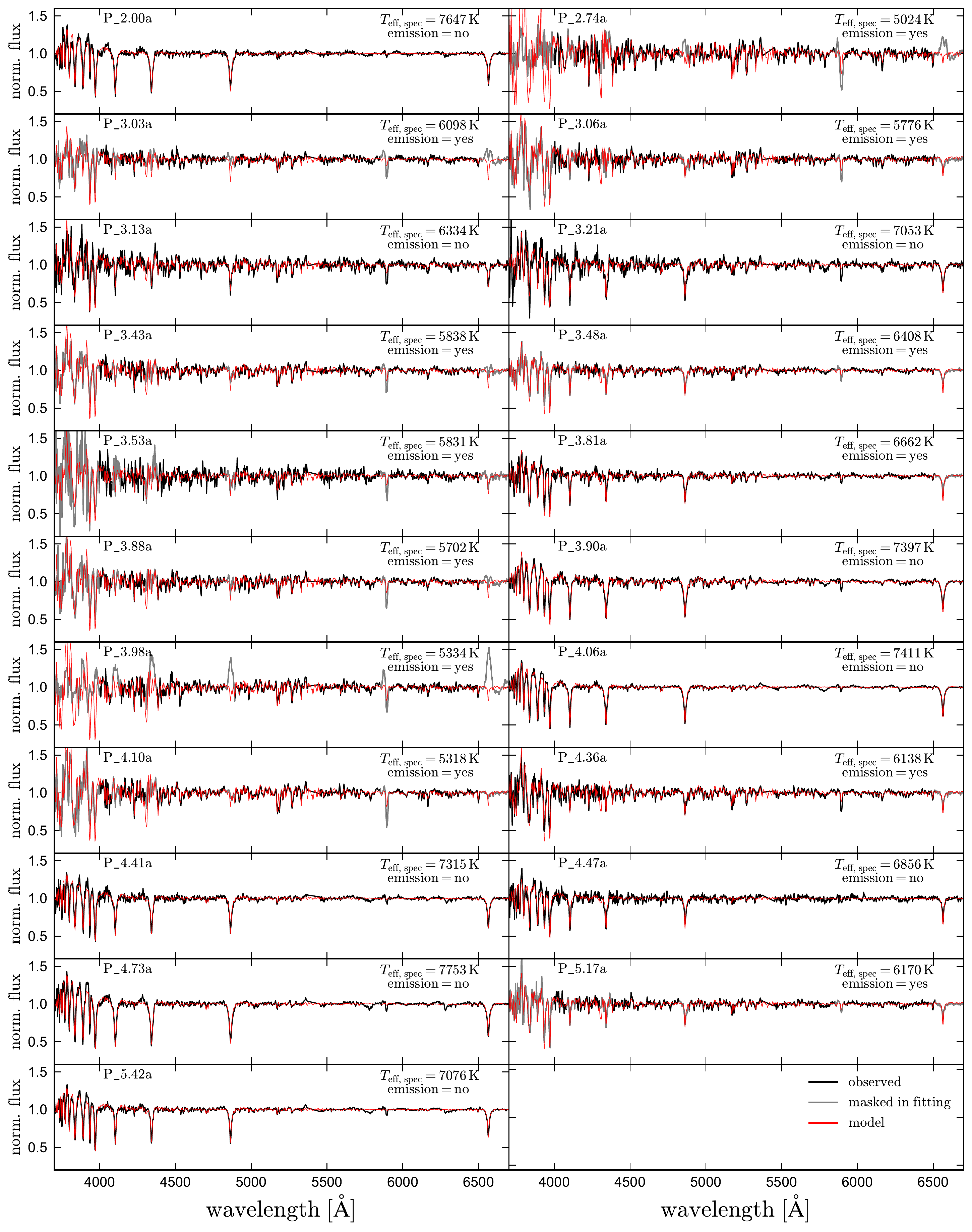}
    \caption{Kast spectra for all targets in the spectroscopic sample. Red lines show the best-fit Kurucz spectral model, with the corresponding effective temperature listed in each panel. Gray regions of the spectra are masked during fitting; these are regions susceptible to contamination from emission lines in targets with evidence of emission. Zoomed-in H$\alpha$ line profiles can be found in Appendix~\ref{sec:halpha}. }
    \label{fig:spectra}
\end{figure*}

Rest-frame coadded spectra for all targets are shown in Figure~\ref{fig:spectra} and are compared to the best-fit model spectra. Both the observed and model spectra are pseudo-continuum normalized, with the pseudo-continuum defined as a running median of the spectrum calculated in a window of width 150\,\AA. 

We implemented {\it The Payne} \citep{Ting_2019, Rix_2016, Elbadry_2018}, a framework for interpolating model spectra, to predict the pseudo-continuum normalized spectra of the proto-WD. We use a 2nd order polynomial spectral model with three labels,  $\vec{\ell}=\left(T_{{\rm eff}},\log g,\left[{\rm Fe/H}\right]\right)$, which we trained on the \texttt{BOSZ} grid of Kurucz model spectra  \citep{Bohlin_2017}. In the relevant part of parameter space, the grid spacing is $\Delta T_{\rm eff} = 250\,\rm K$, $\Delta \log g = 0.5\, \rm dex$, $\Delta [\rm Fe/H] = 0.25\,\rm dex$. Once the spectral model is trained, we fit the coadded spectra using a standard maximum likelihood method, where the likelihood function quantifies the difference between the observed and model spectra assuming Gaussian uncertainties. 

In the systems with the strongest emission lines (e.g. \texttt{P\_3.98a}), emission is obvious in all the Balmer lines and the corresponding absorption lines of the donor are completely filled in. There is also evidence of  emission in the Fraunhofer ``D3'' line at $\approx 5875$\,\AA\, due to neutral helium, and in the Ca H \& K lines. In intermediate cases (e.g. \texttt{P\_3.03a}), the H$\alpha$ emission rises over the continuum, but the core of the donor's absorption line is evident as well. In the weakest cases (e.g. \texttt{P\_3.48a}) the H$\alpha$ line appears only in absorption but is shallower than predicted by the best-fit spectral model. We show zoomed-in profiles of the H$\alpha$ lines of all targets in Appendix~\ref{sec:halpha}.

We caution that there is some ambiguity in interpreting the spectra with the weakest emission, because a too-shallow absorption line could also indicate a mismatch between the spectral model and observed spectra. Spectra with higher resolution and SNR will be required to confirm that the tentatively-detected emission in a few objects with $6000 < T_{\rm eff}/{\rm K} < 7000$ is real. Almost all objects with evidence of a disk in their light curves (eclipses or long-term brightness evolution) also show emission lines. There is one clear exception: the object \texttt{P\_3.13} shows long-term brightness evolution and presumably has a disk, but does not have unambiguous emission lines.

In most cases with evidence for emission, we mask the Balmer lines, He ``D3'' line, and the Ca H \& K lines in fitting. These are the lines most likely to be contaminated by emission, and we find that in most cases other regions of the spectrum are reasonably well-fit by the spectral model. In objects with only weak emission apparent in H$\alpha$, we only mask H$\alpha$. 

We use a flat prior on effective temperature and metallicity, $T_{{\rm eff}}/{\rm kK}\sim\mathcal{U}\left(3,10\right)$ and $\left[{\rm Fe/H}\right]\sim\mathcal{U}\left(-2,0.5\right)$. For surface gravity, we use a prior based on the fact that the stars are nearly or complete Roche lobe filling. From Equation~\ref{eq:rhostar}, we derive 
\begin{equation}
    \label{eq:logg_roche}
\log\left[\frac{g}{{\rm cm\,s^{-2}}}\right]=4.62-\frac{4}{3}\log\left[\frac{P_{{\rm orb}}}{{\rm 4\,hrs}}\right]+\frac{1}{3}\log\left[\frac{M}{0.16\,M_{\odot}}\right]
\end{equation}
The expected mass of ELM WDs and evolved CV donors and at these periods is about $0.16\,M_{\odot}$ \citep[e.g.][]{Rappaport1995, Lin2011, Chen2017}. We therefore adopt a prior $\log g\sim\mathcal{N}\left(\log g_{0},0.1\right)$, where $\log g_{0}=4.62-\frac{4}{3}\log\left[P_{{\rm orb}}/\left({\rm 4\,hrs}\right)\right]$. The dispersion of 0.1 dex corresponds to a factor of 2 uncertainty in mass (e.g. $0.08 < M_{\rm proto\,WD}/M_{\odot} < 0.32$) or to a 10\% uncertainty in Roche lobe filling factor (e.g. 0.9 rather than 1).

The results of our spectral fitting and inspection for emission lines are listed in the first 4 columns of Table~\ref{tab:sed_table}. The best-fit model spectra are compared to the data in Figure~\ref{fig:spectra}; the fits are reasonably good outside of regions contaminated by emission (gray). As discussed in \citet{ElBadry2021}, the formal uncertainties from our full spectral fitting are often unrealistically small, so we inflate them to plausible values given the SNR and degree of contamination from emission lines. We note that our ``final'' effective temperatures and surface gravities are respectively derived from the SED (Section~\ref{sec:ang_diameter}) and the ellipsoidal distortion amplitude (Section~\ref{sec:combined_fits}); these are likely more reliable than the values derived from the spectra alone. 

One striking feature in Figure~\ref{fig:spectra} is that many objects show excess sodium absorption in the Fraunhofer ``D'' lines at $\approx$5890\,\AA.\,This is true for all objects with $T_{\rm eff}\lesssim 7,000\,\rm K$, the temperature range where the spectra are sensitive to sodium. We have verified that this absorption is not interstellar, but is actually associated with the donor: the line clearly shifts in velocity over the course of the orbit. Such enhancement has also been observed in the spectra of other CVs proposed to have evolved donors \citep[e.g.][]{Thorstensen_2002}. A tempting explanation is that the sodium-enhanced material underwent high-temperature burning in the core of the secondary\footnote{In this scenario, ${\rm ^{23}Na}$ is produced by radiative proton capture on $^{22}{\rm Ne}$. The temperatures reached near the end of core-H burning in stars with masses $\gtrsim 1.2 M_{\odot}$ ($T \gtrsim 2.5\times 10^7\,\rm K$) are high enough to produce sodium through this reaction, though the reaction rates are low compared those in later evolutionary stages.} while it was on the main sequence and subsequent stripping of the star's envelope exposed this material. Spectra with higher resolution and SNR will test this scenario, which also predicts that other elements, especially C, N, and O, should have anomalous abundances. 

\subsection{Radial velocities}
\label{sec:orbits}

\begin{figure*}
    \centering
    \includegraphics[width=\textwidth]{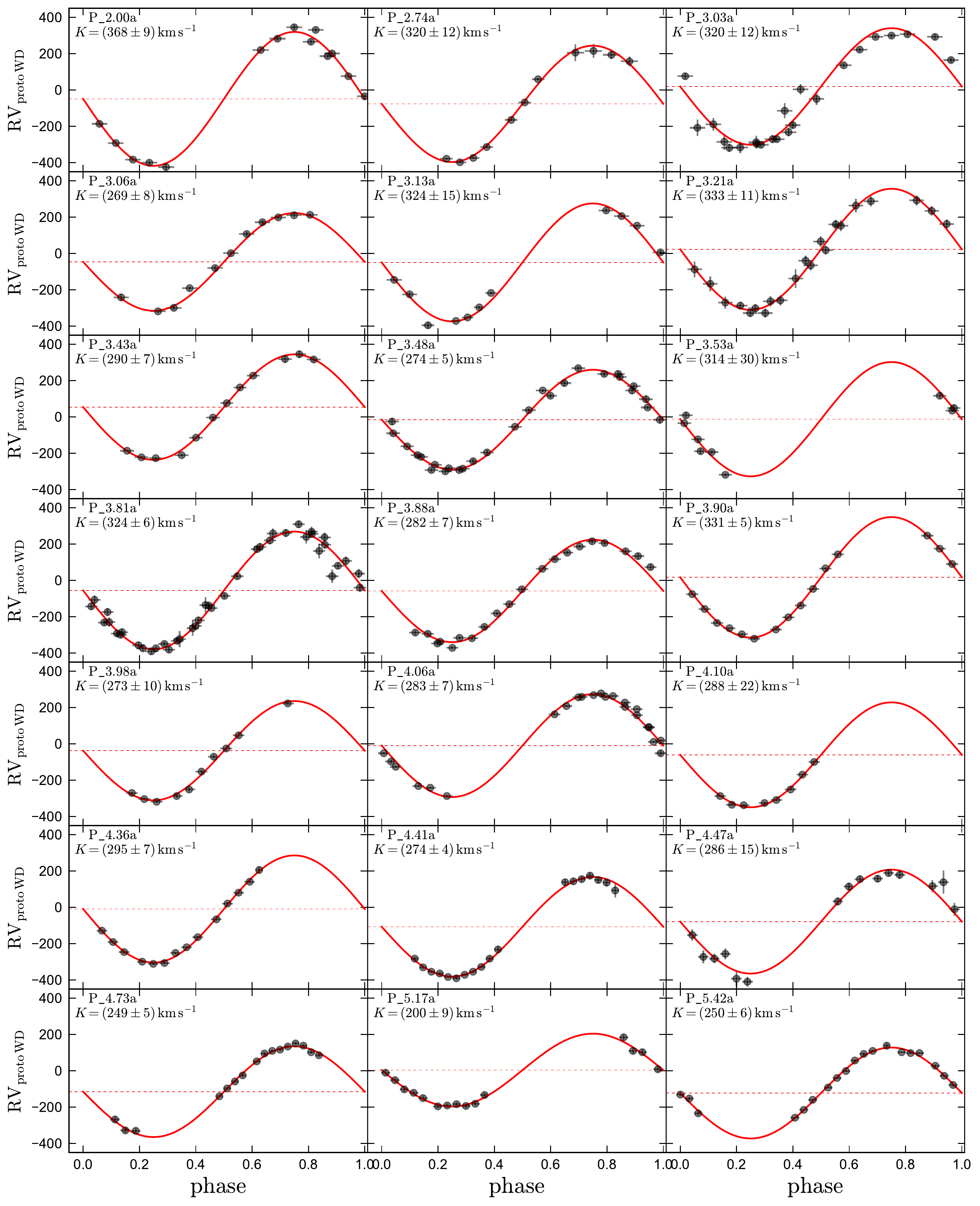}
    \caption{Kast RVs for all targets. Points with errorbars show measured RVs; red line shows the best-fit orbital solution. Dashed lines show the corresponding systemic velocity. The measured RVs cover at least half an orbit and allow for a well-constrained orbital solution for almost all systems.}
    \label{fig:RVs}
\end{figure*}

We fit the RVs of all targets in the spectroscopic sample with a Keplerian model, following the procedure described in \citet{ElBadry2021}. In brief, we fix the orbital period to the value measured from the light curve, and the eccentricity to 0 (since tidal circularization is efficient at such short periods). The measured RVs and best-fit Keplerian orbits are plotted in Figure~\ref{fig:RVs}. Our RVs cover more than half of the orbit in nearly all cases (the only exceptions being  \texttt{P\_3.53a} and \texttt{P\_4.10a}), so the RV semi-amplitudes are well-constrained. In several cases (most drastically, \texttt{P\_3.03a}), the measured RVs are poorly fit by a Keplerian orbit, in the sense that the difference between the model and data is larger than expected from the reported RV uncertainties. We suspect that this is simply due to instability in the Kast wavelength solution, which is known to drift due to instrument flexure. Our inclusion of a scatter term in the fit accounts for such scatter in a controlled way, effectively inflating the observational uncertainties until the reduced $\chi^2$ is of order unity.

The orbital semi-amplitude and period constrain the mass-function, $f_m$, which represents the minimum possible mass of the object the proto-WD is orbiting. It is given by
\begin{align}
    \label{eq:fm}
    f_{m}	&=M_{{\rm WD}}\left(\frac{M_{{\rm WD}}}{M_{{\rm WD}}+M_{{\rm proto\,WD}}}\right)^{2}\sin^{3}i=\frac{P_{{\rm orb}}K_{{\rm proto\,WD}}^{3}}{2\pi G} \\
	&=0.47\,M_{\odot}\times\left(\frac{P_{{\rm orb}}}{4\,{\rm hr}}\right)\left(\frac{K_{{\rm proto\,WD}}}{300\,{\rm km\,s^{-1}}}\right)^3,
\end{align}
where $M_{\rm WD}$ represents the mass of the unseen companion, here presumed to be a WD. $f_m$ is the mass that the companion would have if the proto-WD were a massless test particle and the inclination were 90 degrees, so the actual mass of the companion can only be higher.

\begin{table}
	\centering
	\caption{Orbital solutions from fitting RVs. $K_{\rm proto\,WD}$ is the RV semi-amplitude of the secondary, $\gamma$ is the center-of-mass RV, and $f_m$ is the mass function of the unseen companion (Equation~\ref{eq:fm}). The period and conjunction time are fixed to the light curve-derived values, and the eccentricity is set to 0. The RV zeropoint of the Kast spectra is uncertain at the $30\,\rm km\,s^{-1}$ level; this should be treated as an additional systematic error in interpreting $\gamma$ (but not $K_{\rm proto\,WD}$) values.}
	\label{tab:rvs}
	\begin{tabular}{lccc} 
		\hline
		  ID &  $K_{\rm proto\,WD}$ & $\gamma $ & $f_m$  \\
		&  [$\rm km\,s^{-1}$] &  [$\rm km\,s^{-1}$] & [$M_{\odot}$]   \\
		\hline
		\texttt{P\_2.00a} & $368 \pm 9 $ & $-48 \pm 8 $  & $0.43 \pm 0.03$ \\
        \texttt{P\_2.74a} & $320 \pm 12 $ & $-76 \pm 10 $  & $0.39 \pm 0.05$ \\
        \texttt{P\_3.03a} & $320 \pm 12 $ & $19 \pm 9 $  & $0.44 \pm 0.05$ \\
        \texttt{P\_3.06a} & $269 \pm 8 $ & $-46 \pm 9 $  & $0.26 \pm 0.02$ \\
        \texttt{P\_3.13a} & $324 \pm 15 $ & $-49 \pm 12 $  & $0.48 \pm 0.07$ \\
        \texttt{P\_3.21a} & $333 \pm 11 $ & $22 \pm 7 $  & $0.50 \pm 0.05$ \\
        \texttt{P\_3.43a} & $290 \pm 7 $ & $53 \pm 6 $  & $0.36 \pm 0.03$ \\
        \texttt{P\_3.48a} & $274 \pm 5 $ & $-14 \pm 4 $  & $0.31 \pm 0.02$ \\
        \texttt{P\_3.53a} & $314 \pm 30 $ & $-12 \pm 22 $  & $0.49 \pm 0.14$ \\
        \texttt{P\_3.81a} & $324 \pm 6 $ & $-55 \pm 4 $  & $0.56 \pm 0.03$ \\
        \texttt{P\_3.88a} & $282 \pm 7 $ & $-58 \pm 5 $  & $0.38 \pm 0.03$ \\
        \texttt{P\_3.90a} & $331 \pm 5 $ & $16 \pm 3 $  & $0.61 \pm 0.03$ \\
        \texttt{P\_3.98a} & $273 \pm 10 $ & $-37 \pm 8 $  & $0.35 \pm 0.04$ \\
        \texttt{P\_4.06a} & $283 \pm 7 $ & $-9 \pm 6 $  & $0.40 \pm 0.03$ \\
        \texttt{P\_4.10a} & $288 \pm 22 $ & $-60 \pm 18 $  & $0.43 \pm 0.10$ \\
        \texttt{P\_4.36a} & $295 \pm 7 $ & $-9 \pm 5 $  & $0.48 \pm 0.03$ \\
        \texttt{P\_4.41a} & $274 \pm 4 $ & $-107 \pm 3 $  & $0.39 \pm 0.02$ \\
        \texttt{P\_4.47a} & $286 \pm 15 $ & $-78 \pm 9 $  & $0.46 \pm 0.08$ \\
        \texttt{P\_4.73a} & $249 \pm 5 $ & $-115 \pm 5 $  & $0.32 \pm 0.02$ \\
        \texttt{P\_5.17a} & $200 \pm 9 $ & $3 \pm 7 $  & $0.18 \pm 0.02$ \\
        \texttt{P\_5.42a} & $250 \pm 6 $ & $-122 \pm 3 $  & $0.37 \pm 0.03$ \\
		\hline
	\end{tabular}
\end{table}

Table~\ref{tab:rvs} lists the measured RV semi-amplitudes, center-of-mass velocities, and mass functions for all objects in the spectroscopic sample. The mass functions range from $0.18$ to $0.61 M_{\odot}$, all consistent with WD companions. We constrain the masses of the companions in Section~\ref{sec:combined_fits}.

\subsection{Spectral energy distributions and angular diameters}
\label{sec:ang_diameter}

\begin{figure*}
    \centering
    \includegraphics[width=\textwidth]{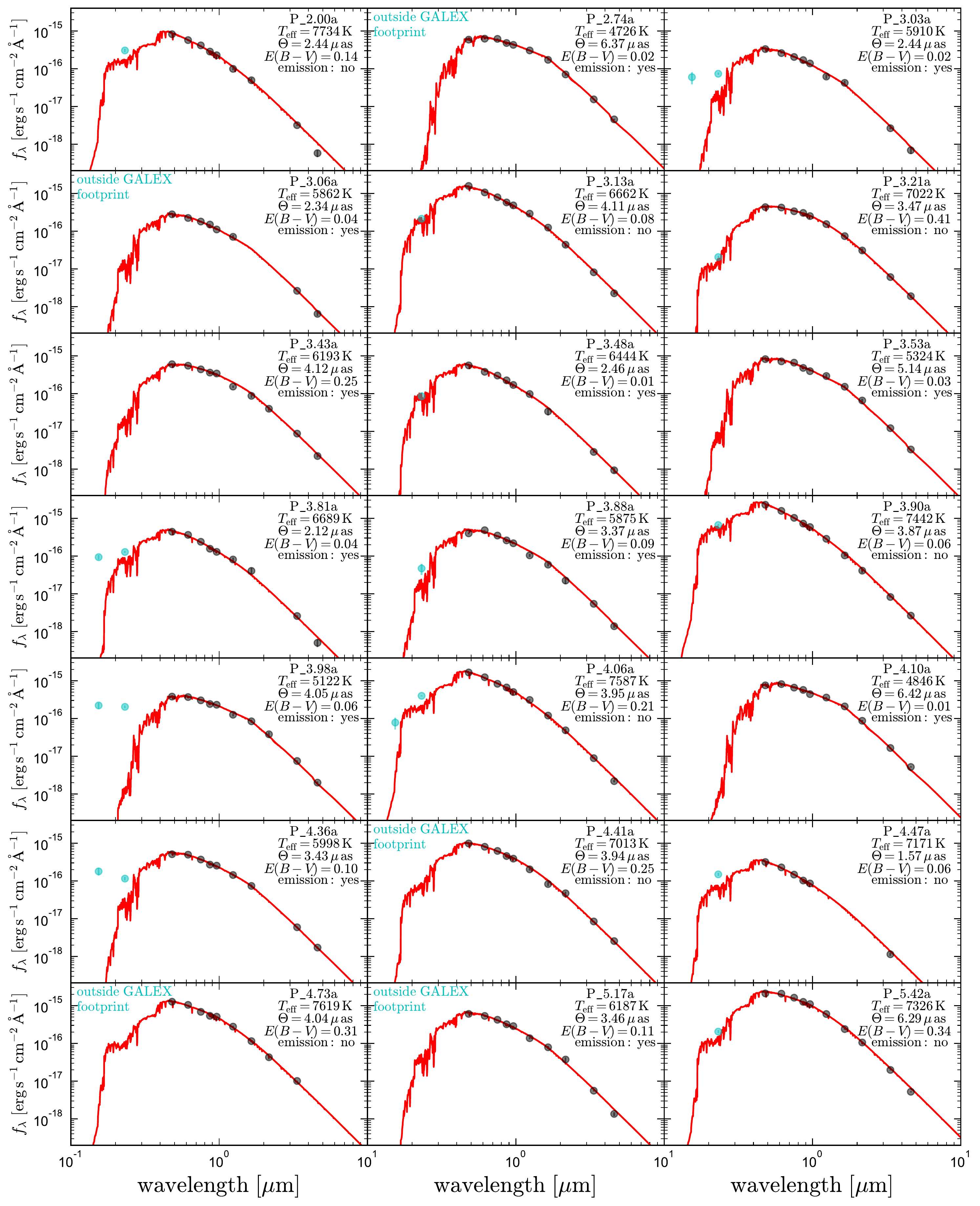}
    \caption{Spectral energy distributions of all targets. Gray points show observed optical-to-IR broadband fluxes from Pan-STARRS, 2MASS, and WISE. Red lines show the best-fit stellar models for the proto-WD, with the effective temperature, angular diameter, and reddening for each object listed. Cyan points show UV fluxes from GALEX. These are not included in the fits, because the WD companion and/or accretion disk are often expected to contribute significantly there.}
    \label{fig:sed_fits}
\end{figure*}

We retrieved available UV-to-IR photometry for targets in the spectroscopic sample and used it to constrain the donors' effective temperatures, angular diameters, and color excesses. We use photometry for all targets from Pan-STARRS1 \citep[][]{Kaiser2002}, 2MASS \citep{Skrutskie_2006}, and WISE \citep{Wright_2010}.\footnote{For most sources, photometry is taken from the ALLWISE catalog \citep{Cutri2021}. In a few cases where ALLWISE contains no reliable photometry, we instead take photometry from the unWISE catalog \citep{Lang2014}. } The resulting spectral energy distributions (SEDs) are shown in Figure~\ref{fig:sed_fits}, with the reported apparent magnitudes converted to flux densities using the appropriate zeropoints for each survey. We also show photometry from GALEX \citep{Martin2005} where available. We do not use it in fitting, because we expect the WD companion  and/or accretion disk to often contribute significantly in the UV. 

In most cases, the SED can independently constrain the effective temperature and angular diameter: the wavelength at which the SED begins to turn over constrains $T_{\rm eff}$, and the normalization of the SED constrains $\Theta$, the angular diameter. Particularly when IR photometry is available, the resulting constraints on the angular diameter and temperature are quite precise \citep[e.g.][]{Blackwell1977}. The effective temperatures of objects in our sample, $4,500\lesssim T_{\rm eff}/{\rm K} \lesssim 8,000$, lead to SEDs that peak between 3,700 and 6,500 \AA, so the flattening of the SED near the peak is always probed by the Pan-STARRS  {\it grizy} photometry. 

We predict bandpass-averaged mean magnitudes of the proto-WD using empirically-calibrated theoretical models from the BaSeL library \citep[v2.2;][]{Lejeune1997, Lejeune1998}. We fix $\log g = 4.75$ and $Z=0.014$ in calculating model magnitudes, since varying $\log g$ and metallicity {\it at fixed angular diameter} has only minor effects on the SED. We assume a \citet{Cardelli_1989} extinction law with total-to-selective extinction ratio $R_V =3.1$. We use \texttt{pystellibs}\footnote{\href{http://mfouesneau.github.io/docs/pystellibs/}{http://mfouesneau.github.io/docs/pystellibs/}} to interpolate between model spectra, and \texttt{pyphot}\footnote{\href{https://mfouesneau.github.io/docs/pyphot/}{https://mfouesneau.github.io/docs/pyphot/}}  to calculate synthetic photometry. These calculations treat the proto-WD as spherical, neglecting the effects of tidal distortion and gravity darkening.\footnote{Using \texttt{Phoebe} \citep{prsa2005}, we find that the phased-average mean flux of a tidally distorted star with given effective temperature and equivalent radius is typically within 5\% of the flux from a non-distorted with the same $T_{\rm eff}$ and radius. The sign of the difference depends on inclination, with face-on systems being brighter and edge-on systems being fainter. Neglecting this effect leads to a $\sim 2$\% systematic uncertainty in the inferred radii, which for all objects in our sample is smaller than the uncertainty due to distance uncertainties.}

Some care must be taken in interpreting the observed magnitudes, since the expected brightness of these sources depends on when they are observed. The magnitudes reported by Pan-STARRS and WISE represent the mean of many exposures from different visits (typically 12 for Pan-STARRS and 20-80 for WISE) and are therefore expected to reflect the phase-averaged mean magnitude reasonably accurately.  We add 0.03 mag in quadrature to the reported photometric uncertainties to account for the random sampling of observations in phase as well as photometric calibration uncertainties.  

Unlike the Pan-STARRS and WISE data, the 2MASS magnitudes were in most cases calculated from only one visit and thus represent the source brightness at a  random phase. Because the 2MASS exposures were obtained $\approx 20$ years before the ZTF data, it is not always possible to accurately calculate the phase at which they were obtained. Rather than attempting to correct them for phase variability, we therefore simply inflate the 2MASS uncertainties by adding 0.1 mag in quadrature to the reported photometric uncertainties. This has little effect on our constraints, because both the 2MASS and WISE data are in the Rayleigh-Jeans tail of the SED for all sources, and WISE data are available for all objects.

We do not attempt to model an accretion disk or companion WD in our fits of the SED. Because emission lines are weak or absent for most sources and the Kurucz model spectra do not significantly over-predict the depth of most absorption lines, we expect that the fraction of the optical flux contributed by the disk is generally small. The companion WD's contributions are also expected to be small (few percent) for plausible masses and temperatures. Still, in systems with emission, it is generally not possible to exclude disks that contribute to the optical flux at the 10-20\% level.  We discuss the effects of unrecognized flux from a disk in Section~\ref{sec:possible_dilution}.

We adopt priors on the color excess for each target based on the 3D dust map from \citet{Green2019}; these values are listed in Table~\ref{tab:observables}. We convert the reported $g-r$ color excess a to $B-V$ color excess using $E(B-V)=0.98 E(g-r)$, as is appropriate for a \citet{Cardelli_1989} extinction law with $R_V = 3.1$. The effective temperature constraints obtained from fitting normalized spectra (Table~\ref{tab:sed_table}) are also adopted as priors. For the angular diameter, we use a flat prior, $\Theta_{{\rm proto\,WD}}\,\left[{\rm \mu\,as}\right]\sim\mathcal{U}\left(0,10\right)$. Here $\Theta_{{\rm proto\,WD}}=2R_{{\rm proto\,WD}}/d$ is the angular diameter of the proto-WD donor, where $R_{\rm proto\,WD}$ is its radius and $d$ is the distance to the system. We fit the SEDs using \texttt{emcee} \citep{emcee2013} to sample from the posterior, monitoring chains for convergence. The resulting constraints on the effective temperatures and angular diameters of the proto WDs are listed in Table~\ref{tab:sed_table}. 

The best-fit model SED for each system is compared to the observed data in Figure~\ref{fig:sed_fits}. The data used in the fit is shown in gray; UV photometry from GALEX is shown in cyan where available but is not considered in fitting. Some of the sources without GALEX photometry are outside the mission's survey footprint; others are within the footprint but were not detected in the UV. Several systems have NUV but not FUV detections; most of these are in fields that GALEX observed after the failure of the FUV detector. The best-fit effective temperatures, angular diameters, and color excesses are listed in each panel; we also note whether emission is evident in the spectrum. Constraints are tabulated in Table~\ref{tab:sed_table}.

The SED fits are generally good. The inferred effective temperatures are in most cases consistent with those inferred from spectral fitting (the median absolute difference is 190 K), and the best-fit $E(B-V)$ values are similar to those predicted by the \citet{Green2019} dust map. None of the sources show evidence of IR excess. Most of the sources detected by GALEX have UV excess, meaning the observed UV fluxes are larger than predicted by the model for the proto-WD alone. This excess can originate either from the WD companion (irrespective of whether it is accreting), or from the inner regions of the accretion disk, if one exists. Most, but not all, of the systems with UV excess also show emission in their spectra. This is expected, since accretion both gives rise to emission lines and keeps the WD hot \citep[e.g.][]{Townsley2009}. On the other hand, a hot WD companion does not guarantee that there is ongoing accretion: for plausible companion WD masses and luminosities in our sample, it would take the WD of order 100 Myr to cool enough that it would no longer be detected in the UV.

\subsubsection{Effects of possible unaccounted flux from the disk or companion WD}
\label{sec:possible_dilution}
Besides emitting in the UV, the WD companion and/or accretion disk could also contribute in the optical, complicating the interpretation of our inferred angular diameters. Given the low accretion rates expected for evolved CVs \citep[see][]{ElBadry2021} and the large masses we infer dynamically (Section~\ref{sec:combined_fits}), we generally expect the companion WDs to contribute negligibly in the optical. The situation is more complicated for disks, which, depending on accretion rate and  stability, can have a wide range of temperatures and SED shapes \citep[e.g.][]{Warner_2003}. For normal CVs, it would {\it not} be a reasonable approximation to neglect the disk, which typically dominates in the optical. This is not the case for objects in our sample, whose spectra are clearly dominated by the donors and whose optical-to-IR SEDs are reasonably well-fit by pure stellar models. 

Nevertheless, the angular diameters we derive are, strictly speaking, upper limits: if the disk contributes additional flux to the SED that is not accounted for, the donor angular diameter will typically be overestimated. If the donor effective temperature is well-constrained and an unaccounted-for disk contributes $f$ times the flux of the donor, then the donor angular diameter will be overestimated by a factor of $\sqrt{1+f}$, and the mass by a factor $\left(1+f\right)^{3/2}$ (e.g. Equation~\ref{eq:rhostar}). Neglecting a 10\% (20\%) flux contribution from a disk would thus lead to a radius over-estimate of 5\% (10\%) and a mass over-estimate of 15\% (31\%). 

There are three objects in the sample for which there is fairly strong evidence that the disk's emission is not negligible in the optical: \texttt{P\_3.06a}, \texttt{P\_3.98a}, and \texttt{P\_4.10a}. In the first case, the light curve shows a strong secular brightness change due to a change in the disk structure (Figure~\ref{fig:evolution_light_curve}). In the 2nd, there are relatively strong emission lines in the spectrum (Figure~\ref{fig:spectra}), and the scatter is significantly larger in the $g$ band than in the $r$ band. In the third, there is a significant asymmetry in the light curve, most likely due to a disk (Figure~\ref{fig:light_cuves}). For these three objects, we explicitly interpret the inferred proto-WD radii and masses (Section~\ref{sec:combined_fits}) as upper limits. 

\subsection{Orbital ephemerides}
\label{sec:ephemerides}
We constrain the orbital ephemeris of each target in the spectroscopic sample by fitting available light curve data from several surveys with a Fourier model, following the same procedure described for the ZTF data in Section~\ref{sec:lightcurves}. The highest-quality light curves are from ZTF. However, these only cover $\sim$3 years. A significantly tighter constraint on the orbital period can be obtained by including light curve data with a longer time baseline. To this end, we retrieved light curves for all targets, where available, from the Catalina Sky Survey \citep[CSS;][]{Drake_2009} and the All-Sky Automated Survey for Supernovae \citep[ASAS-SN;][]{Kochanek_2017}, and we included normalized light curves from all surveys in a joint fit.\footnote{ We note that most of our targets are also detectable as variable sources with TESS light curves, which have low SNR but high cadence. These do not significantly improve the ephemerides, but may be useful for characterizing pulsations and/or flickering in future work, and for extending the search outside the ZTF footprint.} Where CSS data are available, the full light curves span 15-16 years (25 to 70 thousand orbits). We remove outbursts and other outliers in fitting. CSS data are deep enough for all our targets within its footprint to be detected. ASAS-SN data only rarely include defections  fainter than $V\approx 17$, so we only include ASAS-SN data in the fit for targets with $G< 16.75$.

The resulting ephemerides are listed in Table~\ref{tab:ephemeris}. The conjunction times represent the time when the proto-WD would be eclipsed (if viewed edge-on) by its companion. The conjunction times we report are chosen to lie roughly at the median observation time of the data included in fitting, in order to reduce covariance between $T_0$ and $P_{\rm orb}$.

\begin{table}
	\centering
	\caption{Orbital ephemerides from light curve fitting. The conjunction time, $T_0$, is the time corresponding to phase 0, when the WD passes in front of the proto-WD.}
	\label{tab:ephemeris}
	\begin{tabular}{lll} 
		\hline
		 ID &  $P_{\rm orb}$ & $T_0$  \\
		&  [days] & [MHJD UTC]   \\
		\hline
		\texttt{P\_2.00a} & 0.08331301(2)  & 56410.887(1) \\
        \texttt{P\_2.74a} & 0.11414278(4)  & 56489.326(2) \\
		\texttt{P\_3.03a} & 0.12642666(6)  & 56473.433(1) \\
        \texttt{P\_3.06a} & 0.1274832(2)   & 58693.285(1) \\
        \texttt{P\_3.13a} & 0.13056512(8)  & 56431.158(2) \\
        \texttt{P\_3.21a} & 0.13375228(7)  & 56454.278(2) \\
        \texttt{P\_3.43a} & 0.1429875(2)   & 58718.598(1) \\
		\texttt{P\_3.48a} & 0.14479504(6)  & 56350.953(1) \\
        \texttt{P\_3.53a} & 0.1468830(3)   & 58717.743(1) \\
		\texttt{P\_3.81a} & 0.15863390(7)  & 56437.637(1) \\
        \texttt{P\_3.88a} & 0.1615714(3)   & 58717.825(1) \\
		\texttt{P\_3.90a} & 0.1624549(1)   & 58339.258(1) \\
        \texttt{P\_3.98a} & 0.16582294(9)  & 56355.722(2) \\
		\texttt{P\_4.06a} & 0.16898413(4)  & 56411.414(1) \\
        \texttt{P\_4.10a} & 0.1708102(4)   & 58718.202(2) \\
        \texttt{P\_4.36a} & 0.1817430(2)   & 58699.314(1) \\
        \texttt{P\_4.41a} & 0.1838691(2)   & 58717.946(2) \\
        \texttt{P\_4.47a} & 0.18604299(9)  & 56351.560(2) \\
        \texttt{P\_4.73a} & 0.1969063(2)   & 58724.817(1) \\
        \texttt{P\_5.17a} & 0.2153735(2)   & 56346.919(2) \\ 
		\texttt{P\_5.42a} & 0.2258108(1)   & 56548.226(2) \\

		\hline
	\end{tabular}
\end{table}

\subsection{Light curve models}
\label{sec:light_curve_models}

We next explore constraints on the orbital inclination and the Roche lobe filling factor of the proto-WD, $f_{\rm proto\,WD}=R_{\rm proto\,WD}/R_{\rm Roche\,lobe}$, that can be obtained from the light curves. We first consider the case of light curves that only show evidence of ellipsoidal variability (no eclipses) and are not known to be mass-transferring. 

The amplitude of ellipsoidal variability in a tidally distorted star depends on the inclination, Roche lobe filling factor, mass ratio, and effective temperature (through the gravity and limb-darkening laws). For our sample, the effective temperature is well-constrained from spectra and SEDs. The mass ratio is not known, but given the large mass functions and our prior that the donors are proto-WDs, it is reasonable to assume $M_{\rm WD}/M_{\rm proto\,WD}\gg 1$. In this limit, the dependence on mass ratio is weak. 

\begin{figure*}
    \centering
    \includegraphics[width=\textwidth]{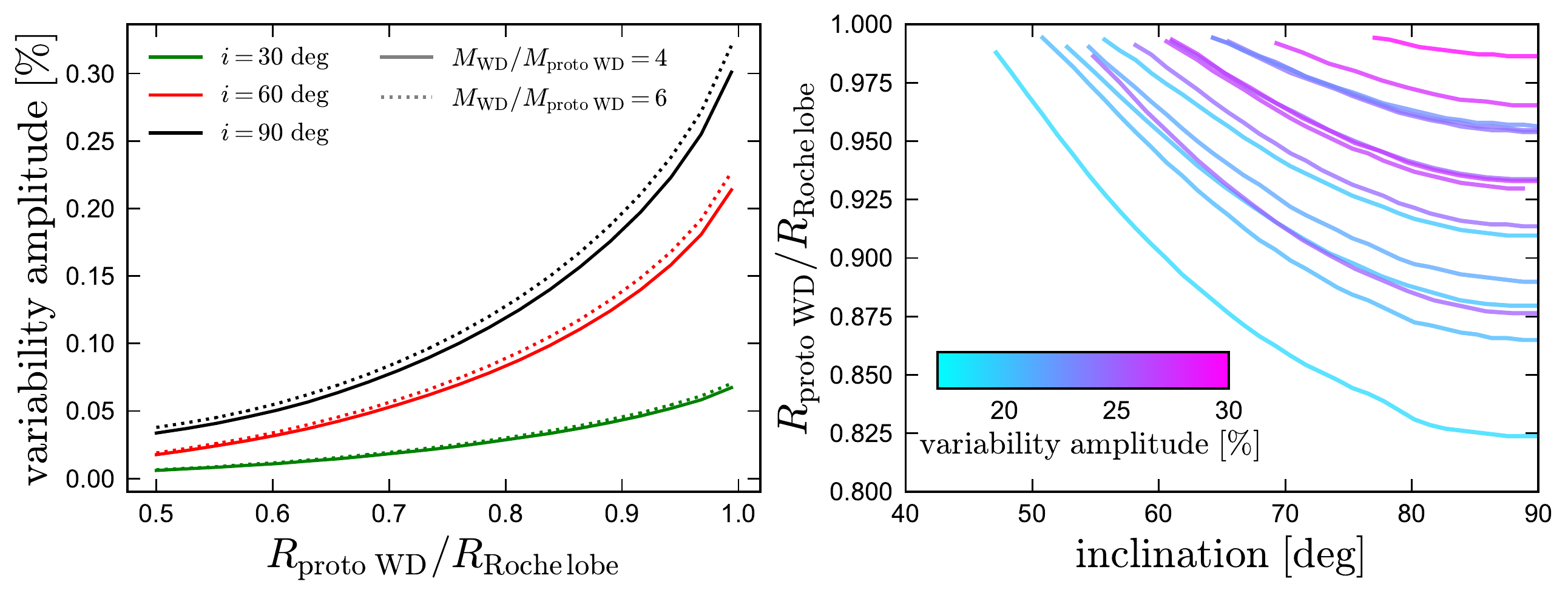}
    \caption{Left: predicted peak-to-peak ellipsoidal variability amplitude for a 7000\,K proto-WD viewed at three different inclinations. We compare predictions for two different mass ratios plausible for our sample, which are very similar.  Only binaries with $R_{\rm proto\,WD}/R_{\rm Roche\,lobe} \gtrsim 0.8$ and inclinations $i \gtrsim 45\,\rm deg$ reach the variability amplitudes greater than 15\% required to enter our sample.
    Right: Roche lobe filling factor of the proto white dwarf as a function of inclination, as implied by the observed ellipsoidal variability amplitude. The systems with the highest variability amplitude (magenta) must be nearly Roche filling and have high inclination. Systems with lower variability amplitude are consistent with a wider range of inclinations and filling factors. Variability amplitude also depends on effective temperature (through temperature-dependent  gravity-darkening), leading to some non-monotonicity in this plane. We only consider systems here in which the variability can be well-described by ellipsoidal variation, excluding systems with eclipses.  }
    \label{fig:inc_vs_roche_factor}
\end{figure*}

This is illustrated in the left panel of Figure~\ref{fig:inc_vs_roche_factor}, which shows the predicted $g-$band variability amplitude for a proto-WD with temperature and mass ratio characteristic of our sample and a range of inclinations and filling factors. We calculate light curves with \texttt{Phoebe} \citep{prsa2005, prsa2016, horvat2018} for two different mass ratios. For a typical proto-WD mass of $\approx 0.15 M_{\odot}$, these correspond to WD companion masses of $0.6$ and $0.9 M_{\odot}$. 
In the limit of $M_{\rm WD}/M_{\rm proto\,WD} \gg 1$, the ellipsoidal variability amplitude scales roughly as $\propto\sin^{2}i\times f_{\rm proto\,WD}^{3}$ \citep[e.g.][]{ Morris1993, Gomel2021}. All the objects in our sample have peak-to-peak variability amplitudes greater than 15\%; Figure~\ref{fig:inc_vs_roche_factor} shows that this implies a Roche lobe filling factor greater than $\approx 0.8$ and an inclination greater than $\approx 45$ degrees. 

The right panel of Figure~\ref{fig:inc_vs_roche_factor} shows the locus in the inclination--filling factor plane that is implied by the measured variability amplitude for non-eclipsing objects in the spectroscopic sample. We first calculated these assuming $M_{\rm WD}/M_{\rm proto\,WD} = 4.66$, with $M_{\rm proto\,WD}=0.15\,M_{\odot}$ and $M_{\rm WD}=0.7\,M_{\odot}$. After fitting for the masses of both components (Section~\ref{sec:combined_fits}), we updated the mass ratios assumed in calculating the tracks, proceeding iteratively until the solution converged. As one might expect, the objects with the largest variability amplitudes are confined to a small range of plausible inclinations and filling factors, while those with lower amplitudes are consistent with a wider range. 

In fitting for the masses of the proto-WDs, we require the systems that do not have emission lines (and thus may not be undergoing mass transfer) to fall along the tracks shown in the right panel of Figure~\ref{fig:inc_vs_roche_factor}. For systems with emission lines, we assume that there is ongoing mass transfer and thus fix $f_{\rm proto\,WD} = 1$. We still leave the inclination free in these systems, since a disk may change the variability amplitude from what is predicted due to ellipsoidal variation alone.

\subsection{Combined fits}
\label{sec:combined_fits}

We carry out a simultaneous fit for the masses of both components, the Roche lobe filling factor of the proto\,WD, distance, and inclination. The ``inputs'' to the fit are the angular diameter from the SED fit, the RV semi-amplitude from the spectroscopic orbit fit, and the corrected {\it Gaia} parallax (Table~\ref{tab:observables}). Following the procedure described in \citet{ElBadry2021}, we use a Markov chain Monte Carlo method to explore what range of parameter space is consistent with all observables, and the covariances between parameters. 

\subsubsection{Sources with emission lines}
For objects with emission lines, which are presumed to be mass transferring and thus must fill their Roche lobes, we fix $R_{\rm proto\,WD}/R_{\rm Roche\,lobe} = 1$. The likelihood function is then 
\begin{align*}
    \ln L_{\rm emission} = &-\frac{1}{2}\frac{\left(K_{{\rm model}}-K_{{\rm obs}}\right)^{2}}{\sigma_{K,{\rm obs}}^{2}}-\frac{1}{2}\frac{\left(\Theta_{{\rm model}}-\Theta_{{\rm obs}}\right)^{2}}{\sigma_{\Theta,\,{\rm obs}}^{2}} \\
    &-\frac{1}{2}\frac{\left(\varpi_{{\rm model}}-\varpi_{{\rm obs}}\right)^{2}}{\sigma_{\varpi,\,{\rm obs}}^{2}}.
\end{align*}
The three terms compare the proto-WD's predicted RV semi-amplitude and angular diameter, and the system parallax, to the observed constraints, which are respectively taken from Tables~\ref{tab:rvs},~\ref{tab:sed_table}, and~\ref{tab:observables}. The predicted RV semi-amplitude of the proto-WD is
\begin{align}
    \label{eq:Kpred}
    K_{{\rm model}}&=\frac{\left[2\pi G\left(M_{{\rm proto\,WD}}+M_{{\rm WD}}\right)/P_{\rm orb}\right]^{1/3}}{\left(1+M_{{\rm proto\,WD}}/M_{{\rm WD}}\right)}\sin i,
\end{align}
the model parallax is $\varpi_{\rm model}=1/d$, and the angular diameter is $\Theta_{{\rm model}}=2R_{\rm proto\,WD}/d$.

We use flat priors on the masses of both components. We take the distance prior from \citet{BailerJones2020}:
\begin{equation}
    \label{eq:distprior}
    P\left(d\right)=\begin{cases}
\frac{1}{\Gamma\left(\frac{\beta+1}{\alpha}\right)}\frac{\alpha}{L^{\beta+1}}d^{\beta}e^{-\left(d/L\right)^{\alpha}}, & {\rm if}\,d\geq0\\
0, & {\rm otherwise}
\end{cases}
\end{equation}
Here $L$, $\alpha$, and $\beta$ are free parameters that depend on sky position and are provided by \citet{BailerJones2020}. They were calculated by fitting the distance distribution of sources in a mock {\it Gaia} eDR3 catalog \citep{Rybizki2020}; we query the appropriate values of $L$, $\alpha$, and $\beta$ for the sky position of each source. None of our results are sensitive to this prior, since they all have high signal-to-noise parallaxes. The median \texttt{parallax\_over\_error} for the spectroscopic sample is 22. 

For the inclination, we use a $\sin i$ prior (as appropriate for random orbit orientations) between $i_{\rm min}$ and 90 degrees. The minimum allowed inclination $i_{\rm min}$ is listed together with our results in Table~\ref{tab:fitting}. For objects without eclipses, it is the minimum allowed value given the observed ellipsoidal amplitude; i.e., the minimum inclination for each source in the right panel of  Figure~\ref{fig:inc_vs_roche_factor}. We note that dilution of variability from a (non-eclipsing) disk would reduce the variability amplitude, only leading to a tighter constraint on the inclination than we impose. For objects whose light curves show evidence of eclipses, we take $i_{\rm min} = 70$ deg. 

\subsubsection{Sources without emission lines}
Sources without emission lines are presumed to be detached, so we leave the Roche lobe filling factors free in fitting and require the binary to fall along the appropriate locus in inclination/filling factor space. We note (see Appendix~\ref{sec:evolution_long_term}) that at least one object without emission lines (\texttt{P\_3.13a}) may still be mass-transferring.
The likelihood function is: 
\begin{align*}
    \ln L_{\rm no\,emission} = &-\frac{1}{2}\frac{\left[i_{{\rm model}}-i_{{\rm light\,curve\,|}f_{\rm proto\,WD}}\right]^{2}}{\sigma_{i}^{2}}\\ &-\frac{1}{2}\frac{\left(K_{{\rm model}}-K_{{\rm obs}}\right)^{2}}{\sigma_{K,{\rm obs}}^{2}}-\frac{1}{2}\frac{\left(\Theta_{{\rm model}}-\Theta_{{\rm obs}}\right)^{2}}{\sigma_{\Theta,\,{\rm obs}}^{2}} \\
    &-\frac{1}{2}\frac{\left(\varpi_{{\rm model}}-\varpi_{{\rm obs}}\right)^{2}}{\sigma_{\varpi,\,{\rm obs}}^{2}}
\end{align*}
Here $i_{\rm model}$ is the inclination (a free parameter), and $i_{{\rm light\,curve\,|}f_{\rm proto\,WD}}$ represents the inclination that, for a particular Roche lobe filling factor, would produce the observed ellipsoidal amplitude (right panel of Figure~\ref{fig:inc_vs_roche_factor}). We adopt $\sigma_{ i}=2\,\rm deg$, allowing for a small amount of scatter around this constraint. The other terms in the likelihood are the same as in the case with emission lines. We also use the same priors, but add a prior on the Roche lobe filling factor (see Appendix~\ref{sec:rl_prior}) that is motivated by binary evolution calculations and favors filling factors near unity.

\section{Results}
\label{sec:results}
The constraints from the combined fits for all 21 objects with spectroscopic follow-up are listed in Table~\ref{tab:fitting}. We translate the constraints on the Roche lobe filling factor to constraints on the physical radius of the proto WD using the fitting function from \cite{Eggleton_1983}:

 \begin{equation}
    \label{eq:roche_filling}
    R_{{\rm Roche\,lobe}} \approx\frac{0.49q^{2/3}}{0.6q^{2/3}+\ln\left(1+q^{1/3}\right)}\times a,
\end{equation}
where $q=M_{\rm proto\,WD}/M_{\rm WD}$ and $a$ is the semi-major axis, which from Kepler's 3rd law is 
\begin{equation}
    \label{eq:semimajor}
    a= \left(\frac{P_{\rm orb}^{2}G\left(M_{{\rm proto\,WD}}+M_{{\rm WD}}\right)}{4\pi^{2}}\right)^{1/3}.
\end{equation}
We calculate $R_{{\rm proto\,WD}}=f_{\rm proto\,WD}\times R_{{\rm Roche\,lobe}} $ directly from the MCMC samples, thus propagating the uncertainties and parameter covariances. We similarly calculate the surface gravity, $\log g_{{\rm proto\,WD}}=\log\left[GM_{{\rm proto\,WD}}/R_{{\rm proto\,WD}}^{2}\right]$, to facilitate comparison of our sample to objects with spectroscopically measured $\log g$, the luminosity, $L_{\rm proto\,WD}=\left(R_{{\rm proto\,WD}}/R_{\odot}\right)^{2}\left(T_{{\rm eff,\,proto\,WD}}/5780\,{\rm K}\right)^{4}  L_{\odot}$, and the gravitational wave inspiral time \citep{Peters1964},
\begin{equation*}
\tau_{{\rm inspiral}}=\frac{5c^{5}}{256G^{5/3}}\frac{\left(M_{{\rm proto\,WD}}+M_{{\rm WD}}\right)^{1/3}}{M_{{\rm proto\,WD}}M_{{\rm WD}}}\left(\frac{P_{{\rm orb}}}{2\pi}\right)^{8/3}
\end{equation*}

An example of the typical covariances between parameters can be found in \citet[][their Figure 9]{ElBadry2021}. In all cases, distance and the mass and radius of the proto-WD are covariant, with a positive correlation, and $M_{\rm WD}$ is correlated with $M_{\rm proto\,WD}$ and anti-correlated with inclination. For most objects in our sample, the parallax uncertainty is the dominant source of uncertainty in the masses and radii of the proto-WDs. The mass of a proto-WD that fills or nearly fills its Roche lobe can be calculated via Equation~\ref{eq:rhostar} to be 

\begin{equation}
    \label{eq:Mprotowd}
    M_{{\rm proto\,WD}}=0.16\,M_{\odot}\left(\frac{\Theta_{{\rm proto\,WD}}}{3\,\mu{\rm as}}\right)^{3}\left(\frac{P_{{\rm orb}}}{4\,{\rm hr}}\right)^{-2}\left(\frac{\varpi}{{\rm mas}}\right)^{-3}.
\end{equation}
This means that when parallax errors dominate and are small enough to justify a linear error analysis, the expected fractional mass uncertainty is 
\begin{equation}
    \label{eq:frac_mass_uncert}
    \frac{\sigma_{M_{{\rm proto\,WD}}}}{M_{{\rm proto\,WD}}}\approx 3\frac{\sigma_{\varpi}}{\varpi},
\end{equation}
so that a 20\% (5\%) parallax uncertainty translates to a 60\% (15\%) mass uncertainty.

\begin{table*}
	\centering
	\caption{Combined fitting results. $M_{\rm proto\,WD}$, $f_{\rm proto\,WD} = R_{\rm proto\,WD}/R_{\rm Roche\,lobe}$, $M_{\rm WD}$, $d$, and $i$ are fitting parameters; $R_{\rm proto\,WD}$, $\log g_{\rm proto\,WD}$, and $\tau_{\rm inspiral}$ are derived. $i_{\rm min}$ and  $f_{\rm proto\,WD,\, min}$ are calculated from the observed ellipsoidal variability amplitude and set the lower limits of $i$ and  $f_{\rm proto\,WD}$ in fitting.} 
	\label{tab:fitting}
	\begin{tabular}{ccccccccccc} 
		\hline
		 ID & $M_{\rm proto\,WD}$ & $R_{\rm proto\,WD}$ &  $\log g_{{\rm proto\,WD}}$ & $M_{\rm WD}$ & $d$ & $i$ & $f_{\rm proto\,WD}$ & $\tau_{\rm inspiral}$ & $i_{\rm min}$ & $f_{\rm proto\,WD,\, min}$ \\
		& [$M_{\odot}$] & [$R_{\odot}$] & [$\rm cm\,s^{-2}$] & [$M_{\odot}$] & [kpc] & deg & & [Gyr] & [deg] &  \\
\texttt{P\_2.00a} & $0.13_{-0.02}^{+0.03}$ & $0.17_{-0.01}^{+0.01}$ & $5.06_{-0.07}^{+0.10}$ & $0.98_{-0.21}^{+0.18}$ & $0.66_{-0.02}^{+0.03}$ & $55.49_{-5.33}^{+14.14}$ & $0.94_{-0.07}^{+0.05}$ & $0.53_{-0.08}^{+0.09}$ & 46 & 0.82 \\
\texttt{P\_2.74a} & $0.13_{-0.01}^{+0.01}$ & $0.23_{-0.01}^{+0.01}$ & $4.84_{-0.01}^{+0.01}$ & $0.62_{-0.06}^{+0.06}$ & $0.34_{-0.00}^{+0.00}$ & $79.76_{-6.75}^{+7.00}$ & $1.00_{-0.00}^{+0.00}$ & $1.59_{-0.16}^{+0.18}$ & 70 & 1.00 \\
\texttt{P\_3.03a} & $0.17_{-0.05}^{+0.07}$ & $0.27_{-0.03}^{+0.03}$ & $4.81_{-0.05}^{+0.05}$ & $0.77_{-0.11}^{+0.14}$ & $1.02_{-0.10}^{+0.13}$ & $73.10_{-9.92}^{+11.29}$ & $1.00_{-0.00}^{+0.00}$ & $1.45_{-0.49}^{+0.68}$ & 59 & 1.00 \\
\texttt{P\_3.06a} & $0.18_{-0.04}^{+0.05}$ & $0.27_{-0.02}^{+0.03}$ & $4.81_{-0.03}^{+0.04}$ & $0.51_{-0.06}^{+0.06}$ & $1.09_{-0.09}^{+0.10}$ & $79.49_{-6.61}^{+7.09}$ & $1.00_{-0.00}^{+0.00}$ & $1.91_{-0.53}^{+0.67}$ & 70 & 1.00 \\
\texttt{P\_3.13a} & $0.18_{-0.02}^{+0.02}$ & $0.28_{-0.01}^{+0.01}$ & $4.81_{-0.02}^{+0.02}$ & $0.74_{-0.08}^{+0.09}$ & $0.63_{-0.02}^{+0.03}$ & $83.66_{-2.75}^{+4.65}$ & $1.00_{-0.01}^{+0.00}$ & $1.49_{-0.22}^{+0.26}$ & 76 & 0.98 \\
\texttt{P\_3.21a} & $0.24_{-0.05}^{+0.06}$ & $0.31_{-0.02}^{+0.02}$ & $4.84_{-0.03}^{+0.04}$ & $0.95_{-0.09}^{+0.10}$ & $0.83_{-0.06}^{+0.06}$ & $71.56_{-3.18}^{+7.29}$ & $0.99_{-0.01}^{+0.01}$ & $1.02_{-0.24}^{+0.30}$ & 64 & 0.95 \\
\texttt{P\_3.43a} & $0.12_{-0.02}^{+0.02}$ & $0.26_{-0.01}^{+0.01}$ & $4.69_{-0.02}^{+0.02}$ & $0.57_{-0.04}^{+0.05}$ & $0.58_{-0.03}^{+0.03}$ & $79.21_{-6.44}^{+7.31}$ & $1.00_{-0.00}^{+0.00}$ & $3.43_{-0.56}^{+0.64}$ & 70 & 1.00 \\
\texttt{P\_3.48a} & $0.11_{-0.03}^{+0.04}$ & $0.26_{-0.02}^{+0.02}$ & $4.67_{-0.04}^{+0.04}$ & $0.53_{-0.06}^{+0.09}$ & $0.97_{-0.08}^{+0.09}$ & $73.90_{-10.57}^{+11.05}$ & $1.00_{-0.00}^{+0.00}$ & $3.89_{-1.06}^{+1.42}$ & 59 & 1.00 \\
\texttt{P\_3.53a} & $0.20_{-0.02}^{+0.02}$ & $0.32_{-0.01}^{+0.01}$ & $4.75_{-0.01}^{+0.01}$ & $0.83_{-0.16}^{+0.19}$ & $0.57_{-0.02}^{+0.02}$ & $79.66_{-6.67}^{+6.96}$ & $1.00_{-0.00}^{+0.00}$ & $1.69_{-0.29}^{+0.37}$ & 70 & 1.00 \\
\texttt{P\_3.81a} & $0.24_{-0.09}^{+0.16}$ & $0.35_{-0.05}^{+0.07}$ & $4.73_{-0.07}^{+0.07}$ & $0.96_{-0.13}^{+0.18}$ & $1.54_{-0.22}^{+0.29}$ & $78.34_{-7.71}^{+7.76}$ & $1.00_{-0.00}^{+0.00}$ & $1.59_{-0.71}^{+1.11}$ & 67 & 1.00 \\
\texttt{P\_3.88a} & $0.12_{-0.02}^{+0.02}$ & $0.28_{-0.01}^{+0.02}$ & $4.62_{-0.02}^{+0.02}$ & $0.59_{-0.04}^{+0.05}$ & $0.78_{-0.03}^{+0.04}$ & $79.64_{-6.60}^{+6.96}$ & $1.00_{-0.00}^{+0.00}$ & $4.56_{-0.78}^{+0.90}$ & 70 & 1.00 \\
\texttt{P\_3.90a} & $0.22_{-0.03}^{+0.03}$ & $0.33_{-0.01}^{+0.01}$ & $4.73_{-0.04}^{+0.05}$ & $1.17_{-0.13}^{+0.10}$ & $0.80_{-0.02}^{+0.03}$ & $64.85_{-4.42}^{+10.19}$ & $0.97_{-0.03}^{+0.03}$ & $1.61_{-0.19}^{+0.21}$ & 57 & 0.96 \\
\texttt{P\_3.98a} & $0.32_{-0.06}^{+0.08}$ & $0.40_{-0.03}^{+0.03}$ & $4.74_{-0.03}^{+0.03}$ & $0.84_{-0.12}^{+0.19}$ & $0.91_{-0.06}^{+0.07}$ & $70.36_{-12.32}^{+13.16}$ & $1.00_{-0.00}^{+0.00}$ & $1.52_{-0.39}^{+0.50}$ & 53 & 1.00 \\
\texttt{P\_4.06a} & $0.12_{-0.02}^{+0.02}$ & $0.28_{-0.01}^{+0.01}$ & $4.63_{-0.05}^{+0.08}$ & $0.81_{-0.13}^{+0.12}$ & $0.65_{-0.02}^{+0.02}$ & $59.48_{-5.30}^{+12.19}$ & $0.95_{-0.05}^{+0.04}$ & $4.10_{-0.45}^{+0.51}$ & 54 & 0.89 \\
\texttt{P\_4.10a} & $0.25_{-0.02}^{+0.03}$ & $0.37_{-0.01}^{+0.01}$ & $4.69_{-0.01}^{+0.01}$ & $0.81_{-0.13}^{+0.15}$ & $0.54_{-0.02}^{+0.02}$ & $79.51_{-6.58}^{+7.05}$ & $1.00_{-0.00}^{+0.00}$ & $2.12_{-0.34}^{+0.41}$ & 70 & 1.00 \\
\texttt{P\_4.36a} & $0.26_{-0.04}^{+0.05}$ & $0.39_{-0.02}^{+0.02}$ & $4.66_{-0.02}^{+0.02}$ & $0.92_{-0.09}^{+0.14}$ & $1.06_{-0.05}^{+0.06}$ & $74.15_{-10.16}^{+10.74}$ & $1.00_{-0.00}^{+0.00}$ & $2.22_{-0.43}^{+0.51}$ & 60 & 1.00 \\
\texttt{P\_4.41a} & $0.20_{-0.02}^{+0.03}$ & $0.35_{-0.01}^{+0.01}$ & $4.64_{-0.04}^{+0.04}$ & $0.82_{-0.09}^{+0.07}$ & $0.83_{-0.03}^{+0.03}$ & $64.02_{-3.90}^{+10.60}$ & $0.97_{-0.03}^{+0.03}$ & $3.20_{-0.40}^{+0.44}$ & 54 & 0.91 \\
\texttt{P\_4.47a} & $0.20_{-0.08}^{+0.14}$ & $0.36_{-0.05}^{+0.07}$ & $4.62_{-0.07}^{+0.07}$ & $0.82_{-0.14}^{+0.18}$ & $2.15_{-0.32}^{+0.43}$ & $73.03_{-2.77}^{+6.83}$ & $0.99_{-0.01}^{+0.01}$ & $3.26_{-1.51}^{+2.47}$ & 63 & 0.95 \\
\texttt{P\_4.73a} & $0.23_{-0.03}^{+0.05}$ & $0.38_{-0.01}^{+0.01}$ & $4.65_{-0.06}^{+0.08}$ & $0.85_{-0.13}^{+0.11}$ & $0.87_{-0.03}^{+0.03}$ & $57.45_{-5.32}^{+12.30}$ & $0.94_{-0.06}^{+0.04}$ & $3.25_{-0.44}^{+0.49}$ & 50 & 0.86 \\
\texttt{P\_5.17a} & $0.15_{-0.03}^{+0.04}$ & $0.37_{-0.03}^{+0.03}$ & $4.48_{-0.03}^{+0.03}$ & $0.44_{-0.07}^{+0.11}$ & $1.00_{-0.07}^{+0.08}$ & $67.57_{-11.87}^{+14.90}$ & $1.00_{-0.00}^{+0.00}$ & $9.74_{-2.52}^{+3.17}$ & 51 & 1.00 \\
\texttt{P\_5.42a} & $0.18_{-0.02}^{+0.02}$ & $0.40_{-0.01}^{+0.01}$ & $4.50_{-0.02}^{+0.02}$ & $0.71_{-0.06}^{+0.06}$ & $0.59_{-0.01}^{+0.01}$ & $68.34_{-4.01}^{+8.65}$ & $0.98_{-0.02}^{+0.02}$ & $6.62_{-0.62}^{+0.72}$ & 63 & 0.95 \\
\hline
		\hline
	\end{tabular}
\end{table*}

\subsection{Comparison to the ELM survey}
\label{sec:elm_comparison}

\begin{figure}
    \centering
    \includegraphics[width=\columnwidth]{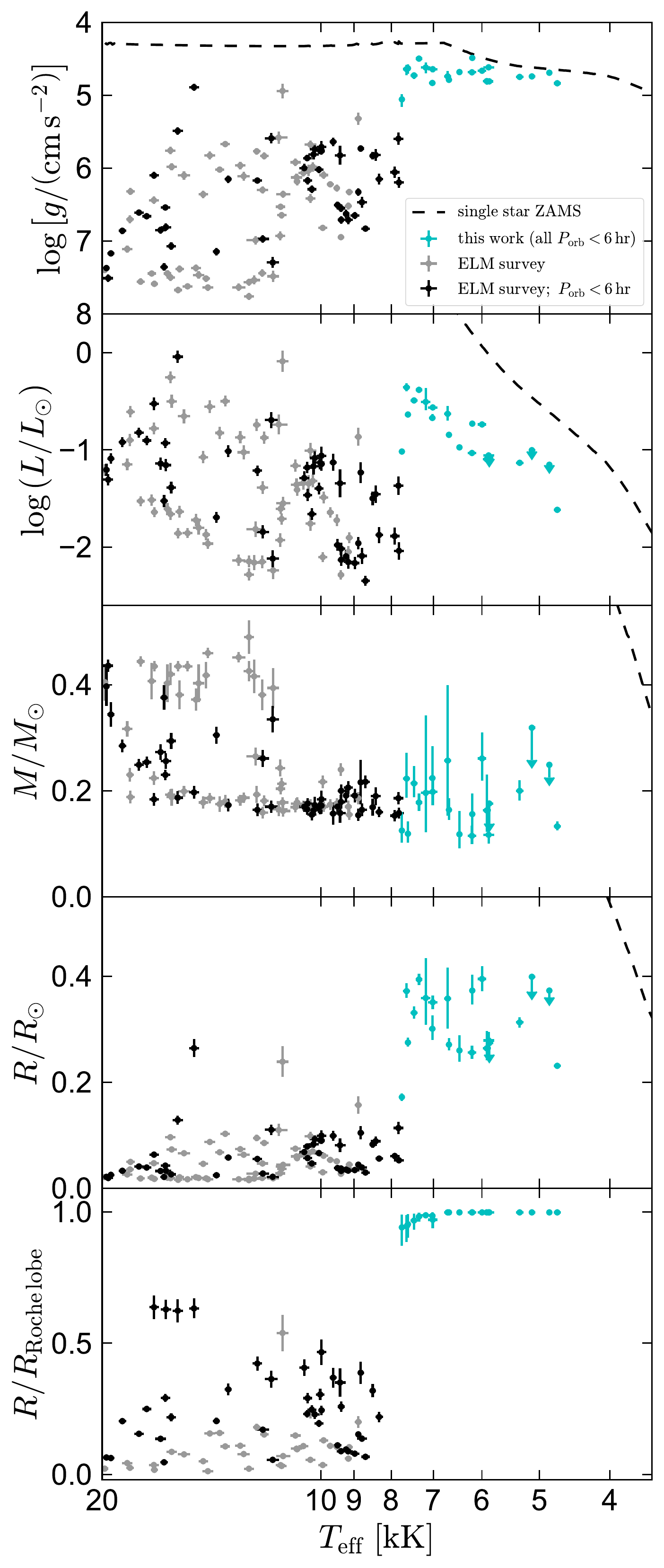}
    \caption{Comparison of objects in our spectroscopic sample (cyan) to WDs and ELM WDs discovered by the ELM survey. Black point show ELM survey targets with $P_{\rm orb} < 6$ hours, the period range investigated by our survey. Gray points show longer periods. Our targets have cooler temperatures, lower surface gravities, larger radii, and larger Roche lobe filling factors than the ELM survey targets. }
    \label{fig:elm_comp}
\end{figure}

We can now put the proto-ELM WDs in our sample in the context of fully-fledged ELM WDs. 
Figure~\ref{fig:elm_comp} compares the 21 objects from our spectroscopic sample with the objects discovered by the ELM survey. Parameters for these objects were taken from \citet{Brown2020}. We plot all objects classified as WDs, including some with estimated masses above the $0.3 M_{\odot}$ threshold \citet{Brown2020} used to distinguish ELMs from normal WDs. We also separate ``short'' period binaries with $P_{\rm orb} < 6$ (the threshold used in our selection) and binaries with longer periods. For reference, we also plot the zero-age main sequence for single stars with solar metallicity, which is taken from MIST models \citep{Choi_2016}.

The top panel shows surface gravity. For the ELM survey, this is measured spectroscopically; for our sample, it is calculated from our mass and radius constraints. The objects in our sample and those from the ELM survey occupy opposite, essentially non-overlapping regions of $T_{\rm eff} - \log g$ space: our candidates have $T_{\rm eff}\lesssim 8,000\,\rm K$ and $\log g \lesssim 5$, while the ELM survey candidates have $T_{\rm eff}\gtrsim 8,000\,\rm K$ and $\log g \gtrsim 5$. This is in some sense by design, since the color selection adopted by the ELM survey was not intended to be sensitive to cooler temperatures. In color-color space, most of our targets overlap with the (more numerous) populations of main-sequence A and F stars, but {\it Gaia} distances place them below the main sequence. 

The luminosities (2nd panel), radii (4th panel), and Roche lobe filling factors (bottom panel) of objects in our sample are all larger than those of typical objects from the ELM survey.\footnote{We calculate radii for the objects in the ELM survey sample from their reported $\log g$ and mass. In calculating Roche lobe radii and filling factors, we need an estimate of the mass of the companion WD and its uncertainty; for this, we use the mass functions reported by \citet{Brown2020} and assume a $\sin i$ distribution of inclinations. We exclude ELMs with possible period aliases from this panel.} Indeed, none of the ELM survey systems have large enough Roche lobe filling factors that one would expect them to have large enough ellipsoidal variability amplitude to enter our sample (e.g., left panel of  Figure~\ref{fig:inc_vs_roche_factor}).\footnote{Curiously, there is one object in the ELM survey sample, {\it Gaia} eDR3 3616216816596857984, that {\it must} be near Roche filling, because its light curve shows large-amplitude ellipsoidal variability and thus enters our sample (\texttt{P\_2.71a}; Figure~\ref{fig:other_targets}). However, the reported parameters yield only $R/R_{\rm Roche\,lobe} = 0.38\pm 0.04$ for this system. We conclude that the reported $\log g = 5.73\pm 0.05$ must be too large; a value of $\log g = 4.93 \pm 0.1$ is needed produce the observed variability amplitude.} Figure~\ref{fig:elm_comp} shows an apparently strong dichotomy in $R/R_{\rm Roche\,lobe}$ between our sample and the ELM survey objects, but we caution that this is at least partly a selection effect, since objects that do not nearly fill their Roche lobes would not pass our selection criteria. 

The masses of the proto WDs in our sample (3rd panel of Figure~\ref{fig:elm_comp}) appear, for the most part, to be an extension of the low-mass mode of the ELM survey's mass distribution to lower temperatures; they are all consistent with being between 0.1 and $0.25\,M_{\odot}$. We note that our masses are constrained dynamically from the measure radii and ellipsoidal variability amplitudes, while the masses of the ELM survey objects are measured by comparing measurements of $T_{\rm eff}$ and $\log g$ to evolutionary models. It is not obvious that either of these approaches is better, but their limitations are different. 

The mass distribution of the ELM survey sample is bimodal, with one mode at $M \lesssim 0.2\,M_{\odot}$ and one at $M \gtrsim 0.4 M_{\odot}$. This has been interpreted as a consequence of CNO flashes experience by intermediate mass He WDs, which have been predicted to accelerate their evolution \citep[e.g.][]{vanKerkwijk2005, Brown2016}. However, the  lifetimes of low-mass proto-WDs appear to decrease monotonically with WD mass \citep[e.g.][]{Istrate2014}, so the full explanation may be more complicated. In any case, all the objects in our sample seem to have masses more consistent with the lower-mass population. We discuss the mass distribution further in Section~\ref{sec:discussion}.

The median mass we infer for the WD companions is $0.81 M_{\odot}$, with a dispersion of $0.18 M_{\odot}$. This is similar to the mass distribution inferred from companions to WDs from the ELM survey, which has a mean of $0.76 M_{\odot}$ and a dispersion of $0.25 M_{\odot}$ \citep{Brown2016}. It is also similar to the mass distribution of WDs in normal CVs, which has mean value of $0.83 M_{\odot}$ and a dispersion of $0.23 M_{\odot}$ \citep[e.g.][]{Zorotovic2011, Pala2020}. All of these values are larger than the mean mass observed for isolated WDs, which is about $0.64 M_{\odot}$ \citep{Falcon2010}. 

From the inferred masses of the proto-WDs and companions, we find gravitational radiation inspiral times ranging from 0.5 to 10 Gyr, with a median of 2.2 Gyr. That is, systems similar to those in our sample have had ample opportunities to reach the period minimum and/or merge. The actual inspiral times are likely shorter for the still-mass transferring systems, since magnetic braking likely still accelerates the angular momentum loss in these systems.

\subsection{Moving On: the CV-to-ELM transition}
\label{sec:cv_to_elm}

\begin{figure*}
    \centering
    \includegraphics[width=\textwidth]{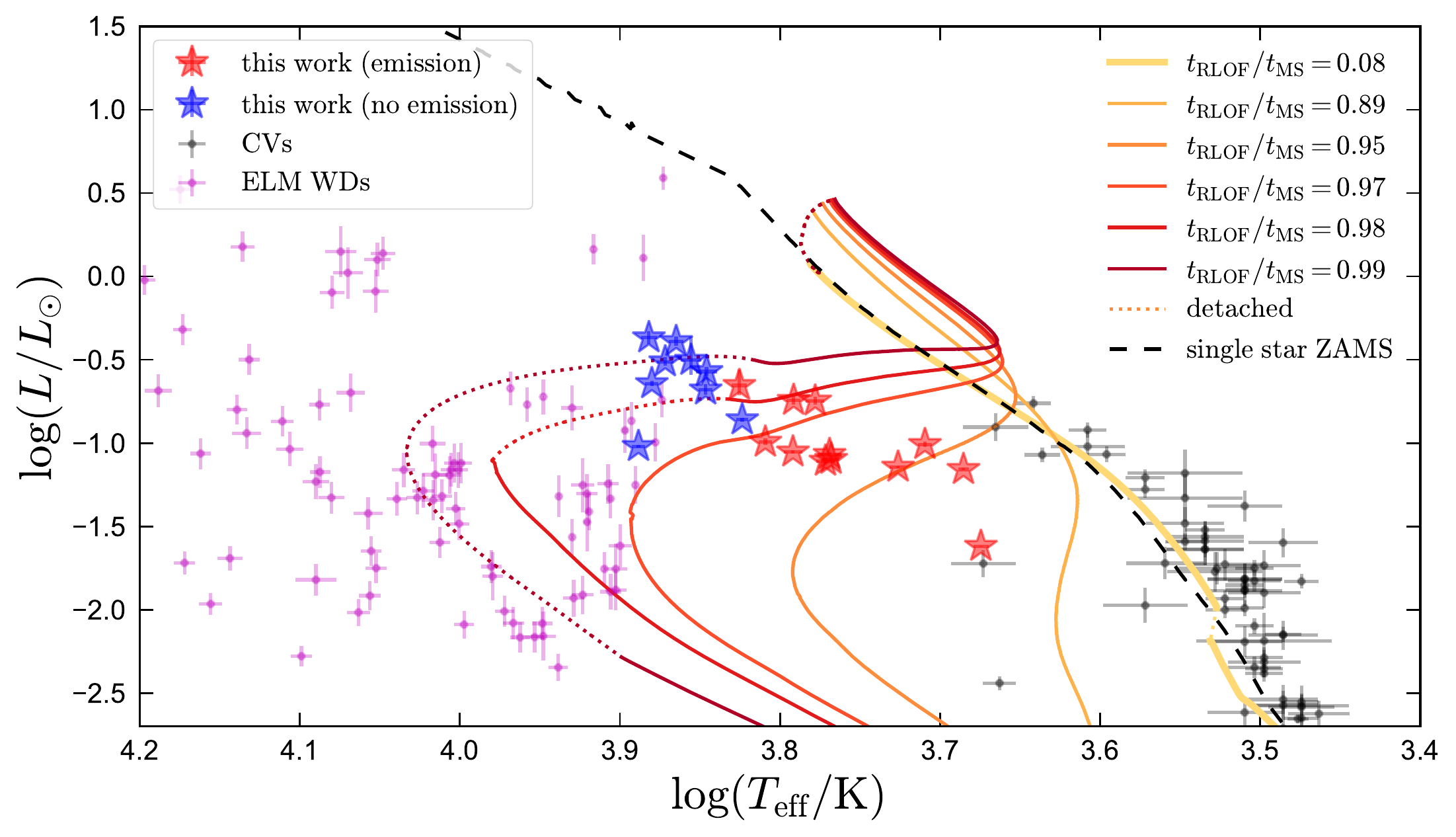}
    \caption{HR diagram. Blue and red stars show targets from our spectroscopic sample with emission lines (likely still mass-transferring) and without them (likely detached). Magenta points show ELM WD candidates from the literature, and gray points show known CV donors, most of which fall close to the main sequence (dashed black line). Colored lines show evolutionary tracks for donor stars in CVs, with the thick yellow line corresponding to an unevolved CV donor, and darker tracks to models in which mass transfer begins increasingly near the end of the main sequence. Dotted lines are detached; solid lines have ongoing mass transfer. Objects in our sample are transitional between CV donors and ELM WDs.}
    \label{fig:hrd}
\end{figure*}

Figure~\ref{fig:hrd} compares, on the HR diagram, the proto-WDs in our sample to known ELM WDs from the compilation of \citet{Pelisoli2019} (a majority of which are also from the ELM survey) and to known cataclysmic variable donors from \citet{Knigge_2006}. These samples are described in more detail in \citet{ElBadry2021}. We also show MESA binary evolution models \citep{Paxton_2011, Paxton_2013, Paxton_2015, Paxton_2018, Paxton_2019} calculated by \citet{ElBadry2021}, which show the evolution of an initially $1.1 M_{\odot}$ donor that begins transferring mass to a $0.9 M_{\odot}$ WD at various points in its evolution. The $t_{\rm RLOF}/t_{\rm MS}$ values noted in the legend indicate the fraction of the donor star's normal main-sequence lifetime that has elapsed when mass transfer begins. The yellow track shows a model in which mass transfer begins before the donor has undergone significant nuclear evolution -- the most common scenario for CVs. In this case, the donor stays close to the main sequence as it shrinks. The other tracks show models in which mass transfer begins near the end of the donor's main sequence evolution, with darker colors corresponding to progressively more evolved donors. The most evolved models eventually become detached (dotted lines). 

The donors in our spectroscopic sample fall between the CV donors and the literature ELM WDs, in a region of the HR diagram that is spanned by the set of MESA models we show. The transition between objects with and without emission lines in our sample occurs at a similar effective temperature as detachment in the MESA models. In the models, detachment occurs when the donor loses its convective envelope and magnetic braking becomes inefficient, slowing the inspiral of the binary. This typically occurs a temperatures between 6,500 and 7,000 K. In the model with $t_{\rm RLOF}/t_{\rm MS} = 0.95$, the binary has reached sufficiently short periods ($\sim$2 hours) when the donor loses its convective envelope that angular momentum losses from gravitational waves are sufficient to keep the system in Roche lobe contact.

The basic picture of detachment occurring at temperatures where magnetic braking becomes inefficient seems to be supported by the observed systems. The transition from mass-transferring CVs to detached ELMs is then related to the {\it Kraft break} \citep[][]{Kraft1967}; i.e., the abrupt change in typical rotation periods of main sequence stars at effective temperatures near 6,500 K. This break is thought to occur because stars warmer than this lack convective envelopes and thus cannot generate a magnetic field at the boundary between the convective and radiative layers. This makes removal of spin angular momentum (and orbital angular momentum, in tidally synchronized binaries) via magnetic braking inefficient. The details of this picture are somewhat murky, because significant magnetic fields {\it are} observed to exist in stars without radiative/convective boundaries \citep[e.g.][]{Donati2009}. However, the existence of the Kraft break provides good empirical evidence that magnetic braking is less efficient above $T_{\rm eff}\approx 6,500\,\rm K$, and the mass transferring-to-detached transition we observe likely occurs through a similar process. The exact temperature threshold at which donors in our sample become detached is still unclear, because faint disks may not lead to detectable emission lines. One object with $T_{\rm eff}\approx 6,662\,\rm K$ (\texttt{P\_3.13a}) has no emission lines but shows clear light curve evolution suggestive of a disk; another, with $T_{\rm eff} \approx 7,022\,\rm K$ (\texttt{P\_3.21a}) shows more tentative evidence of evolution (Appendix~\ref{sec:evolution_long_term}). 

\subsubsection{Are sdA stars evolved CVs?}
There are three objects, plotted in magenta in Figure~\ref{fig:hrd}, that appear to have similar effective temperatures and higher luminosities than the warmest objects in our sample. These targets, which are clustered near $\log\left(T_{{\rm eff}}/{\rm K}\right)=3.9$ and $\log\left(L/L_{\odot}\right)=0.5$, are ``sdA'' stars identified as ``probable'' proto-ELM WDs by \citet{Pelisoli2018}. We think it is unlikely, however, that these are the same type of objects in our sample. 

The IDs of these objects, as named by \citet{Pelisoli2018}, are J0455-0432, J0904+0343, and J1626+1622. Their reported velocity semi-amplitudes are $60\pm 23$, $48\pm 2$, and $93\pm 19\, \rm km\,s^{-1}$. These all correspond to mass functions below $0.03 M_{\odot}$, much lower than any of the targets in our sample (Table~\ref{tab:rvs}). Inspecting the ZTF light curves of these targets, we do not find evidence of ellipsoidal variability (or other variability with amplitude $\gtrsim$0.01 mag). For the object J0455-0432, the reported $\log g$ and orbital period are inconsistent with a normal WD companion, because in this scenario the proto-WD would not fit inside its Roche lobe. The other two objects have longer orbital periods -- so it is possible that a WD companion might escape photometric detection -- but it seems unlikely that the orbits of all three objects would be nearly face on, as would be required for the low mass functions to be normal WDs. 

Indeed, of the 15 sdA stars classified as confirmed and probable proto-ELM WDs by \citet{Pelisoli2018}, just 3 have mass functions above $0.05\,M_{\odot}$. Those that do have mass functions suggestive of normal-mass WD companions have orbital periods longer than 8 hours and, based on their light curves, are not close to Roche lobe filling. This same appears to be true for the larger sample of sdA proto-ELM candidates produced by \citet{Pelisoli2019b}, of which only 1 of 50 presented large-amplitude ellipsoidal variation \citep[see][]{ElBadry2021}. We therefore conclude that most sdA stars identified thus far from spectroscopic surveys are not the same class of objects in our sample. 

\subsection{Comparison to other evolved CVs}
\label{sec:other_evolved}

Most of the CVs from the literature plotted in Figure~\ref{fig:hrd} have temperatures and spectral types only modestly different from those of main-sequence stars of similar mass. However, there are by now about a dozen CVs that have been proposed to have evolved donors, which are warmer than prescribed by the CV donor sequence at their orbital period. 

\begin{figure*}
    \centering
    \includegraphics[width=\textwidth]{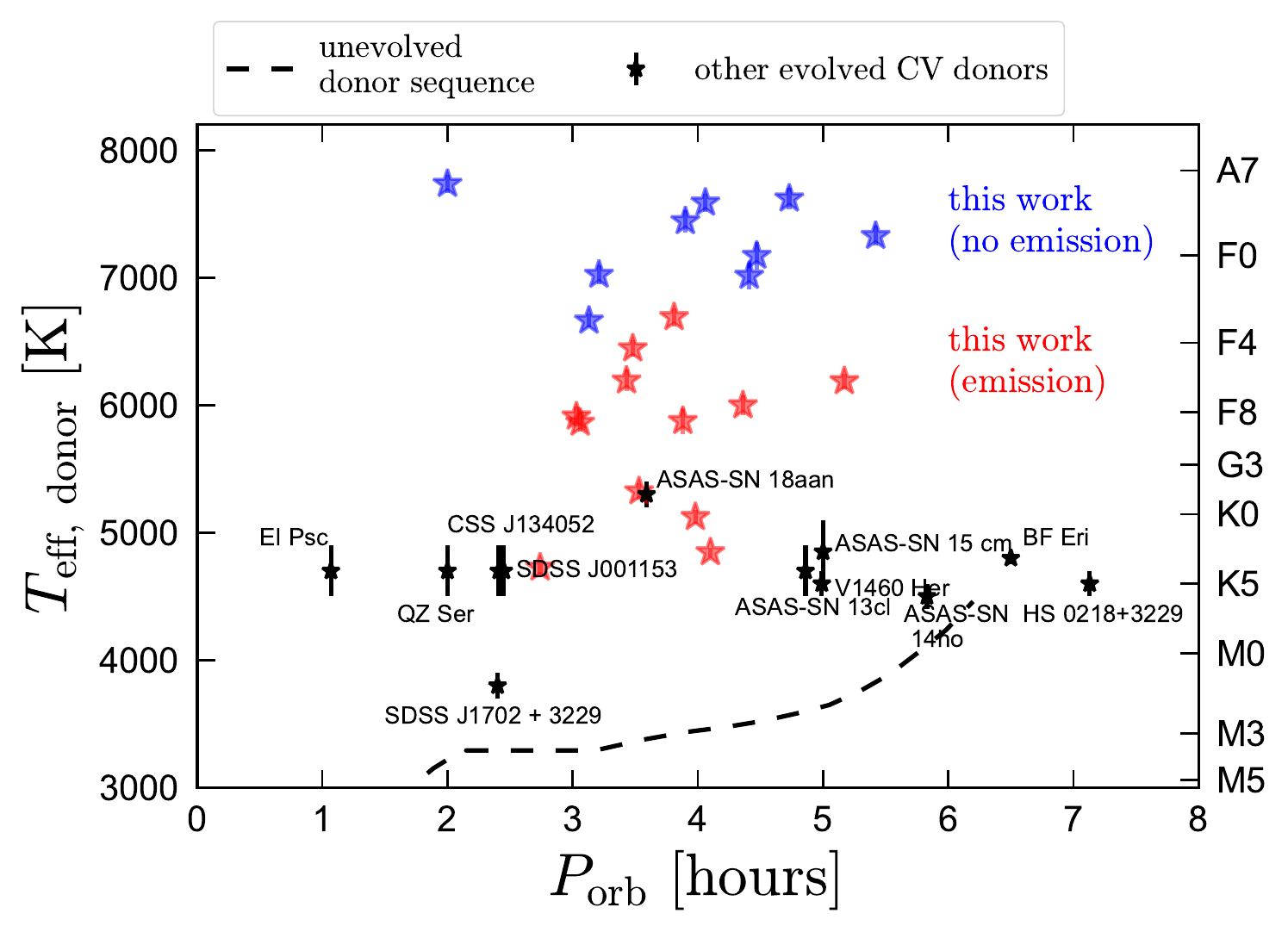}
    \caption{Donor effective temperature versus orbital period. Objects from this study with (red) and without (blue) emission lines are shown with colored stars. Previously known CVs suspected to have evolved donors are shown in black and labeled.  Dashed line shows the semi-empirical CV donor sequence from \citet{Knigge_2006}, along which most normal CV donors are found. Most of the objects in our sample are hotter than any previously known CV donors. Objects with $T_{\rm eff} < 7,000\,\rm K$ have emission lines and likely are still undergoing mass transfer; those with higher temperatures do not. }
    \label{fig:Porb_vs_teff}
\end{figure*}

In Figure~\ref{fig:Porb_vs_teff}, we compare the objects in our sample to these objects, and to the canonical donor sequence for unevolved donors (black dashed line). The other objects are QZ Ser \citep{Thorstensen_2002b}, EI Psc (also called 1RXS J232953.9+062814; \citealt{Thorstensen_2002}), SDSS J170213.26 + 322954.1 \citep{Littlefair_2006}, BF Eridani \citep{Neustroev2008}, CSS J134052.0 + 151341 \citep{Thorstensen_2013}, HS 0218+3229 \citep{RodriguezGil2009}, SDSS J001153.08-064739.2 \citep{Rebassa_2014},  ASAS-SN 13cl \citep{Thorstensen2015}, ASAS-SN 15cm \citep{Thorstensen2016}, ASAS-SN 14ho \citep{Gasque2019}, V1460 Her \citep{Ashley2020}, and ASAS-SN 18aan \citep{Wakamatsu2021}. To our knowledge, this represents a near-complete inventory of currently known evolved CV donors, though new systems are regularly being discovered. EI Psc and QZ Ser, two of the first CVs discovered to have unusually evolved donors, were also included in the \citet{Knigge_2006} sample plotted in Figure~\ref{fig:hrd}; they are the two discrepantly warm objects at low luminosity. 

A majority of the objects in our sample, including those that have emission lines and thus are likely still undergoing mass transfer, are hotter than any previously known CV donors. The hottest 8 objects, which have $T_{\rm eff} > 7000\,\rm K$, do not have emission lines that are detectable in our spectra. This suggests that they either are fully detached, or are undergoing mass transfer at an even lower rate than the objects with emission lines.

\begin{figure}
    \centering
    \includegraphics[width=\columnwidth]{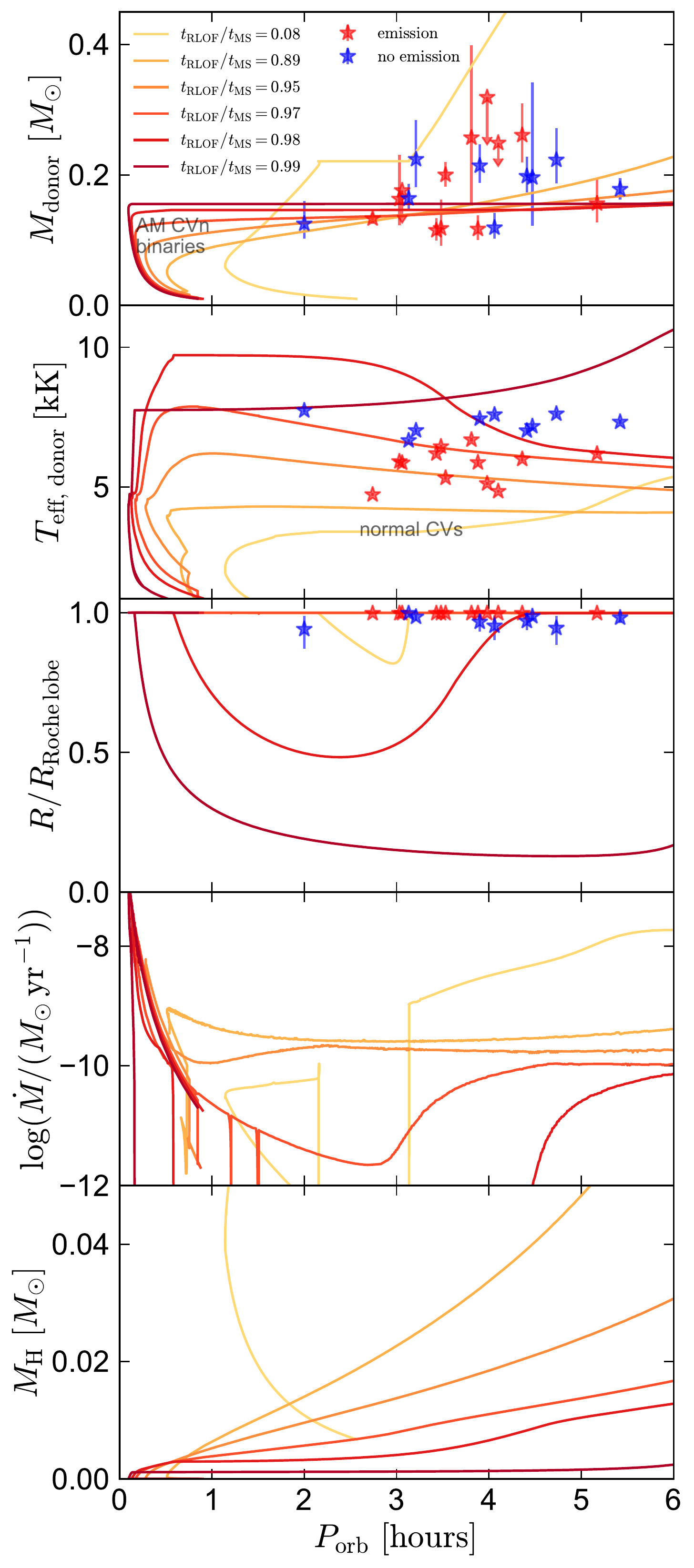}
    \caption{Comparison of our observed targets (red and blue) to MESA binary evolution models for evolved CVs. These are the same models shown in Figure~\ref{fig:hrd}. As a function of orbital period, we show the donor mass, effective temperature, Roche lobe filling factor, mass loss rate, and remaining hydrogen mass. The models predict that evolved CV donors at these periods should have masses between 0.1 and $0.2 M_{\odot}$, low mass transfer rates compared to normal CVS, and at most a few $\times\,0.01\,M_{\odot}$ of hydrogen remaining. }
    \label{fig:porb_masss}
\end{figure}

\subsection{Evolutionary history}
\label{sec:evol_hist}

To gain additional insight into the formation history and future evolution of the systems in our sample, we plot additional properties of the MESA models shown in Figure~\ref{fig:hrd} in Figure~\ref{fig:porb_masss}, now as a function of orbital period.

Compared to normal CVs, models for evolved CVs evolve to much shorter minimum periods, eventually reaching periods between 5 and 30 minutes, where they could be observed as AM CVn binaries \citep[e.g.][]{Podsiadlowski_2003}.  At the orbital periods represented in our sample, $2 < P_{\rm orb}/{\rm hr} < 6$, all models that match the temperatures of the observed systems have masses $0.1 < M_{\rm proto\,WD}/M_{\odot} < 0.2$. 

The observed effective temperatures are best matched by models with $0.9 < t_{\rm RLOF}/t_{\rm MS} < 1$; this depends only weakly on the initial mass of the donor. For all evolved models, the mass transfer rates at periods of 3.1 to 6 hours (above the period gap for normal CVs) are much lower than in normal CVs; this is probably why the emission lines observed in our sample are weaker than in normal CVs and outbursts are less frequent. 

The unevolved CV model (yellow, with $t_{\rm RLOF}/t_{\rm MS} = 0.08$) becomes detached at orbital periods $2.1 \lesssim P_{\rm orb}/{\rm hr} < 3.1$. This is the classical ``period gap'', which is thought to occur when the donor becomes fully convective and magnetic braking becomes inefficient \citep[e.g.][]{Spruit1983}. The evolved models, however, do not generically become detached at these periods: because they are mostly helium, they do not become fully convective. Within our spectroscopic sample, there are 3 objects with evidence of mass transfer and $2.1 \lesssim P_{\rm orb}/{\rm hr} < 3.1$. The sample without spectroscopic follow-up contains another 9 systems in this period range; based on their colors and light curves, at least 7 of them are likely mass transferring. 

The bottom panel of Figure~\ref{fig:porb_masss} shows the total remaining hydrogen mass of the donor in the MESA models. Typical values for evolved donors that have not yet reached the WD cooling track (i.e., those that are still evolving toward higher $T_{\rm eff}$) are in the range of $0.003 < M_{\rm H}/M_{\odot} < 0.03$. The hotter, more evolved donors are predicted to have lower H masses. These values are high enough that spectra would show strong hydrogen features above the period minimum. {\it Below} the period minimum, the predicted H masses drop below $10^{-10}M_{\odot}$, where no hydrogen features would be likely to be detected. The H envelope masses also set the lifetime of the bloated proto-ELM WD phase, since objects with large H envelopes can sustain shell burning longer \citep[e.g.][]{Istrate2014}.

\section{Discussion}
\label{sec:discussion}

\subsection{Donor masses}
\label{sec:masses_discussion}
The distribution of donor masses found in our fitting is relatively narrow, with most systems having best-fit donor masses between $0.1$ and $0.2\,M_{\odot}$ and a few having best-fit masses between $0.2$ and $0.3\,M_{\odot}$.
Our MESA evolutionary models (Figure~\ref{fig:porb_masss}) predict that all CV donors sufficiently evolved to reach $T_{\rm eff}\gtrsim 4,500$\,K at $P_{\rm orb} \lesssim 6$ hours should have masses lower than $\approx 0.2\,M_{\odot}$, with the most common donor mass being about $0.15 M_{\odot}$.

The dispersion in inferred donor masses is somewhat larger than expected from our MESA models. In particular, several objects have inferred masses $M>0.2 M_{\odot}$, larger at the $2\sigma$ level than predictions of any of the models. The most banal explanation would be that these objects simply have slightly overestimated masses. As described in Section~\ref{sec:possible_dilution}, this could occur if an accretion disk contributed significantly to the SED, leading to an overestimate of the angular diameter, and hence, the radius and mass. Such unaccounted-for contributions from the disk are more likely to affect systems with detected emission lines (red points in Figure~\ref{fig:porb_masss}). The few systems with the highest best-fit masses indeed show evidence of a disk in both their spectra and light curves; for these, it seems quite plausible that the best-fit masses could be overestimated. 

However, there are also a few systems that do not show emission lines and are more massive than predicted by the MESA models. The most significant cases are \texttt{P\_3.90a} ($M_{\rm proto\,WD} =0.21\pm 0.03 M_{\odot}$) and \texttt{P\_4.73a} ($M_{\rm proto\,WD} =0.20_{-0.02}^{+0.03} M_{\odot}$). We now explore whether such masses can be reconciled with the evolutionary scenario of recently detached evolved CVs.

\subsubsection{How robust are the model predictions for proto-WD masses ?}
To investigate the sensitivity of donor masses in the MESA calculations to the input physics and initial parameters of the component stars, we varied the MESA calculations from those described in \citet{ElBadry2021} in the following ways:

\begin{enumerate}
    \item Increase the initial donor mass from $1.1 M_{\odot}$ to $1.5 M_{\odot}$.
    \item Decreasing the initial metallicity from $Z=0.02$ to $Z=0.001$.
    \item Increasing the exponential overshooting parameter from $f_{\rm ov}=0.014$ to 0.028.\footnote{The mixing diffusion coefficient in the overshoot region is defined to be $D_{{\rm ov}}=D_{{\rm conv}}\exp\left[-2z/\left(f_{{\rm ov}}H_{p}\right)\right]$. Here $D_{\rm conv}$ is the diffusion coefficient in the convective core, $z$ is the radial distance above the edge of the convective core, and $H_p$ is the pressure scale height at the edge of the convective core.}
\end{enumerate}

\begin{figure}
    \centering
    \includegraphics[width=\columnwidth]{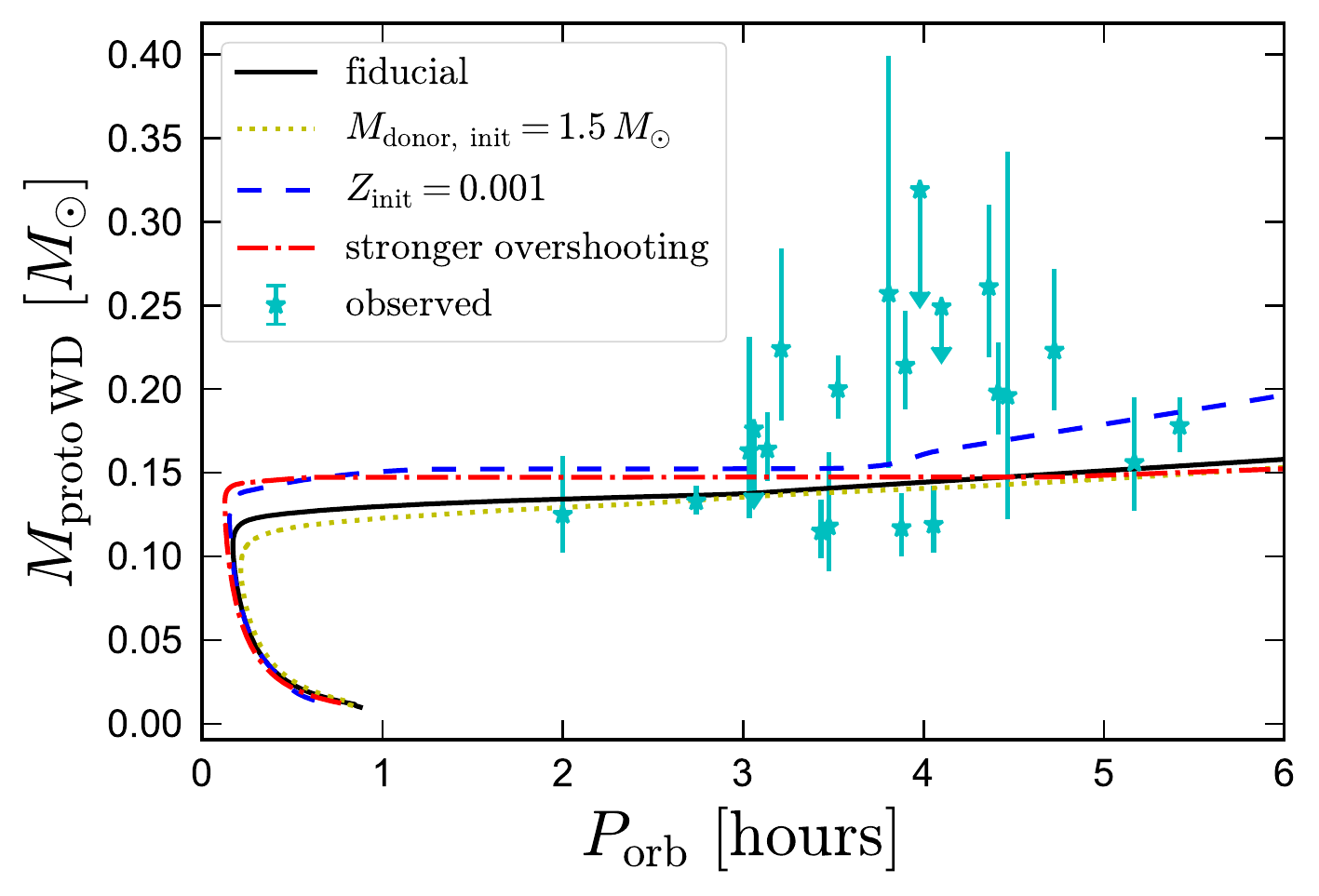}
    \caption{Comparison of observed proto-WDs to MESA calculations with a range of initial conditions and input physics. All model calculations have initial periods chosen such that Roche lobe overflow begins just before the end of the donor's main-sequence evolution ($t_{\rm RLOF}/t_{\rm MS} \approx 0.98$). The predicted proto-WD masses at these periods depend only weakly on modeling assumptions, as they are set primary by the core-mass vs radius relation for objects with degenerate cores. }
    \label{fig:diff_models_mesa}
\end{figure}

In all calculations (including the fiducial model, which is the one labeled $t_{\rm RLOF}/t_{\rm MS} =0.97$ in Figures~\ref{fig:hrd} and~\ref{fig:porb_masss}), we fine-tune the initial period such that the secondary first overflows its Roche lobe at $\approx 97\%$ of its main sequence lifetime.
The results of these experiments are shown in Figure~\ref{fig:diff_models_mesa}. Increasing the initial donor mass has basically no effect on the mass that remains when the binary reaches $P_{\rm orb}$ < 6 hours. Increasing the size of the overshooting region and decreasing the initial metallicity both slightly increase the mass of the core that survives at short periods, but the difference is quite modest, only $\sim 0.02 M_{\odot}$ at fixed period. Varying other parameters, such as the mass of the accreting WD or the convective mixing length (``$\alpha$'') have similarly small effects. 

For WD + MS binaries above the bifurcation limit (i.e., those with initial separations wide enough that mass transfer begins as the donor ascends the giant branch; see \citealt{Podsiadlowski_2003}) that go through stable mass transfer, the relationship between final orbital period and WD mass is reasonably well understood \citep[e.g.][]{Rappaport1995}. In such systems, the orbit expands as the donor loses mass, so the final separation is set by the maximum radius of a giant with a given degenerate core mass. For our systems, which are below the bifurcation limit, the relation is somewhat less clear, because there need not necessarily be a simple relation between core mass and radius. Nevertheless, our calculations do suggest the dispersion in donor mass at fixed period should be small for binaries formed through stable mass transfer. Calculations by other authors \citep[e.g.][]{Lin2011, Chen2017, Li2019} find similar masses, all in the range of 0.14 to 0.18, for low-mass white dwarfs formed by stable mass transfer with $P_{\rm orb}\lesssim 6$ hours.

\subsubsection{Could the more massive proto-WDs have formed via common envelope ejection?}
It is also worth considering whether the overmassive, already-detached systems in our sample could have formed via common-envelope evolution (CEE; which likely allows a significantly wider range of proto-WD masses at these periods;  e.g. \citealt{Li2019}) rather than stable mass transfer. Systems which begin mass transfer on the giant branch with donor masses that are initially significantly larger than the mass of the accreting WD have too-high accretion rates to sustain stable mass transfer and instead spiral to short periods through a common envelope phase. It has been argued \citep[e.g.][]{Brown2016} that a dominant fraction of all low-mass WDs form through this channel and only become detached at short ($P_{\rm orb} \lesssim 1\,\rm hour$) periods. 
The details of what occurs during CEE are not well understood, and so it is commonly invoked to explain the existence of ELM WDs whose properties cannot be easily explained by stable mass transfer. At periods of a few hours, the mass distribution of WDs from the ELM survey suggests that perhaps half of the observed systems formed via stable mass transfer and the rest formed through CEE \citep[e.g.][]{Li2019}. 

It is unclear whether a proto-WD formed through CEE might spend an appreciable amount of time nearly Roche lobe filling, or whether enough of the proto-WD's envelope is ejected during CEE that the proto-WD is expected to emerge already having shrunk well inside its Roche lobe. However, there are other cases in the literature where CEE is assumed to have occurred and the filling factor is still large \citep[e.g.][]{Irrgang2021}. Formation through CEE would likely also allow for a wider range of envelope masses (and thus effective temperatures) during the proto-WD phase \citep[][]{Calcaferro2018}.

\subsection{Selection function, space density, and birth rate}
\label{sec:selection_function}

For population modeling, it will be important to know what fraction of evolved CV-like objects that our survey is sensitive to. As described in Section~\ref{sec:sample_selection}, there are several conditions that must be met for an evolved CV to enter our spectroscopic sample:
\begin{enumerate}
    \item The object must appear in the {\it Gaia} catalog. This should not significantly affect the selection function, because {\it Gaia} eDR3 is basically complete down to our magnitude limit of $G < 18$ \citep[e.g.][]{Boubert2020, Fabricius2021, GaiaCollaboration2021}, except in very crowded regions.
    
    \item Apparent magnitude $G < 18$. For typical objects in the spectroscopic sample with $M_{\rm G}\approx 7$, this corresponds to a distance limit of 1.6 kpc, neglecting extinction. The limit is, however, fuzzy, both because our sample contains objects with a range of absolute magnitudes, and because extinction varies with sky position and distance. 

    \item Fractional parallax uncertainty better than 20\% (i.e., \texttt{parallax\_over\_error > 5}). For a star with $G = 18$, the typical reported parallax uncertainty in {\it Gaia} eDR3 is $\sigma_{\varpi} = 0.120\,\rm mas$. For $G=16$, it is  $0.042\,\rm mas$ \citep{Lindegren2021b}.

    From the mean parallax uncertainty as a function of apparent magnitude, one can calculate the expected \texttt{parallax\_over\_error} for a source of given absolute magnitude as a function of distance \citep[e.g.][]{El-Badry2019twin, Rix2021}. For sources with $M_{\rm G} \gtrsim 6.5$, the cut of  \texttt{parallax\_over\_error > 5} has no significant effect on the selection function, because most such sources will automatically pass the cut if they have $G < 18$. For intrinsically brighter sources (of which our sample contains only a few), the cut can become relevant. For example, at $M_{\rm G}=5$ (brighter than any source in our sample), a cut of $G < 18$ corresponds to a distance limit of $\approx 4$ kpc, but  \texttt{parallax\_over\_error > 5} corresponds to a limit of $\approx 2.6$ kpc. 
    
    \item CMD position well below the main sequence: we only considered targets whose observed magnitudes (without extinction correction) place them below a crude cut in color -- absolute magnitude space (Figure~\ref{fig:sample_selection}). Our final light-curve selected sample (unlike the parent sample) does not hug the dividing line at the upper edge of the selection region. This suggests that our selection probably does not exclude a large fraction of similar systems, but extending the search region closer to the main sequence is needed to confirm this. In the absence of disk contamination, our MESA models suggest that systems with the most evolved donors reach higher temperatures and thus move farther blueward of the main sequence (Figure~\ref{fig:hrd}). 
    Like the magnitude limit, our CMD selection introduces a bias against systems with high extinction. However, the reddening vector in the {\it Gaia} CMD is nearly parallel to the selection region boundary (Figure~\ref{fig:sample_selection}), so this effect is modest. 
    
    \item Good ZTF light curve: Requirement of a ZTF light curve limits the sample to declination $\delta \gtrsim -28$ deg. This cut removed $\approx 40\%$ of sources passing the cuts above. Basically all sources with uncontaminated photometry within the ZTF footprint have light curves, but as of DR4, approximately $12\%$ of candidates had fewer than 40 uncontaminated epochs and therefore were not included in our sample. Requiring a good ZTF light curve thus removes $\approx 50\%$ of otherwise acceptable sources. 
    
    \item ``Well behaved'' ellipsoidal light curve with peak-to-peak variability amplitude >15\%. For systems that are Roche lobe filling, this amplitude limit corresponds to an inclination limit of $\sin i \gtrsim 0.5$, or $i \gtrsim 45$ degrees. For a $\sin i$ inclination distribution, this excludes 29\% of binaries. For recently detached systems, the fraction of binaries we can expect to detect depends on the Roche lobe filling factor (Figure~\ref{fig:inc_vs_roche_factor}). We should not expect to detect {\it any} systems with filling factors less than 0.8. For filling factors of 0.9, we expect to detect only systems with $i \gtrsim 65$ degrees, or about 42\% of systems. 
    
    The ``well-behaved'' requirement is more difficult to quantify. Our search would not be sensitive to CVs with large-amplitude, irregular, and frequent brightness variations, as are frequently observed in normal CVs and several related classes of compact binaries (e.g., bottom left and center panels of Figure~\ref{fig:various_lcs}). Such variability would likely thwart our period search. On the other hand, occasional outbursts that occur on top of an otherwise periodic light curve (e.g. Figure~\ref{fig:outburst_light_curve}) or modest changes in the mean magnitude due to changes in disk structure (e.g. Figure~\ref{fig:evolution_light_curve}) do not prevent sources from being detected. Fortunately, most evolved CVs at $P_{\rm orb} < 6$\,hours are expected to have low accretion rates, and thus, faint disks and infrequent outbursts. We therefore expect most evolved CVs, and all proto-ELM WDs that recently become detached, to have well-behaved light curves to which our search is sensitive. 
    
    \item Spectroscopic follow-up: We have thus far obtained multi-epoch spectra for 21 of 51 light-curve selected targets. The objects we followed-up are biased somewhat toward bright absolute magnitudes: the spectroscopic sample includes only one object with $M_{\rm G} > 8$, while the full sample includes 13. For objects with $M_{\rm G} < 8$, the $M_{\rm G}$ distribution of the spectroscopic sample is representative of that of the full light-curve selected sample; there, we have observed 53\% of all objects. 
    
\end{enumerate}

\begin{figure}
    \centering
    \includegraphics[width=\columnwidth]{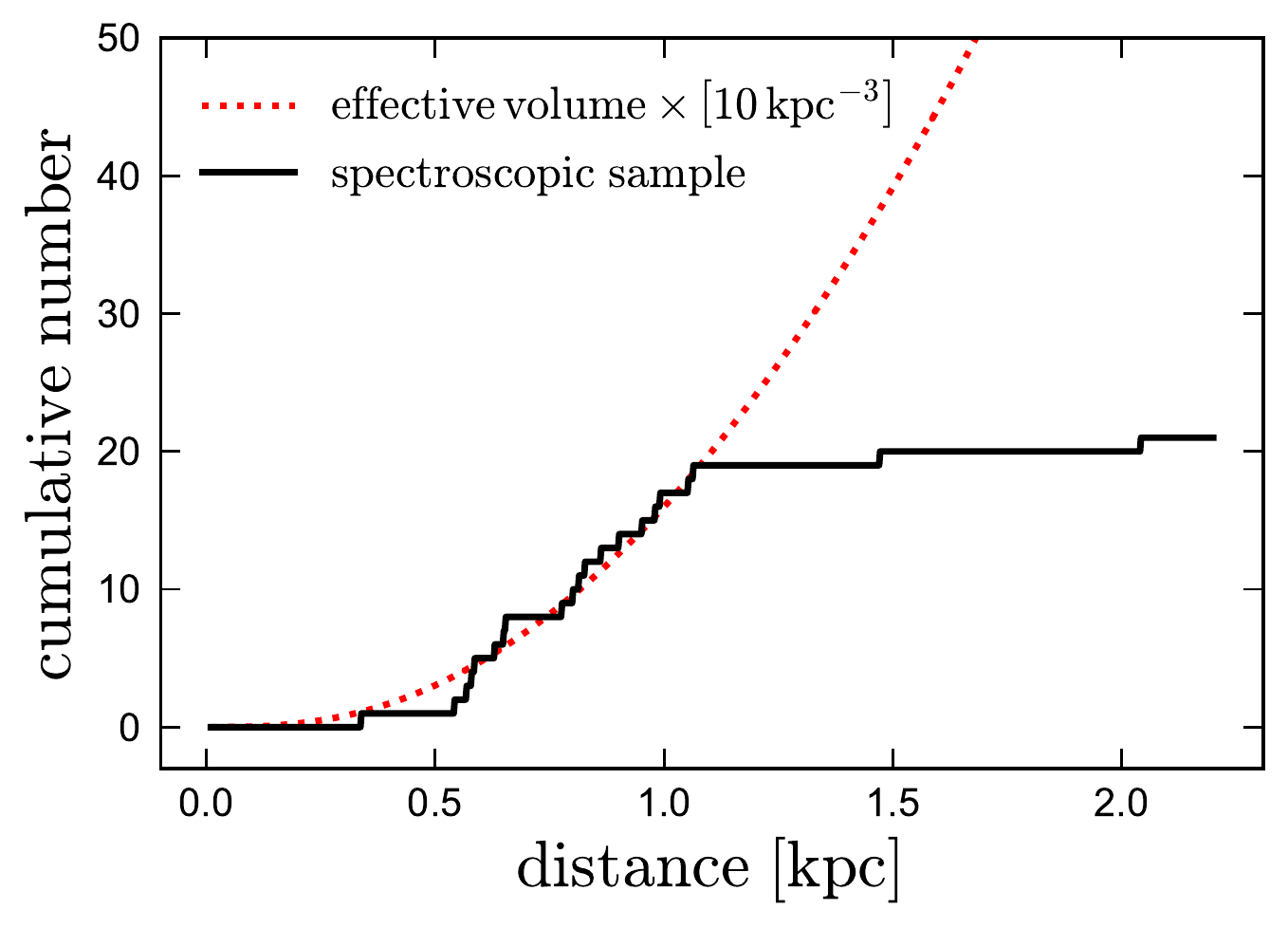}
    \caption{Dotted red line is proportional to the effective stellar volume contained in a sphere of given radius, accounting for the exponential vertical density structure of the Galactic disk (Equation~\ref{eq:vtilde}). Black line shows the cumulative distance distribution of our spectroscopic sample. A space density of $10\,\rm kpc^{-3}$ makes these roughly agree within 1\,kpc, beyond which they strongly diverge. This suggests that within 1 kpc, the completeness of our sample does not decrease much with distance. The true space density is about 6 times larger, $\sim 60\,\rm kpc^{-3}$, because we are not sensitive to all systems and have not spectroscopically followed-up all objects.}
    \label{fig:space_density}
\end{figure}

\subsubsection{Space density}
As a simple diagnostic of how the survey's completeness varies with distance, we compare in Figure~\ref{fig:space_density} the cumulative distance distribution of sources in the spectroscopic sample to the effective stellar volume enclosed in a given distance limit. We define the latter as the expected number of stars in a sphere of radius $d_{\rm max}$ divided by the stellar density at the disk midplane:
\begin{equation}
	\label{eq:vtilde}
\tilde{V} =2\pi\int_{0}^{d_{{\rm max}}}e^{-z/h_{z}}\left(d_{{\rm max}}^{2}-z^{2}\right)dz\\
\end{equation}
where $h_z=0.3\,\rm kpc$ is the disk scale height and $z$ the vertical coordinate. This expression asymptotes to $4\pi d_{\rm max}^3/3$ in the limit of $d_{\rm max} \ll h_z$; otherwise, it ``corrects'' for the decreasing density above and below the disk at large $d_{\rm max}$.

Figure~\ref{fig:space_density} shows that the number of objects in our spectroscopic sample approximately tracks Equation~\ref{eq:vtilde} out to $\approx 1$\,kpc when we scale it by an assumed midplane space density of $10\,\rm kpc^{-3}$. This suggests that the completeness is relatively constant out to this distance, and then drops precipitously. Accounting for the objects that fail to enter our sample because of lack of good light curves ($\sim$\,50\%), incomplete spectroscopic follow-up ($\sim$\,50\%), and low inclination ($\sim$\,30\%) suggests that about 1 out of 6 evolved CVs and proto WDs with $8 \lesssim M_{G} \lesssim 5$ enters our sample, and thus that the true space density is about $60\,\rm kpc^{-3}$. This estimate of course brushes over many complexities of the selection function; we defer more detailed modeling to future work. 
For context, $60\,\rm kpc^{-3}$ is a factor of 2.7--5 lower than the space density of ELM WDs discovered by the ELM survey \citep[160-300 in the same units;][]{Brown2016}, about a factor of 80 lower than the space density of CVs \citep[$4800\pm 800$;][]{Pala2020}, about 8 times lower than the space density of AM-CVn binaries \citep[$500\pm 300$;][]{Carter2013},
and 75,000 times lower than the space density of all WDs \citep[$(4.49 \pm 0.38)\times 10^6$;][]{Hollands2018}. 

\subsubsection{Birth Rate}
Evolutionary interpretation of these densities depends critically on the {\it lifetime} of the proto-ELM phase, and on the lifetimes of other types of binaries to which they are linked. Our MESA models predict observable lifetimes ranging from 200 Myr to 2 Gyr.\footnote{Here we define the lifetime as the time between when the donor first reaches $T_{\rm eff}>4500\,\rm K$ at $P_{\rm orb} < 6\,\rm hr$ (i.e., when it becomes significantly hotter than a normal CV donor) and when its Roche lobe filling factor falls below 0.9 (when it would no longer be easily detectable via ellipsoidal variability).} More-evolved models that become detached have longer lifetimes than less-evolved models that do not, because magnetic braking accelerates the evolution of the less-evolved models. We adopt $\tau = 500\,\rm Myr$ as a typical lifetime in estimating the birth rate. 

To convert the measured space density to a birth rate, we first estimate the total number of evolved CVs and proto-ELM WDs in the Milky Way. We model the Milky Way as an exponential disk with vertical scale height $h_z = 0.3\,\rm kpc$ and radial scale length  $h_R = 3\,\rm kpc$. This yields a total number 
\begin{align}
    \label{eq:Ntot}
N_{{\rm tot}}&=n_{0}\int_{0}^{2\pi}\int_{-\infty}^{{\rm \infty}}\int_{0}^{\infty}\exp\left(-\frac{R-R_{{\rm Sun}}}{h_{R}}-\frac{\left|z\right|}{h_{z}}\right)R\,{\rm d}R\,{\rm d}z\,{\rm d}\phi\\&=4\pi n_{0}h_{z}h_{R}^{2}\exp\left[R_{{\rm Sun}}/h_{R}\right]\\&\approx488\,{\rm kpc^{3}\times}n_{0}.
\end{align}
Here $n_0 = 60\,\rm kpc^{-3}$ is our inferred midplane space density in the solar neighborhood and $R_{\rm Sun}=8\,\rm kpc$ is the distance from the Solar neighborhood to the Galactic center. We then divide the total number of systems by the typical lifetime, $\tau = 500\,\rm Myr$, yielding a Milky Way-integrated birth rate of 
\begin{equation}
    \label{eq:birth_rate}
    \mathcal{R}_{{\rm Milky\,Way}}=\frac{N_{{\rm tot}}}{\tau}\approx59\,{\rm Myr}^{-1}.
\end{equation}
The uncertainty on this estimate is probably about a factor of 2, with comparable contributions from the uncertainty in the population-averaged lifetime of evolved CVs and from uncertainty in the space density.

This birth rate is a factor of $\sim$ 50 lower than the inferred merger rate for ELM WD binaries, $\mathcal{R}\approx3,000\,{\rm Myr^{-1}}$ \citep{Brown2016}. This is expected, since the period distribution observed by the ELM survey suggests a large majority of ELM WDs are born at short periods ($P_{\rm orb} \sim 0.5\,\rm hr$) via common envelope ejection and quickly merge. On the other hand, the birth rate of AM CVn binaries is $\mathcal{R}\approx130\,{\rm Myr^{-1}}$ \citep{Carter2013, Brown2016}, a factor of 2.2 larger than our inferred birth rate of evolved CVs and consistent with it when plausible uncertainties on both rates are accounted for. 

We conclude that the evolved CV channel contributes only a small fraction of all ELM WDs ever formed. However, since ELM WDs formed at short periods via common envelope ejection quickly merge, it contributes a significant fraction of all ELM WD binaries existing at a given time. The similarity between the birth rates of evolved CVs and AM-CVn binaries suggests that the evolved CV channel may contribute significantly to the formation of AM-CVn systems. 

\section{Conclusions}
\label{sec:conclusions}

We have initiated a spectroscopic survey for compact binaries containing a Roche-filling, low-mass proto-white dwarf secondary. These objects fill in the evolutionary gap between evolved CV donors and ELM WDs formed via stable mass transfer.
 
Our search is sensitive to systems on both sides of the CV-to-ELM WD transition: systems with still-ongoing mass transfer, and systems that are detached WD + proto-ELM WD binaries in which mass transfer has recently ceased. Thus far, we have obtained spectroscopic follow-up of about 40\% of all candidates brighter than $G = 18$ that satisfy our search criteria. The sample we have assembled is the first systematic sample of its kind. It contains {\it all} known systems with donor temperatures above $\sim$5,000\,K and triples the number of spectroscopically confirmed evolved CVs. Including systems which have not yet been spectroscopically followed-up (Appendix~\ref{sec:nospec_sample}), it expands the evolved CV population by a factor of $\sim$6.
Our main results are as follows:
\begin{enumerate}
    
    \item {\it Target selection}: We select targets that fall below the main sequence in the {\it Gaia} color-magnitude diagram (Figure~\ref{fig:sample_selection}) and have ZTF light curves dominated by ellipsoidal variation (Figure~\ref{fig:light_cuves}). This selection yields binaries in which the donor star is hotter and more luminous, and the mass transfer rate is lower, than in normal CVs with similar periods. We identified 51 candidates and have obtained multi-epoch spectra for 21 of them. The selection strategy is very efficient: 100\% of the sources we have obtained spectra for are indeed evolved CVs or their recently-detached ELM WD descendants. Our final sample is well-separated from related populations of ELM WDs, hot subdwarfs, and normal CV donors in the color-magnitude diagram, constituting a distinct population.
    
    \item {\it Donor properties}: Joint analysis of light curves (Figure~\ref{fig:light_cuves}), low-resolution spectra (Figure~\ref{fig:spectra}), radial velocities (Figure~\ref{fig:RVs}), and broadband spectral energy distributions (Figure~\ref{fig:sed_fits}) allows us to obtain tight constraints on the proto-WD masses, radii, and effective temperatures. Our inferred proto-WD masses are typically quite low, $M_{\rm proto-WD}\approx 0.15\,M_{\odot}$ (Figure~\ref{fig:porb_masss}). The masses of the WD companions are high but typical for CVs, $M_{\rm WD}\approx 0.8 M_{\odot}$ on average. These proto-WDs are cooler and more bloated than objects observed by the ELM survey (Figure~\ref{fig:elm_comp}). On the HR diagram, they fall between known populations of ``normal'' CV donors and detached ELM WDs (Figure~\ref{fig:hrd}). 
    
    \item {\it Comparison to other CVs}: All the donors in our sample are significantly warmer than typical CV donors at the same orbital period, and most of them are warmer than any previously characterized evolved CVs (Figure~\ref{fig:Porb_vs_teff}). Without a well-informed prior or multi-epoch spectroscopy, the objects in our sample likely would not have been characterized as CVs at all, because most have not been observed in outburst and only show weak emission lines.
    
    \item {\it Mass transfer rates}: More than half of the systems have ongoing mass transfer, as evidenced by weak emission lines (Figure~\ref{fig:spectra}), eclipses of the donor by an accretion disk (Figure~\ref{fig:light_cuves}), and irregular variations in brightness (Figure~\ref{fig:evolution_light_curve}). Only two of the systems with evidence of ongoing accretion exhibit dwarf nova-like outbursts (Figure~\ref{fig:outburst_light_curve}), indicating that the recurrence timescale of such outbursts is long ($\gtrsim 10$ years), if they occur at all. The systems that do not show evidence of ongoing accretion must have terminated mass transfer relatively recently, as their proto-WDs all nearly fill their Roche lobes (Figure~\ref{fig:inc_vs_roche_factor}).
    
    \item {\it Formation histories}: Most of the systems in our sample can be well described by binary evolution models in which a relatively massive ($1 \lesssim M/M_{\odot} \lesssim 2 $) donor overflows its Roche lobe and begins transferring mass to a WD companion near the end of its main-sequence evolution. Mass transfer in these models must begin before the donor star starts to ascend the giant branch (``below the bifurcation period''), or the final orbital period will be too long. 

    In these models, the transition from mass-transferring CV to detached ELM WD occurs when the effective temperature of the donor increases to above 6500--7000\,K and the donor star loses its radiative envelope. This presumably leads to a weakening, or at least a significant rearrangement, of the donor's magnetic field, which is thought to originate at the boundary between the radiative and convective layers \citep[e.g.][]{Spruit1983}. This picture is broadly supported by the spectroscopic properties of the donors in our sample: those with $T_{\rm eff} \lesssim 7000\,\rm K$ have emission lines, likely indicative of ongoing mass transfer, and those with hotter temperatures do not (Figure~\ref{fig:Porb_vs_teff}).
    
    \item {\it Space density and birth rate}: After accounting for the selection function of our survey, we infer a local space density of evolved CVs $n_0 \sim 60\,\rm kpc^{-3}$ and a Galaxy-integrated formation rate $\mathcal{R}\sim 60\,\rm Myr^{-1}$ (Section~\ref{sec:selection_function}). This birth rate is a factor of $\sim$ 50 lower than the merger rate for ELM WD binaries, likely because a large majority of ELM WDs are born at short periods  and quickly merge. It is a factor of $\sim 2.2$ lower than the inferred birth rate of AM CVn binaries.

\end{enumerate}

Objects in our sample are prime candidates to eventually evolve to short periods and become AM CVn binaries (Figure~\ref{fig:porb_masss}), either after a period of detachment (for the most evolved systems), or remaining mass transferring to short periods (less evolved systems; e.g. Figure~\ref{fig:hrd}). Previous work from the ELM survey has shown that a majority of ELM WDs do not form AM CVn systems \citep{Brown2016}. However, the extreme mass ratios in our sample (median $M_{\rm proto\,WD}/M_{\rm WD} = 0.24$) should make them more likely to evolve to short periods without merging \citep[e.g.][]{Marsh2004}. The fate of extreme mass ratio WD binaries after mass transfer begins remains unclear \citep[e.g.][]{Shen2015}.

More work is required for a more complete understanding of the evolved CV/proto-ELM WD population. Spectra with higher resolution and SNR from 10m-class telescope will allow us to investigate the nature and phase-dependence of emission lines in the systems with weak emission, and to measure detailed abundance patterns of the donors. HST/COS spectroscopy of the systems that are brightest in the UV will allow us to measure the temperatures of the accreting WDs, yielding constraints on the time-averaged accretion rate. High-cadence photometry will shine light on the origin of the photometric scatter found in the light curves of a majority of the ZTF light curves (Figure~\ref{fig:light_cuves}), which for many sources is still consistent with coming from disk flickering, longer-term changes in disk structure, or pulsations in the donor.  

\section*{Acknowledgements}
We thank the anonymous referee for constructive comments, Boris Gänsicke, Tom Marsh, Ingrid Pelisoli, and John Thorstensen for helpful discussions, and Geoff Tabin and In-Hei Hahn for their hospitality during the writing of this paper.
We thank the staff at Lick observatory for their assistance in obtaining follow-up spectra.

Based on observations obtained with the Samuel Oschin 48-inch Telescope at the Palomar Observatory as part of the Zwicky Transient Facility project. ZTF is supported by the National Science Foundation under Grant No. AST-1440341 and a collaboration including Caltech, IPAC, the Weizmann Institute for Science, the Oskar Klein Center at Stockholm University, the University of Maryland, the University of Washington, Deutsches Elektronen-Synchrotron and Humboldt University, Los Alamos National Laboratories, the TANGO Consortium of Taiwan, the University of Wisconsin at Milwaukee, and Lawrence Berkeley National Laboratories. Operations are conducted by COO, IPAC, and UW.
This work has made use of data from the European Space Agency (ESA) mission {\it Gaia} (\url{https://www.cosmos.esa.int/gaia}), processed by the {\it Gaia} Data Processing and Analysis Consortium (DPAC, \url{https://www.cosmos.esa.int/web/gaia/dpac/consortium}). Funding for the DPAC has been provided by national institutions, in particular the institutions participating in the {\it Gaia} Multilateral Agreement.

\section*{Data Availability}

The data underlying this article are available upon reasonable request to the corresponding author. For convenience, we include as supplemental data a digital table that combines all data from Tables~\ref{tab:observables}-\ref{tab:fitting}.

\bibliographystyle{mnras}


\appendix

\section{Roche lobe filling factor prior}
\label{sec:rl_prior}

In our fitting in Section~\ref{sec:combined_fits}, we require a prior on the proto-WD's Roche lobe filling factor, $f=R_{\rm proto\,WD}/R_{\rm Roche\,lobe}$, which is often strongly covariant with the orbital inclination. One option would be to adopt a uniform prior between $f_{\rm min}$, the minimum filling factor that can explain the observed ellipsoidal variability amplitude, and 1, essentially implying that all allowed filling factors are equally probable. However, our MESA evolutionary models predict that donors contract more slowly when the filling factor is near 1. This suggests that our prior of $f$ should give more weight to value of $f$ near unity. 

\begin{figure*}
    \centering
    \includegraphics[width=\textwidth]{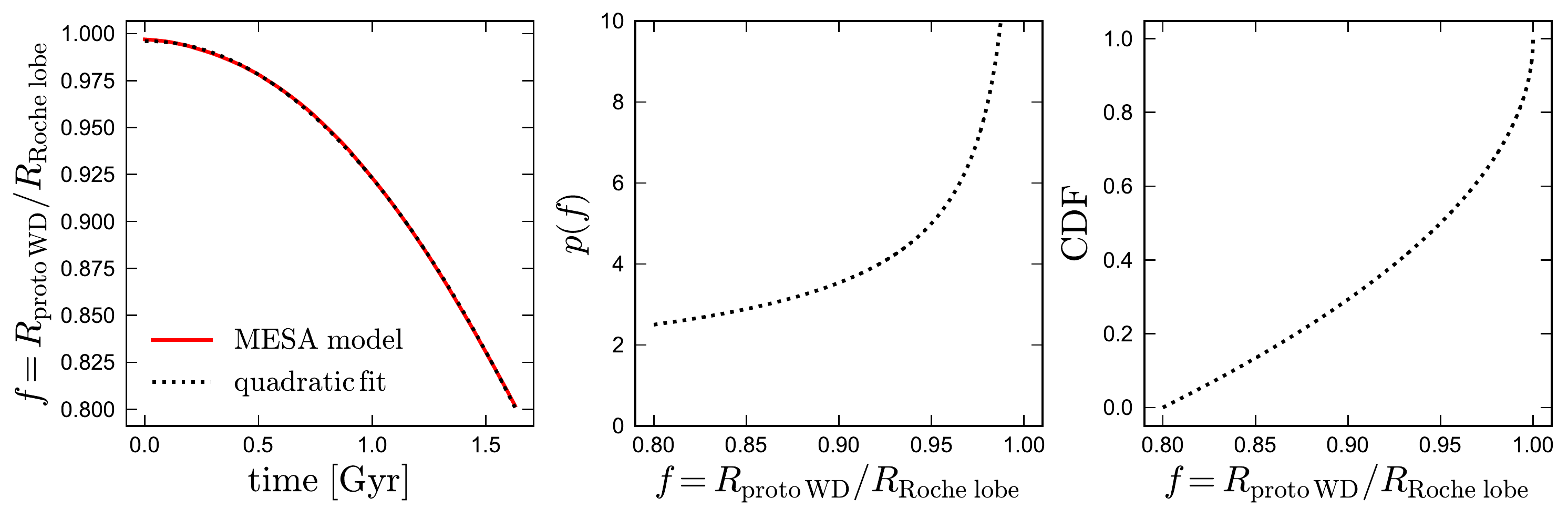}
    \caption{Left: Roche lobe filling factor of the MESA binary model with $t_{\rm RLOF}/t_{\rm MS}=0.98$. Dotted black line shows a quadratic fit, which we use to mitigate numerical noise. The model contracts less steeply just after mass transfer ceases (when $f\approx 1$), and the contraction accelerates over time. This means that the distribution of filling factors for randomly observed detached systems is expected to be weighted toward $f=1$. Middle panel shows the distribution of filling factors expected given the quadratic fit in the left panel and a minimum filling factor of 0.8 (Equation~\ref{eq:analytic}). Right panel shows the corresponding cumulative distribution function.  }
    \label{fig:prior_rl}
\end{figure*}

This is illustrated in Figure~\ref{fig:prior_rl}. The left panel shows the MESA model with $t_{\rm RLOF}/t_{\rm MS} = 0.98$, whose effective temperature evolution provides a good match for several objects in our sample, just after it becomes detached. The red line shows the model, and the dotted black line shows a quadratic fit in the form $f = 1 - at^2$, which we use to avoid numerical noise in calculating derivatives. It is clear that $f(t)$ steepens with decreasing $f$, meaning e.g. that the model spends more time contracting from $f=1$ to $f=0.98$ than from $f=0.9$ to $0.88$. 

The probability of observing such a binary at a given $f$ is
\begin{equation}
    \label{eq:dpdf}
    p\left(f\right)=\frac{{\rm d}p}{{\rm d}f}=\frac{{\rm d}p}{{\rm d}t}\frac{{\rm d}t}{{\rm d}f},
\end{equation}
where $\frac{{\rm d}p}{{\rm d}t}$ represents the (presumed constant) probability of observing the system per time interval. For $f(t)=1-at^2$, this yields 

\begin{equation}
    \label{eq:analytic}
    p\left(f\right)=\begin{cases}
\frac{1}{2\sqrt{\left(f_{\rm min}-1\right)\left(f-1\right)}}, & f\geq f_{{\rm min}}\\
0, & f<f_{{\rm min}}
\end{cases},
\end{equation}
with no dependence on $a$. Thus, for any $f(t)$ that decrease quadratically near detachment, Equation~\ref{eq:analytic} provides a reasonable approximation for $p(f)$. This is plotted in the middle panel of Figure~\ref{fig:prior_rl}, where we chose $f_{\rm min} = 0.8$. The function diverges for $f\to 1$, but its integral converges (right panel). We adopt Equation~\ref{eq:analytic} as our prior on $f$ for all systems which have no clear emission lines and thus appear to be detached.

Adoption of this prior has a relatively weak effect on our results; for example, none of the inferred proto-WD masses change by more than $0.02 M_{\odot}$ when we instead adopt a flat prior. The change is, however, systematic: in all cases, adopting the prior leads to a higher inferred Roche lobe filling factor, and thus, a lower inferred proto-WD mass. 

\section{H$\alpha$ emission}
\label{sec:halpha}

\begin{figure*}
    \centering
    \includegraphics[width=\textwidth]{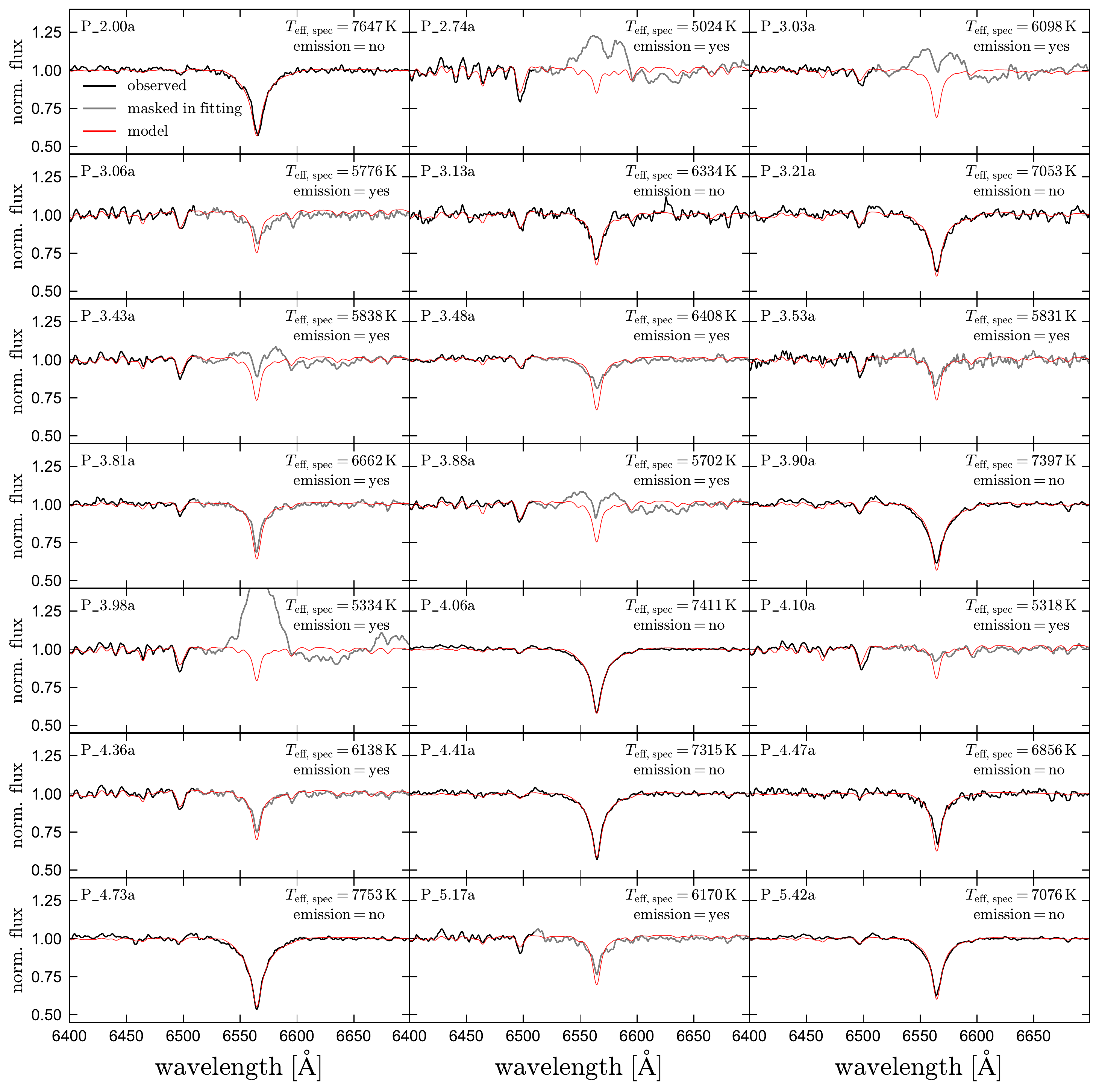}
    \caption{Zoom-in on the H$\alpha$ line for all objects in our spectroscopic sample (full spectra shown in Figure~\ref{fig:spectra}). Objects for which the best-fit Kurucz model is significantly deeper than the observed spectrum likely contain H$\alpha$ emission. The model spectrum is fit to the full wavelength range shown in Figure~\ref{fig:spectra}.}
    \label{fig:halpha}
\end{figure*}

Figure~\ref{fig:halpha} highlights the H$\alpha$ lines of objects in the spectroscopic sample. 

\section{Light curve evolution and scatter}
\label{sec:evolution_long_term}

\begin{figure*}
    \centering
    \includegraphics[width=\textwidth]{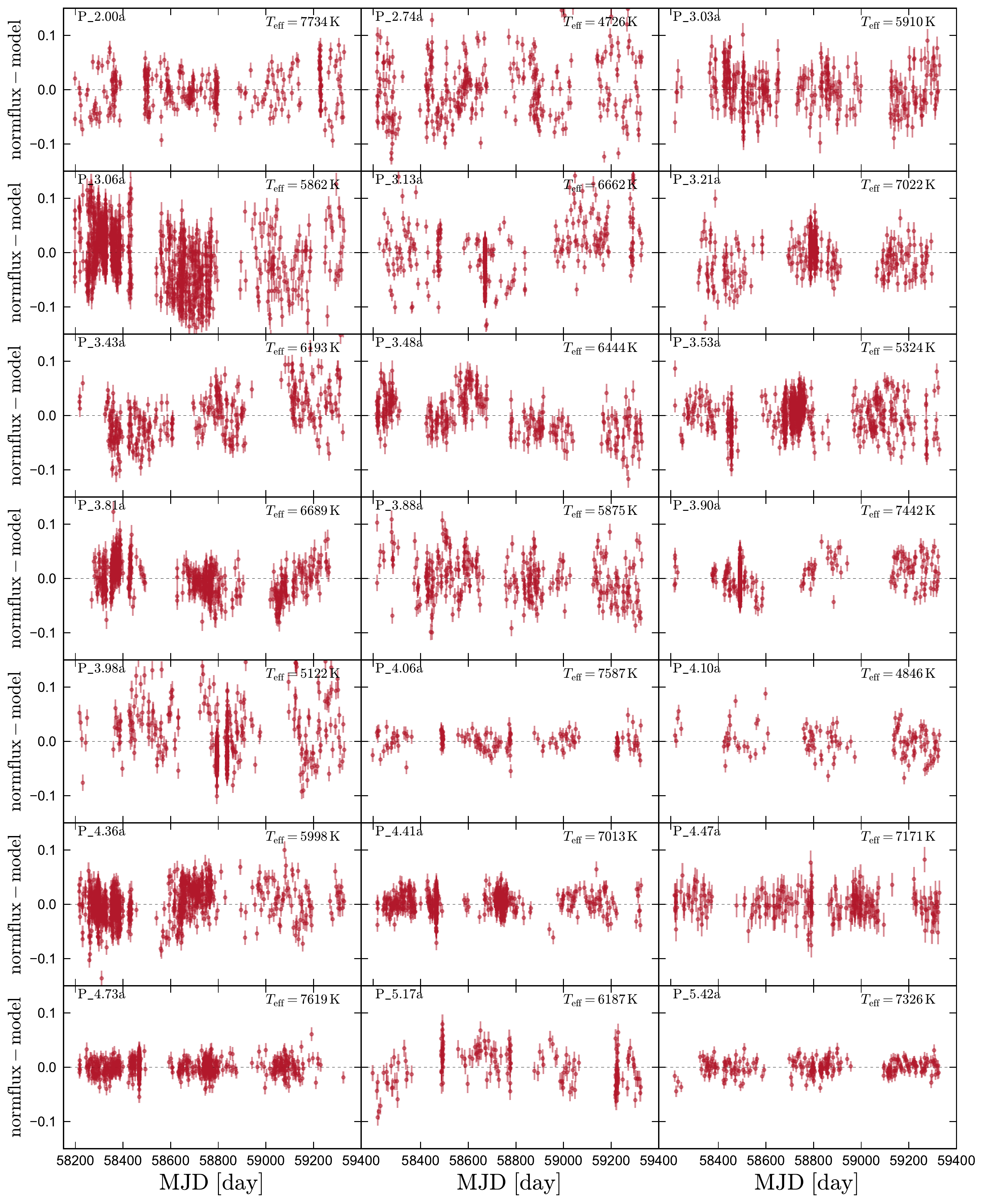}
    \caption{ZTF $r-$band light curves for all objects in the spectroscopic sample, with the mean Fourier model for the full light curve subtracted. About half the light curves (\texttt{P\_3.06a}, \texttt{P\_3.13a}, \texttt{P\_3.43a}, \texttt{P\_3.48a}, \texttt{P\_3.81a}, \texttt{P\_3.98a}, \texttt{P\_4.36a}, \texttt{P\_5.17a}) show evidence of evolution on long (few months to years) timescales, typically with few-percent changes in brightness.}
    \label{fig:long_term_evol}
\end{figure*}

To search for long-timescale brightness evolution, we subtracted a Fourier model for the periodic variability from the ZTF $r-$ band light curves. The resulting flux-space residuals are shown in Figure~\ref{fig:long_term_evol} for all objects in the spectroscopic sample. For many sources, this reveals clear evolution in the phase-averaged mean brightness. The source with the most obvious variations is \texttt{P\_3.06a} (see Figure~\ref{fig:evolution_light_curve}), with a maximum range of $\sim$7\% ($\sim$0.07 mag). Evolution is somewhat more subtle (not accompanied by a clear change in light curve shape) in most other sources, but few-percent variations are common in the sample.

All the sources with clear evolution have donors with $T_{\rm eff}\lesssim 7,000\,\rm K$, consistent with a scenario in which long-term evolution is due to a disk and the hotter donors are detached. There are also several cooler sources that have emission lines and/or eclipses but do not show evolution; these source likely have more stable disks. Two sources (\texttt{P\_3.13a} and \texttt{P\_3.21a}) show evidence of brightness evolution (only tentative for \texttt{P\_3.21a}) but no obvious emission lines. 

\begin{figure*}
    \centering
    \includegraphics[width=\textwidth]{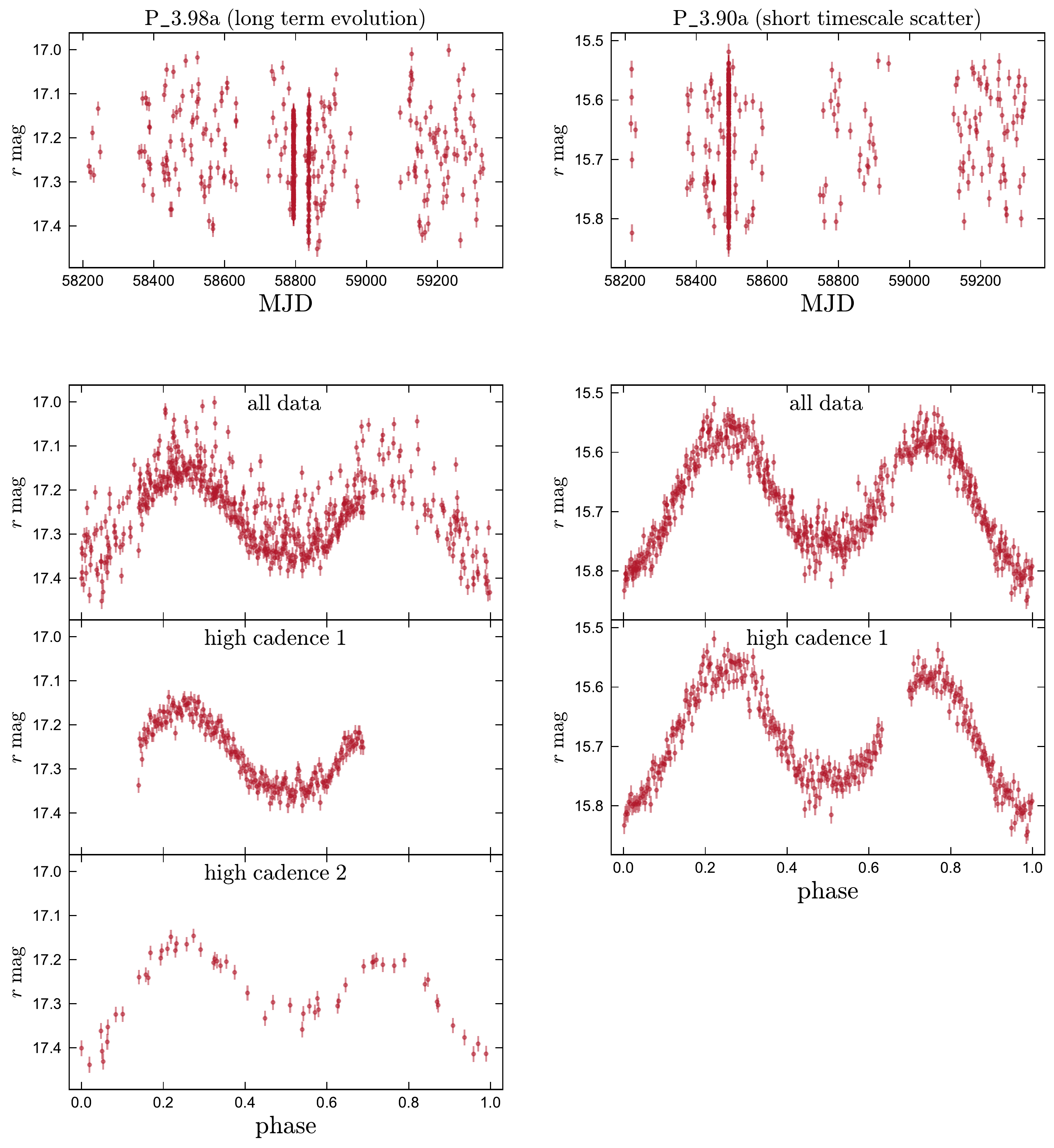}
    \caption{$r-$band light curves for two objects observed by the ZTF high-cadence survey, which provides continuous observations over a  few-hour window. Top panels show the full light curves; the high-cadence windows are evident as vertical stripes with many datapoints. Bottom panels show the full phased light curves, as well as phased light curves of only the data taken at high cadence during a single few-hour window. In \texttt{P\_3.98a} (left), the scatter within the high-cadence windows is much less (though still nonzero) than in the full light curve, suggesting that most of its scatter is due to longer-timescale disk evolution. The high-cadence light curve also shows evidence of disk flickering.  In \texttt{P\_3.90a} (right) the scatter in the  high-cadence window is similar (about 2\%) to that in the full light curve, suggesting it is dominated by short-timescale variations such as pulsations or disk flickering. }
    \label{fig:highcadence}
\end{figure*}

To assess evolution on shorter timescales, we inspected light curves of the 7 sources in the spectroscopic sample with continuous observations over a few-hour period through the ZTF high-cadence survey. Examples for two sources are shown in Figure~\ref{fig:highcadence}. Most sources with significant scatter in their full phased light curves have significantly less scatter when only the high-cadence observations are considered (e.g., \texttt{P\_3.98a} in Figure~\ref{fig:highcadence}). One object, \texttt{P\_3.90a}, has similar scatter in the high-cadence data and in the full light curve, suggesting the excess variability must occur primarily on short timescales.

\section{Light curve-selected targets without spectroscopic follow-up}
\label{sec:nospec_sample}

\begin{figure*}
    \centering
    \includegraphics[width=\textwidth]{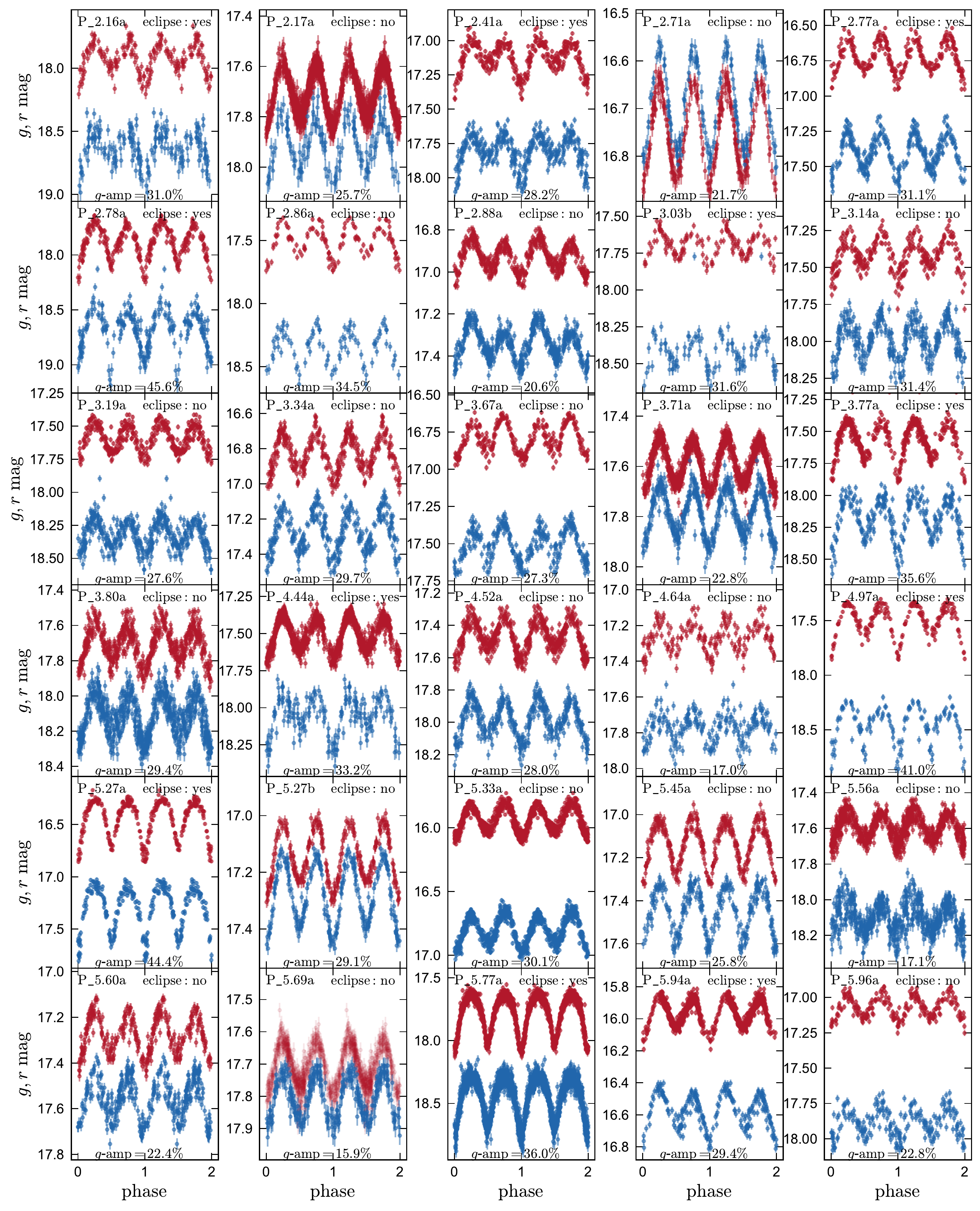}
    \caption{Phased ZTF light curves of the 30 objects identified in our initial sample selection (Section~\ref{sec:sample_selection}) that have not (yet) been followed-up spectroscopically. The sample contains classical ellipsoidal variables, eclipsing systems (presumably with disks), and systems with additional scatter due to long-term evolution and/or pulsations.  }
    \label{fig:other_targets}
\end{figure*}

As described in Section~\ref{sec:sample_selection}, our light curve + CMD selection yielded 51 objects, of which we have thus far spectroscopically followed-up 21. Basic parameters of the objects for which we have not obtained spectra are listed in Table~\ref{tab:nospec}, and their ZTF light curves are shown in Figure~\ref{fig:other_targets}.

\begin{table*}
	\centering
	\caption{Objects without spectroscopic follow-up. Parameters are defined as in Table~\ref{tab:observables}.}
	\label{tab:nospec}
	\begin{tabular}{clllccccccc} 
		\hline
		ID & {\it Gaia} eDR3 ID & $l$ & $b$ &  $G$ & $P_{\rm orb}$ &  $\varpi_{\rm corrected}$ &  $E(g-r)$ & $g$-amplitude & $g$-scatter  & eclipse? \\
		& & [deg] & [deg] & [mag] & [hours] &  [mas] & [mag] & [\%] & [\%]  &  \\
		\hline
\texttt{P\_2.16a} & 6831016738348476800 & 31.196798 & -43.921607 & 17.98 & 2.16 & $1.81 \pm 0.17$ & $0.04 \pm 0.01$ & $31.0 \pm 3.1$ & 6.6 & yes \\
\texttt{P\_2.17a} & 4086789647048435968 & 16.841985 & -8.809333 & 17.62 & 2.17 & $0.92 \pm 0.13$ & $0.15 \pm 0.01$ & $25.7 \pm 1.8$ & 0 & no \\
\texttt{P\_2.41a} & 2443219535337313280 & 96.348334 & -67.601219 & 17.18 & 2.41 & $2.10 \pm 0.09$ & $0.06 \pm 0.02$ & $28.2 \pm 1.5$ & 4.8 & yes \\
\texttt{P\_2.71a} & 3616216816596857984 & 330.929484 & 50.744110 & 16.66 & 2.71 & $0.86 \pm 0.08$ & $0.00 \pm 0.01$ & $21.7 \pm 0.6$ & 0 & no \\
\texttt{P\_2.77a} & 5153771948438163712 & 198.467868 & -59.571062 & 16.85 & 2.77 & $2.46 \pm 0.06$ & $0.02 \pm 0.01$ & $31.1 \pm 1.6$ & 3.7 & yes \\
\texttt{P\_2.78a} & 6909438962819026560 & 41.729647 & -32.275404 & 17.89 & 2.78 & $1.71 \pm 0.15$ & $0.11 \pm 0.01$ & $45.6 \pm 2.3$ & 7.6 & yes \\
\texttt{P\_2.86a} & 3652272306837720448 & 354.915285 & 52.489703 & 17.54 & 2.86 & $2.36 \pm 0.10$ & $0.03 \pm 0.01$ & $34.5 \pm 2.8$ & 3.8 & no \\
\texttt{P\_2.88a} & 2859983434955681536 & 113.362300 & -33.515517 & 16.92 & 2.88 & $1.31 \pm 0.08$ & $0.04 \pm 0.00$ & $20.6 \pm 0.7$ & 1.8 & no \\
\texttt{P\_3.03b} & 4368353340609239936 & 20.456594 & 20.691756 & 17.87 & 3.03 & $1.38 \pm 0.13$ & $0.38 \pm 0.00$ & $31.6 \pm 5.3$ & 9.9 & yes \\
\texttt{P\_3.14a} & 3106121585420422784 & 214.951623 & -1.486974 & 17.49 & 3.14 & $1.48 \pm 0.09$ & $0.10 \pm 0.02$ & $31.4 \pm 2.0$ & 4.9 & no \\
\texttt{P\_3.19a} & 3373007043559875328 & 191.954484 & 4.467368 & 17.62 & 3.19 & $1.37 \pm 0.12$ & $0.16 \pm 0.04$ & $27.6 \pm 1.9$ & 5.9 & no \\
\texttt{P\_3.34a} & 2467074539612904832 & 159.793481 & -65.773407 & 16.87 & 3.34 & $1.59 \pm 0.07$ & $0.00 \pm 0.00$ & $29.7 \pm 1.5$ & 4.2 & no \\
\texttt{P\_3.67a} & 2453564290487329280 & 165.460401 & -73.264722 & 16.97 & 3.67 & $1.82 \pm 0.07$ & $0.00 \pm 0.00$ & $27.3 \pm 1.2$ & 3.4 & no \\
\texttt{P\_3.71a} & 1812627897159845632 & 62.870714 & -13.309963 & 17.56 & 3.71 & $0.72 \pm 0.10$ & $0.09 \pm 0.01$ & $22.8 \pm 0.8$ & 2.0 & no \\
\texttt{P\_3.77a} & 4223953344655923584 & 41.618275 & -19.079074 & 17.53 & 3.77 & $0.89 \pm 0.12$ & $0.15 \pm 0.01$ & $35.6 \pm 7.1$ & 31.6 & yes \\
\texttt{P\_3.80a} & 1715477184326927744 & 117.920994 & 36.733613 & 17.87 & 3.80 & $0.72 \pm 0.08$ & $0.06 \pm 0.01$ & $29.4 \pm 1.2$ & 4.6 & no \\
\texttt{P\_4.44a} & 5717210604722373248 & 234.331125 & 2.454777 & 17.63 & 4.44 & $0.88 \pm 0.09$ & $0.08 \pm 0.01$ & $33.2 \pm 2.3$ & 4.8 & yes \\
\texttt{P\_4.52a} & 3325744016547861888 & 201.735824 & -2.770070 & 17.50 & 4.52 & $1.08 \pm 0.10$ & $0.14 \pm 0.01$ & $28.0 \pm 1.2$ & 3.2 & no \\
\texttt{P\_4.64a} & 3910299549545984000 & 257.535495 & 62.412238 & 17.26 & 4.64 & $1.08 \pm 0.11$ & $0.03 \pm 0.01$ & $17.0 \pm 2.0$ & 4.7 & no \\
\texttt{P\_4.97a} & 1763012778552710528 & 61.499213 & -16.496899 & 17.60 & 4.97 & $1.49 \pm 0.11$ & $0.06 \pm 0.01$ & $41.0 \pm 3.3$ & 6.5 & yes \\
\texttt{P\_5.27a} & 2477944857025864192 & 150.912529 & -67.405362 & 16.67 & 5.27 & $1.99 \pm 0.37$ & $0.00 \pm 0.00$ & $44.4 \pm 1.2$ & 5.9 & yes \\
\texttt{P\_5.27b} & 2652639593773429120 & 65.931043 & -49.944372 & 17.14 & 5.27 & $0.72 \pm 0.09$ & $0.07 \pm 0.01$ & $29.1 \pm 0.7$ & 0 & no \\
\texttt{P\_5.33a} & 2184121338243194624 & 93.438488 & 8.305524 & 16.02 & 5.33 & $2.67 \pm 0.03$ & $0.04 \pm 0.01$ & $30.1 \pm 0.6$ & 2.8 & no \\
\texttt{P\_5.45a} & 3743862209546227584 & 327.579445 & 76.471956 & 17.14 & 5.45 & $0.47 \pm 0.07$ & $0.00 \pm 0.01$ & $25.8 \pm 1.0$ & 2.4 & no \\
\texttt{P\_5.56a} & 997691559644734080 & 159.807366 & 19.137537 & 17.71 & 5.56 & $0.87 \pm 0.11$ & $0.11 \pm 0.01$ & $17.1 \pm 1.3$ & 4.7 & no \\
\texttt{P\_5.60a} & 3740297730288413056 & 344.568767 & 71.347303 & 17.28 & 5.60 & $0.52 \pm 0.09$ & $0.06 \pm 0.04$ & $22.4 \pm 1.2$ & 3.5 & no \\
\texttt{P\_5.69a} & 4241883875351568128 & 42.892677 & -12.409634 & 17.66 & 5.69 & $0.54 \pm 0.09$ & $0.14 \pm 0.01$ & $15.9 \pm 0.8$ & 0 & no \\
\texttt{P\_5.77a} & 1419776515721297536 & 80.834579 & 37.177558 & 17.86 & 5.77 & $0.88 \pm 0.11$ & $0.07 \pm 0.01$ & $36.0 \pm 0.7$ & 4.3 & yes \\
\texttt{P\_5.94a} & 3035296754374910976 & 226.832469 & 3.810814 & 15.99 & 5.94 & $1.51 \pm 0.05$ & $0.05 \pm 0.02$ & $29.4 \pm 1.3$ & 2.5 & yes \\
\texttt{P\_5.96a} & 3874949971539651840 & 231.629329 & 47.310078 & 17.14 & 5.96 & $1.39 \pm 0.10$ & $0.00 \pm 0.01$ & $22.8 \pm 1.9$ & 4.9 & no \\
		\hline
	\end{tabular}
\end{table*}

\section{Overlap with other surveys}
\label{sec:overlap}

We cross-matched the 51 light curve-selected candidates with the final sample of low-mass white dwarfs from the ELM survey \citep{Brown2020}. This yielded only one object in common, \texttt{P\_2.71a} (J1401-0817 in their catalog). Unlike essentially all other objects discovered by the ELM survey, it has a light curve dominated by large-amplitude ellipsoidal variability, with peak-to-peak amplitude of 28\% (see light curve in Appendix~\ref{sec:nospec_sample}). This suggests that mass transfer only very recently ceased.
It is immediately evident from Figure~\ref{fig:other_targets} that the object is hotter than any of the other targets in our sample, because it is the only one with $g - r < 0$. This explains why most of our targets were not observed by the ELM survey: they are too cool to fall within its color cuts \citep{Brown2012}. 

We also cross-matched our sample with the 5,762 ELM WD candidates identified by \citet{Pelisoli2019}. Here too, \texttt{P\_2.71a} was the only object that appeared in both samples. This is probably also a consequence of the CMD cuts \citet{Pelisoli2019} employed, which most of our objects are too red to pass.

We also searched SIMBAD \citep{Wenger2000} for previous studies of all 51 candidates. Within our spectroscopic sample, this yielded two matches: the object \texttt{P\_2.74a} was classified as an eclipsing binary by the LINEAR survey \citep{Palaversa2013}, and \texttt{P\_3.48a} was (incorrectly) classified as a Delta Scuti variable by the Catalina survey \citep{Drake2014}.

Several objects in the light curve selected sample without spectroscopic follow-up also had previous classifications: \texttt{P\_2.71a} (the same object observed by the ELM survey) was spectroscopically classified as low-mass WD based on its SDSS spectrum by \citet{Kepler2015}. \texttt{P\_2.41a} was classified as an evolved CV, consistent with our classification, by \citet{Rebassa_2014}, and was studied in detail. Six objects were classified as eclipsing, contact, or ellipsoidal binaries by the Catalina survey  \citep{Drake2014, Marsh2017}; these are \texttt{P\_5.27a}, \texttt{P\_5.27b}, \texttt{P\_5.45a}, \texttt{P\_5.60a}, \texttt{P\_5.77a}, and \texttt{P\_5.96a}. Given their CMD position, we suspect they are all evolved CVs.

\subsection{Known evolved CVs not in our sample}
To better understand the selection function of our sample, we also investigated the known evolved CVs that are not among the 51 light curve-selected candidates. We considered all previously known systems in Figure~\ref{fig:Porb_vs_teff} with $T_{\rm eff,\,donor} > 4000\,\rm K$. Of the 11 such systems, only SDSS J001153.08-064739.2 \citep{Rebassa_2014} is in our sample (as  \texttt{P\_2.41a}). Two systems, BF Eri and HS 0218+3229, have orbital periods $P_{\rm orb} > 6$ hours, longer than our sample limit. Two systems, V1460 Her and ASAS-SN 13cl, fall outside our selection region in the CMD, as they are closer to the main sequence than the upper limit of our selection region. One system, ASAS-SN 15cm, is fainter than our magnitude limit of $G<18$, and another, ASAS-SN 14ho, is farther south than our declination limit of $\delta > -28$ deg. One object, CSS J134052.0 + 151341, has ellipsoidal variability amplitude less than our limit of $>15\%$, likely because it is almost face-on (\citealt{Thorstensen_2013} found an inclination of 27 degrees). The last two systems, QZ Ser and EI Psc, were rejected in our light curve search because their ZTF light curves are not dominated by ellipsoidal variation, but show complex and irregular variability. This is likely a consequence of significant changes in the structure of their disks over the $\sim 3$ year ZTF baseline, since their light curves do show ellipsoidal variability over shorter timescales \citep{Thorstensen_2002b, Thorstensen_2002}. 

We conclude that while our search strategy should not be expected to find every CV with an evolved donor, it has not missed known objects that we would have expected it to detect. The sensitivity of our search is lowest for objects with high accretion rates, in which the disk contributes significantly to the optical photometry, outbursts are more common, and the light curve is more complicated than expected for simple ellipsoidal variability. At the orbital periods represented in our survey, the accretion rate and light contribution of the disk are expected to be lowest for the most evolved donors \citep[e.g.][]{ElBadry2021}. This likely explains why most of the systems in our sample are warmer than the previously known systems not included in it (e.g. Figure~\ref{fig:Porb_vs_teff}).


\bsp	
\label{lastpage}
\end{document}